\documentclass[12pt]{article}
\usepackage[margin=2.5cm]{geometry}  % add also "show frame" options  
\usepackage{enumerate}
\usepackage{float}
\usepackage{bm}
\usepackage{graphicx}
\usepackage{amsmath}
\usepackage{amssymb}
\usepackage{graphicx}
\usepackage{subcaption}
\usepackage{natbib}
\usepackage{comment}
\usepackage[table]{xcolor}  
\usepackage{hyperref}
\hypersetup{
	colorlinks=true,
	citecolor=blue,
}

\usepackage{rotating}
\usepackage{epsfig}%

\newcommand{\nin}{\noindent}

\def\bkR{{\rm I\kern-.17em R}}

\def \1n{1\hskip -3pt \mbox{N}}

\def \Frac {\displaystyle \frac }

\newfont{\bbf}{cmbx12 scaled 1435}
\begin{document}

\setlength{\baselineskip}{.26in}
\thispagestyle{empty}
\renewcommand{\thefootnote}{\fnsymbol{footnote}}
\vspace*{0cm}
\begin{center}

\setlength{\baselineskip}{.32in}
{\bbf The Online LSTM Prediction Models and The Term Structure of Predictions 
}\\

\vspace{0.5in}

\large{Christian Gourieroux}\footnote{University of Toronto, Canada, Toulouse School of Economics and CREST, France, email: {\it Christian.Gourieroux@ensae.fr}}, 
\large{Joann Jasiak}\footnote{York University, Canada, e-mail: {\it jasiakj@yorku.ca}}, 
\large{Cheng Zhong }\footnote{Concordia University, Canada, e-mail: {\it c\_zhong2@live.concordia.ca} \\ 
The first author gratefully acknowledges financial support of the ACPR Chair "Regulation and Systemic Risks" and the ERC DYSMOIA; the 2nd and 3rd authors thank the Natural Sciences and Engineering Research Council of Canada (NSERC: RGPIN-2024-06211 and RGPIN-2021-04144) and Social Sciences and Humanities Research Council (SSHRC: 430-2024-01325).}

\setlength{\baselineskip}{.26in}
\vspace{0.4in}

\today\\

\medskip

\vspace{0.3in}
\begin{minipage}[t]{12cm}
\small
\begin{center}
Abstract \\
\end{center}
We examine and illustrate the properties of the Long Short Term Memory (LSTM) prediction models applied online to stationary inputs. We discuss the effects of: a) the network architecture that defines the admissible prediction functions, b) the objective function used to measure the prediction accuracy, and c) the Stochastic Gradient Descent (SGD) algorithm. We also study the prediction optimality, the term structure of predictions, the filtered memory dynamic. We also introduce a set of diagnostic tools for memory-based prediction models that focus on the estimated dynamics of the hidden states.

\bigskip

\bigskip


\end{minipage}

\end{center}
\renewcommand{\thefootnote}{\arabic{footnote}}
\newpage
\section{Introduction}

This paper examines and illustrates the properties of Neural Network (NN)-based prediction models applied to stationary time series with an emphasis on the Long Short Term Memory (LSTM) prediction models\footnote{The LSTM has initially been introduced for nonstationary inputs: "(We) focus on time-varying inputs as opposed to stationary inputs" [\citet{HochreiterSepp1997LSM}, Section 2].}. 
 We discuss the effects of: a) the network architecture that defines the admissible prediction functions, b) the objective function used to measure the prediction accuracy, and c) the Stochastic Gradient Descent (SGD) algorithm. 
 We study the prediction optimality, the term structure of predictions, and the filtered memory dynamics. We also introduce a set of diagnostic tools for memory-based prediction models that focus on the estimated dynamics of the hidden states.

The analysis of the prediction model is performed in the context of economic and financial time series, and does not necessarily concern applications to image analysis, music, speech recognition, text translations, etc. The economic and financial time series are often characterized by a limited number of observations, depending on the application. For example, univariate macroeconomic data can consist of hundreds of observations, but financial data can comprise thousands or even hundreds of thousands of observations on intraday stock prices, for example.
The economic and financial prediction models may have a structural interpretation, while the NN prediction models are mainly descriptive.

The objective of the paper is to discuss the LSTM-based prediction optimality and to introduce diagnostic tools to help evaluate the performance of these prediction models for stationary time series. We define the LSTM-prediction model and describe its three components: the LSTM unit, the objective function and the SGD algorithm. Regarding the predictive performance of the LSTM, we define the associated sets of admissible prediction functions and point out the sub-optimality of predictions. Next, we derive the term structure of such sub-optimal predictions. We consider two alternative approaches. The first approach computes the LSTM-based predictions in parallel, i.e., horizon by horizon, using either task-specific or multitask prediction criteria. These predictions are shown to violate the Law of Iterated Expectations (LIE). The second approach is based on consecutive one-step predictions and its limitations depend on the model of interest. 

The properties of the LSTM filters are explored based on a set of stationary Data Generating Processes (DGP) to show if the filters can capture the long memory effects and which hidden state variables are able to do so. 
The focus is on processes commonly used in economics and finance, including the ARMA and GARCH processes.
We study the basic LSTM with possible parallel and recursive set-ups and their adequacy for predicting processes such as ARMA and GARCH. We reveal several weaknesses of the LSTM-based prediction approaches, such as violating the positivity of the conditional volatility and the Law of Iterated Expectations in the term structure of predictions. We also discuss the interpretation of the hidden states $c_t, h_t$  as long memory factors. It appears that, depending on the DGP, these hidden states can capture the memory of the process only up to a short lag. The results depend on the selected objective function,
and automatic aggregation of objective functions (multi-task objective) does not necessarily provide a good solution.
In fact, the lack of robustness of this class of prediction models with respect to the DGP is difficult to examine theoretically. Indeed, the results also depend on the learning algorithm and numerical and statistical consistency properties that are uncertain to hold because of the dynamics of the input series, the lack of identification of the parameters, or the architecture and the objective function.

The paper is organized as follows. Section 2 describes the LSTM prediction model. Section 3 considers the application to the term structure of predictions, distinguishes the direct/recursive, single/multistep approaches, and points out the sub-optimality of predictions and the violation of the LIE. Section 4 explores and illustrates by simulations the properties of the LSTM filters and introduces a set of diagnostic tools. Section 5 concludes. Appendix A describes the recursive LSTM+SGD filter. Additional figures and diagnostic tools are given in Appendix B.

\setcounter{equation}{0}

\section{LSTM-based Prediction Models}

As pointed out in the Introduction, the neural networks predictions are based on three components: i) a parametric prediction model, ii) an objective function measuring the accuracy of the prediction, and iii) a numerical algorithm used to optimize that objective function.

There exists a multiplicity of possible choices of these components, some implemented in the LSTM and others exploited in neural networks such as the GRU (Gated Recurrent Unit) [\citet{ChoEtAl2014RNNEncoderDecoder}]. In this Section, we consider a specific choice of LSTM prediction model, and then discuss the mean square prediction error at horizon 1 and higher used to calculate the objective function, and an online SGD algorithm for numerical stochastic optimization. This allows us to illustrate the complete architecture of the components and their relationships and to reveal the (sub) "optimality" of the parameterized set of prediction functions.

\subsection{The LSTM prediction model}

The LSTM prediction model (or LSTM filter, LSTM unit) has been introduced in \citet{HochreiterSepp1997LSM} to get theoretical prediction functions that depend on the whole past of the observed series, rather than on a finite past. It includes deterministic transition equations that relate a set of state (latent, i.e. hidden) variables (also called neurons) to the observable series plus an additional equation explaining how the prediction of interest depends on these variables.

The state equations are given in equations (\ref{(1)}) to (\ref{(6)}) below for one LSTM filter\footnote{In some applications several LSTM filters can be used in parallel or recursively [see e.g. Sections 3 and 4]}:

%A cell state $c_t$ is introduced to store long-term memory through gate mechanisms . The system equation of one LSTM neuron can be written as follows :

\begin{equation}
    f_t = \sigma(W_{f_1}\cdot y_t + W_{f_2}\cdot h_{t-1} + b_f),
    \label{(1)}
\end{equation}

\begin{equation}
    i_t = \sigma(W_{i_1}\cdot y_t + W_{i_2} \cdot h_{t-1} + b_c),
    \label{(2)}
\end{equation}

\begin{equation}
    \tilde{c}_t = tanh(W_{c_1}\cdot y_t + W_{c_2}\cdot h_{t-1} + b_c),
    \label{(3)}
\end{equation}

\begin{equation}
    c_t = c_{t-1}\odot f_t + \tilde{c}_t \odot i_t,
    \label{(4)}
\end{equation}

\begin{equation}
    o_t = \sigma(W_{o_1} \cdot y_t+ W_{o_2}h_{t-1} + b_o),
    \label{(5)}
\end{equation}

\begin{equation}
    h_t = tanh(c_t)\odot o_t.
    \label{(6)}
\end{equation}

%(\ref{var})

These equations are completed by an output equation providing the prediction of $y_{t+1}$ computed at date $t$\footnote{The equation (2.7) has to be distinguished from equation $ y_{t+1} \sim \mathcal{L} (W_y h_t + b_y)$, where $y_{t+1}$ is drawn in a distribution $\mathcal{L}$ with time-varying parameters $W_y h_t + b_y$, introduced in generative LSTM models to define the dynamic of process ($y_t$) [see, e.g.  \citet{ChoEtAl2014RNNEncoderDecoder}, Section 2, \citet{Graves2014GeneratingSequences}, output layer equation (4) for mixture density modeling].}:

\begin{equation}
    \hat{y}_{t,t+1} = W_y\cdot h_{t} + b_0,
\end{equation}

\nin where  $\sigma(.)$ and $tanh(.)$ represent the logistic activation function, and the hyperbolic tangent activation function (in vector forms), respectively, and $\odot$ is the element-wise (Hadamard) product\footnote{For example, $y =\sigma(z)$ denotes $y_i = \sigma(z_i) = 1/[1+ \exp(-z_i)], \; i=1,...,n$, $y=tanh(z)$ denotes $y_i = tanh(z_i) = [\exp(z_i) - \exp(-z_i)]/[\exp(z_i) + \exp(-z_i)], \, i=1,...,n$.}. 

The choice of these activation functions introduces nonlinear dynamic features while keeping the state variables $f_t, i_t, o_t$ bounded between 0 and 1 (resp. between -1 and 1 for $\tilde{c}_t$, $h_t$, and -2 and 2 for $c_t$). 
Other nonlinearities are introduced by means of the operator $\odot$ and correspond to possible pairwise effects of the state variables.

The symbols of the state variables correspond to terms such as the forget gate for $f_t$, and the input gate for $i_t$. However, these names have no real structural interpretation and are often misleading. Typically, the input at date $t$ is $y_t$. In fact, there are two types of state variables $z_t=(f_t', i_t', \tilde{c}_t', o_t')'$ that correspond to the neurons in the first intermediate layer, and $h_t, c_t$ that are used to transfer the information from date $t-1$ to date $t$ as shown in the propagation scheme in Figure \ref{fig:lstm_time}, which highlights the role of the LSTM unit.

\begin{center}

\begin{figure}[H]
    \centering
        \caption{Propagation scheme in predictive LSTM}

    \label{fig:lstm_time}
\end{figure}

\vspace{-0.5in}

\end{center}

\hspace{2.1in} $y_t$ input \hspace{1in} $y_{t+1}$ input

\hspace{2.3in} $\downarrow$ \hspace{1.7in} $\downarrow$ 

$$ \rightarrow \left( \begin{array}{c} h_{t-1}(\theta) \\ c_{t-1}(\theta) \end{array} \right) \rightarrow \mbox{LSTM}(\theta) \rightarrow \left( \begin{array}{c} h_{t}(\theta) \\ c_{t}(\theta) \end{array} \right) \rightarrow \mbox{LSTM}(\theta) \rightarrow$$

\hspace{2.3in} $\downarrow$ \hspace{1.7in} $\downarrow$ 

\hspace{2.0in} $\hat{y}_{t,t+1}(\theta)$ output \hspace{0.7in} $\hat{y}_{t+1,t+2}$ output

\bigskip
 By including memory states $h_t, c_t$ to transfer the information, the LSTM unit extends the Recurrent Neural Network (RNN) (see Appendix A.1). At date $t$, we obtain the internal propagation scheme illustrated in Figure \ref{fig:lstm_cell}.

\begin{figure}[H]
    \centering
        \caption{Within LSTM propagation scheme}
    \includegraphics[width=1\linewidth]{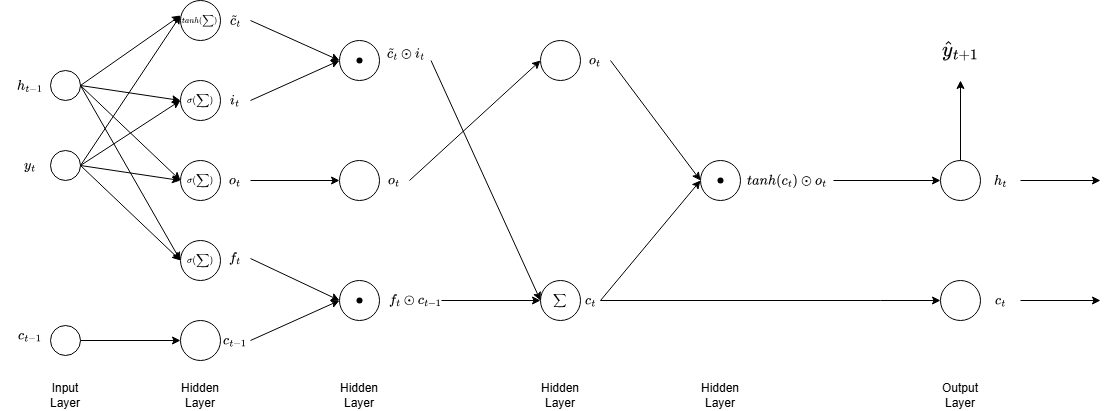}

    \label{fig:lstm_cell}
\end{figure}

The NN prediction models are often (although not necessarily) used in high-dimensional framework with a high-dimensional $y$, high-dimensional state variables, and a large number of parameters. 
Let us denote by $n=dim(y_t)$, and by $m$ the dimension of the hidden states $h$ and $c$. We have:  $dim(h_t)= dim(c_t)= m=dim(i_t)=dim(\tilde{c}_t)= dim(f_t)$, $p=dim(o_t)$. The constraints on the dimensions are due to the layers (\ref{(4)}) and (\ref{(6)}).  Denote by $\theta$ the set of all parameters, including the weights $W$ and biases $b$. The terms "weight" and "bias" have a different meaning than in inferential statistics, as the weights $W$ are not necessarily positive and do not necessarily sum up to 1, and the "intercept" parameters $b$ are not intended for bias adjustment (except for the output bias $b_y$ in our LSTM application). Then, we have:

\begin{equation}
dim(\theta) =  4m(n + m +1) +nm+n.
\label{(8)}
\end{equation}

\medskip
\nin 
Let us now consider the variables $z_t$ that are neither inputs nor outputs, nor memory states:
$z_t(\theta) = [ f_t'(\theta), i_t(\theta)', \tilde{c}(\theta)', o_t(\theta)']$. They appear in the LSTM filter in addition to 
$y_t, c_t(\theta), h_t(\theta), \hat{y}_{t,t+1}(\theta)$. Then, the LSTM filter at date $t$ can be summarized as:

\begin{equation} \left\{    
\begin{array}{ccl} 
z_t(\theta) & = & a(y_t, h_{t-1}(\theta)), \\
 h_t(\theta) & = & b(c_{t}(\theta), z_{t}(\theta)), \\
  c_t(\theta) & = & c(c_{t-1}(\theta), z_{t}(\theta)), \\
  \hat{y}_{t,t+1}(\theta) & = & d(h_t(\theta)), \end{array}
\right.,
\label{(9)}
\end{equation}

\nin with initial conditions $h_0, c_0$ for memory states $h$ and $c$.

This sequence can be solved recursively from date 1 to date $t$ to provide a set of prediction functions depending on the observations up to time $t$: $\underline{y_t} =(y_1,...,y_t)$, on the initial values $h_0, c_0$ and on the parameters in $\theta$. They are denoted as follows:

\begin{equation}
\tilde{z}_t( \underline{y_t}, \theta, h_0, c_0), \tilde{h}_t(\underline{y_t}, \theta, h_0, c_0), \tilde{c}_t(\underline{y_t}, \theta, h_0, c_0), \tilde{y}_{t,t+1}(\underline{y_t}, \theta, h_0, c_0).
\end{equation}

\nin For ease of exposition, we omit the symbol $\underline{y_t}$ from now on.

\nin We deduce the following definition:

\medskip
{\bf Definition 1:} The LSTM prediction model (2.1)-(2.9) defines a parametric set of prediction functions parametrized by $\theta, h_0, c_0$.

\medskip
In economic and financial applications, the dimensions $m, n, p$ as well as the number of LSTM cells introduced in more complex parallel and/or recursive architectures are rather small. Then, we do not invoke the universal approximation of neural networks from an infinite number of layers used for nonparametric analysis instead of parametric analysis [\citet{Kolmogorov1956Representation}, \citet{White1992ArtificialNeuralNetworks}, \citet{Sprecher1993UniversalMapping}]. Moreover, this restricted
set of prediction functions has been fixed without taking into account the true dynamics of process $(y_t)$ and does not include, in general, the optimal prediction function that corresponds to $E_0(Y_{t+1}|\underline{Y_t})$, where $E_0(.|.)$ denotes the true conditional expectation. It is called a "fuzzy" model in \citet{Graves2014GeneratingSequences}.
However, we can try to choose a function within this set, i.e., search for some sub-optimal predictions.

\subsection{Sub-Optimal Prediction}

As shown above, $\hat{y}_{t,t+1} (\theta, c_0, h_0)$ provides a restricted class of nonlinear prediction functions with parameter $\theta$, given the initial conditions $c_0, h_0$. Such a class of nonlinear prediction functions is provided for any date $t$. Then, at each date $t$, one can choose a value of $\theta$ that minimizes with respect to $\theta$ the predictive accuracy measure along with the sequence of predictions $\hat{y}_{t,t+1}, t=1,2,...$. 

More precisely, optimal prediction functions are obtained for $\theta^*(c_0, h_0)$ such that:

\begin{equation}
\theta_t^*(c_0, h_0) = Argmin_{\theta} E_0 [ Y_{t+1} - \hat{Y}_{t,t+1} (\theta, c_0, h_0)| \underline{Y_t}]^2,
\label{(10)}
\end{equation}

\nin where $E_0(.|.)$ is the conditional expectation under the true distribution of process $(Y_t)$. However, the true conditional density is unknown. We do not know the form of the optimal prediction function or its compatibility with the predictive LSTM.

To proceed, we can change the prediction criterion  $E_0 [ Y_{t+1} - \hat{Y}_{t,t+1} (\theta, c_0, h_0)| \underline{Y_t}]^2$ by replacing it by, for example, the average predictive accuracy measure (under the implicit assumption of strict stationarity of $(Y_{t+1},
\hat{Y}_{t,t+1})'$. Since:

\begin{equation}
E_0 [ Y_{t+1} - \hat{Y}_{t,t+1} (\theta, c_0, h_0)]^2 = \lim_{T \rightarrow \infty} \frac{1}{T} \sum_{t=0}^{T-1} [ Y_{t+1} - \hat{Y}_{t,t+1} (\theta, c_0, h_0)]^2,
\label{(11)}
\end{equation}

\nin we can replace the theoretical criterion by the historical criterion: $\frac{1}{T}\sum_{t=1}^T [ y_{t} - \hat{y}_{t-1,t} (\theta, c_0, h_0)]^2$, i.e., the Mean Squared Error (MSE). 

However, this optimization is intractable and the solution has no closed form. It is commonly done by a type of Stochastic Gradient Descent (SGD) algorithm that leads to a sequence of values $\theta_t^*(\underline{y_t}, c_0, h_0, \theta_0^*, \alpha, \beta)$ for varying $t$ that depends on an additional starting value $\theta_0^*$ for parameter $\theta$ and other tuning parameters $\alpha, \beta$ driving the learning rate of the SGD algorithm.

Then, the estimated (sub)optimal sequence of prediction functions becomes:
\begin{equation}
\hat{y}_{t,t+1} (\underline{y_t}, \theta_0^*, c_0, h_0, \alpha, \beta) = \tilde{y}_{t,t+1} (\underline{y_t}, \theta_t^*(\underline{y_t}, c_0, h_0,  \theta_0^*, \alpha, \beta) , c_0, h_0), \; \forall t,
\label{(12)}
\end{equation}

\nin in which the control parameters of the two underlying recursive systems have been kept fixed. This leads to a new set of prediction functions depending on $\underline{y_t}=(y_1,...,y_t)$ now parametrized by the control parameters. We will discuss their choice in  Sections 3 and 4. The degree of sub-optimality is difficult to evaluate or even to bound in practice. It inherently depends on the architecture, objective function, learning algorithm and the data generating process [see \citet{SchmidtHieber2020NonparametricRegression} for an analysis in an idealized framework, and Appendix A.3]. 

Let us now describe a typical algorithm for the online derivation of $\theta_t^*$.

\subsection{Online-SGD Applied to LSTM Prediction Model}

Let us now describe the recursive equations of the basic online SGD for $\theta$ when the predictive objective function is:
\begin{equation}
\sum_{t=1}^T [y_t - \hat{y}_{t-1,t} (\theta, c_0, h_0)]^2 \equiv \sum_{t=1}^T l(y_t, \underline{y_{t-1}}, \theta, c_0, h_0).
\label{(13)}
\end{equation}

\nin The recursive equations are:

\begin{equation}
\theta_t^* =  \theta_{t-1}^* + \delta_t \Frac{\partial l (y_t, \underline{y_{t-1}}, \theta^*_{t-1}, c_0, h_0)}{\partial \theta},
\label{(14)}
\end{equation}

\nin with initial condition $\theta_0^*$, where $\delta_t$ is a learning rate of the form $\delta_t = \alpha t ^{- \beta}$ with $\alpha >0$, and $ \beta \in (0.5, 1)$. This form of learning rate ensures the statistical convergence of $\theta_t^*$ when $t \rightarrow \infty$ [\citet{benveniste1987algorithmes,benveniste2012adaptive}, \citet{polyak1992acceleration}].

Below, we present two online filters at date $t$ with either a fixed or online updated parameter $\theta$ and fixed tuning parameters. We start from a LSTM filter used to predict at time $t$ the process $(y_t)$ at horizon 1.

i) In the prediction filter with a given parameter $\theta$, the LSTM model is used recursively. At a date $t$, we start from the initial memory states $h_{t-1}(\theta), c_{t-1}(\theta)$, which are followed by the computations of latent variables $z_{t}(\theta)$ based on the observed values of $y_{t}$, and then of  $h_{t}(\theta), c_{t}(\theta)$. The filter $F_1(\theta)$ is illustrated in Figure~\ref{fig:online_fixed_filter} and provides the prediction $\hat{y}_{t,t+1}(\theta), \forall t$. This prediction is infeasible since the true value of $\theta$ is unknown.

ii) The approach i) is not implemented when $\theta$ is not fixed and instead an online training of $\theta$ is performed. Then, both the estimation and prediction filters are used jointly to update $h_{t},  c_{t}, \tilde{\theta}_{t}$ with the SGD that optimizes the objective function $\sum_{t=1}^T l(y_{t}, \underline{y_{t-1}}, \theta, h_0, c_0)$ with respect to $\theta$. The propagation algorithm is modified and becomes an online filter $F_1$:

 \newpage
\begin{center}
\captionsetup{hypcap=false}
\captionof{figure}{Online filter $F_1(\theta)$}\label{fig:online_fixed_filter}
\end{center}
\begin{center} $y_{t}$ \\
$\downarrow$ 
\end{center}
\begin{equation}
 \rightarrow \left( \begin{array}{c} h_{t-1}(\theta) \\ c_{t-1}(\theta) \end{array} \right) \rightarrow  \left[ \begin{array}{c} LSTM(\theta) \; \mbox{equations} \; (\ref{(1)})-(2.7) \\  \end{array} \right] \rightarrow \left( \begin{array}{c} h_{t}(\theta) \\ c_{t}(\theta) \end{array} \right) 
\nonumber
 \end{equation}
 \begin{center} $\downarrow$ \\
 $\hat{y}_{t,t+1}(\theta)$ \\
\end{center}

\medskip

\medskip

\begin{center}
\captionsetup{hypcap=false}
\captionof{figure}{Online joint filter $F_1$}\label{fig:online_joint_filter}
\end{center}
\vspace{-0.5in}
\begin{center} $y_{t}$ \\
$\downarrow$ 
\end{center}
\begin{equation}
 \rightarrow \left( \begin{array}{c} h_{t-1} \\ c_{t-1} \\ \tilde{\theta}_{t-1} \end{array} \right) \rightarrow  \left[ \begin{array}{c} LSTM \; \mbox{equations} \; (\ref{(1)})-(2.7) \\ SGD \; \mbox{equation} \; (2.14) \\ \end{array} \right] \rightarrow \left( \begin{array}{c} h_{t} \\ c_{t} \\ \tilde{\theta}_{t} \end{array} \right) \nonumber \end{equation}

\medskip
\nin The differences between filters $F_1(\theta)$  and $F_1$ in Figures~\ref{fig:online_fixed_filter} and~\ref{fig:online_joint_filter} have not been discussed in the theoretical or applied literature on deterministic nonlinear filters
applied to time series. This might have caused a confusing interpretation of the results.

\subsection{Discussion}

When the joint filter $F_1$ is used, we get predictions at horizon 1 that  depend on the tuning parameters that include the initial values $h_0, c_0, \theta_0^*$ and the parameters $\alpha, \beta$ of the learning rate. It is known that the prediction results can significantly depend on these tuning parameters [see e.g. \citet{sutskever2013importance} for the dependence with respect to initial values].

It would theoretically be possible to fix the tuning parameters by performing a second optimization with respect to the tuning parameters\footnote{This would require setting additional starting values for the tuning parameters, and so on, which would complicate the analysis even further.}.
In practice, for numerical reasons and/or to avoid overfitting, the choice of the tuning parameters is often done ex-ante and/or by a cross-validation. For example, the starting values $h_0, c_0$ are generally set to zero, whereas the tuning parameters in the SGD algorithm can be fixed by cross-validation, or from a sequence of epochs [see Appendix A.3]. Theoretically, there are no clear rules or intuition on how to fix the starting values.

\setcounter{equation}{0}

\section{Term Structure of Predictions (multi-step forecasts)}

\subsection{Direct and recursive approaches}
The online prediction approach can be applied for predicting at date $t$ for any horizon $k$, i.e., to compute the term structure of predictions, which is a curve in $k$ for univariate time series. However, it is known that the optimal term structure is constrained by the Law of Iterated Expectations (LIE). Let us examine if it is possible to make the online approach based on the LSTM and SGD compatible with this Law. We assume $dim(Y)=1$ for ease of exposition.

We know that:

\begin{equation}
E[Y_{t+2}| \underline{Y_t}] = E [ E(Y_{t+2}| \underline{Y_{t+1}}|\underline{Y_t}],
\label{(17)}
\end{equation}

\nin from the Law of Iterated Expectations. Let us assume that $(Y_t)$ is a Markov process of order 1. Let us also define the nonlinear theoretical filter ${\cal F}_1$. For any square integrable function $g$, we have: $E[g(Y_{t+1})| Y_t] = \tilde{g}(Y_t)$, where $\tilde{g}$
is also square integrable. Then, the conditional expectation defines an operator [nonlinear filter]
${\cal F}_1$ that associates $\tilde{g} = {\cal F}_1(g)$ to $g$. Similar definitions are introduced at larger horizons $k=1,...,K$.

Let ${\cal F}_1,{\cal F}_2,... $ denote the conditional expectation filters at horizons $k=1,2,...$. We have ${\cal F}_2 = {\cal F}_1 \circ {\cal F}_1, {\cal F}_k = {\cal F}_1^{\circ k}$.

By analogy, we can define a LSTM filter $F_k$ with the SGD (see Fig. 4) optimizing $Min_\theta \sum_{t=k+1}^T [y_{t} - \hat{y}_{t-k,t}(\theta)]^2$ instead of $Min_\theta \sum_{t=2}^T [y_{t} -\hat{y}_{t-1,t}(\theta)]^2$. We can use those joint filters ${\cal F}_k $ to obtain at each date (online) the term structure of predictions $(\hat{y}_{t,t+1},....,\hat{y}_{t,t+K})$ and of associated hidden states ($h_{1,t}, h_{2,t},...,h_{K,t}), \, \forall t$.

Then, there exist several combinations of the LSTM filters that do not provide the same prediction results at horizons $k>1$:

\medskip

i) The direct method, which can be interpretede as the local projection method [see \citet{Jorda2005LocalProjections}, and \citet{JordaTaylor2025LocalProjections} for a survey] commonly used in (linear) dynamic macroeconomic models corresponds to the following propagation algorithm which uses two filters $(F_1 +SGD_1, F_2 + SGD_2)$ for horizons 1 and 2, respectively, and where $SGD_k$ is the SGD algorithm for predicting at horizon $k=1,2$:

\newpage

\begin{center}
\captionsetup{hypcap=false}
\captionof{figure}{Local projection approach at horizons 1 and 2}\label{fig:local_projection_scheme}
\end{center}
\vspace{-0.5in}
\begin{center} $y_{t}$ \\
$\downarrow$ 
\end{center}
\begin{equation}
 \rightarrow \left( \begin{array}{c} h_{t-1,1} \\ c_{t-1,1} \\ \theta_{t-1,1} \\  h_{t-1,2} \\ c_{t-1,2} \\ \theta_{t-1,2} \end{array} \right) \rightarrow  \left[ \begin{array}{c} F_1 + SGD_1\\ \vdots \\ \vdots \\ F_2 + SGD_2\\  \end{array} \right] \rightarrow \left( \begin{array}{c} h_{t,1} \\ c_{t,1} \\ \theta_{t,1} \\ h_{t,2} \\ c_{t,2} \\ \theta_{t,2}\end{array} \right) \nonumber \end{equation}
 $$\downarrow $$
$$[\hat{y}_{t,t+1},\hat{y}_{t,t+2}]$$

\medskip
\nin  At each time $t$ the filters $F_1 +SGD_1$ and $F_2+SGD_2$ are applied separately (i.e. in parallel). In particular, the filtered state variables and parameters depend on the prediction horizon. This approach is used for example in \citet{ZhangEtAl2025TransformerLSTM}, Fig 10, Table 5, for wind power prediction at horizons 15, 30 and 45 minutes.

\medskip

\nin ii) The recursive method, which can be viewed as a structural LIE-based method, relies on the same filter $F_1$ used recursively at date $t$. More precisely, at date $t$, we feed input $y_t$ into filter $F_1+SGD_1(y_t)$, where $SGD_1(y_t)$ is the $SGD_1$ algorithm for predicting $\hat{y}_{t,t+1}$. Next, we feed input $\hat{y}_{t,t+1}$ into filter $F_1+SGD_1(\hat{y}_{t,t+1})$, where $SGD_1(\hat{y}_{t,t+1})$ is the $SGD_1$ algorithm for predicting $\hat{y}_{t+1,t+2}$, and so on. Then, at time $t$ we get $h_{1,t}, h_{2,t}, ...$ and predictions $\hat{y}_{t,t+1}, \hat{y}_{t,t+2},...$ that are different from those obtained from the direct approach.

This recursive approach corresponds to the propagation scheme in Figure~\ref{fig:recursive_scheme}, where $\theta_k, \,k=1,2$ denote the set of LSTM parameters at horizon $k$ \footnote{This recursive approach is also used in image analysis where $k$ denotes the distance to the object of interest, e.g. $k=1$ , 2 meters, etc. }.

\medskip The direct and recursive approaches do not provide the same forecasts. 
The parameters estimated by $SGD_1$ and $SGD_2$ under the direct approach are different since the SGD objective functions are different and the prediction model is misspecified. Under the recursive approach, even if the prediction model at horizon 1 is well-specified, the LSTM computes the parameters first from an observation and next from a prediction. 
It has been proposed to combine them, or to select the best of them in the so-called rectify strategy [\citet{BenTaiebHyndman2012RecursiveDirect}].
It is therefore important to distinguish these two propagation algorithms by writing explicitly the analytical formulas of updating equations.

\begin{center}
\captionsetup{hypcap=false}
\captionof{figure}{Recursive approach at horizons 1 and 2}\label{fig:recursive_scheme}
\end{center}
\vspace{-0.4in}
\begin{center}  $y_t$\\
  $\downarrow$ 
\end{center}
$$
\begin{array}{ccc} [h_{1,t-1} & c_{1,t-1} & \theta_{1,t-1}]  \end{array}  \rightarrow  \left[ \begin{array}{c} F_1 + SGD_1 (y_t)  \end{array} \right] \rightarrow  \begin{array}{ccc} [h_{1,t} & c_{1,t} & \theta_{1,t}] \end{array}  $$
\begin{center}  $\downarrow$   \\
$\hat{y}_{t,t+1}$\\
  $\downarrow$ 
\end{center}
$$ \begin{array}{ccc} [h_{2,t-1} & c_{2,t-1} & \theta_{2,t-1}]  \end{array}  \rightarrow  \left[ \begin{array}{c} F_1 + SGD_1 (\hat{y}_{t,t+1})  \end{array} \right] \rightarrow  \begin{array}{ccc} [h_{2,t} & c_{2,t} & \theta_{2,t}] \end{array}$$
 \begin{center} 
 $\downarrow$   \\
$\hat{y}_{t,t+2}$
\end{center}

\subsection{Affine Term Structure Models (ATSM)}

The direct and recursive approaches described in Section 3.1 lead to a full-rank ATSM, in which:

\begin{equation}
\hat{y}_{t,t+k} = w_{y,k} h_{k,t} + b_{y,k}, \;\;\; k=1,...,K , 
\end{equation}

\nin are written in vector form as:

\begin{equation}
(\hat{y}_{t,t+k}) = W_{y} H_{t} + b_{y}, 
\end{equation}

\nin where $W_y = diag(w_{y,k}), H_t = (h_{k,t}), b_y = (b_{y,k})$. Therefore the equation (3.3) can be seen as an affine model of predictions with a number of factors equal to the number of maturities. By analogy to the theory of ATSM applied to interest rates, we can look for a constrained factor model with a smaller number of factors, such as a linear one-factor model:

\begin{equation}
(\hat{y}_{t,t+h}) = \hat{W}_{y,t} h_{t} + \hat{b}_{y,t}, \;\;\; k=1,...,K,  
\end{equation}

\nin where $dim(h_t)=1$ and $W_y$ is of reduced-rank and of dimension $(K,1)$.

Under this restriction (3.4), a direct LSTM/SGD can be trained using the basic LSTM equations (\ref{(1)}) - (\ref{(6)}), and eq. (3.2) instead of equation (2.7) for the prediction formula, and a multi-task objective function $\sum_{k=1}^K \pi_k (y_{t+k} - \hat{y}_{t,t+k} )^2$, where $\pi_k$ are nonnegative weights\footnote{A recursive LSTM/SGD version constrained by the assumption of a single factor is given in Appendix A.}.

This constrained direct approach has three limitations. First, the estimated (trained) models depend on the selected weights $\pi_k$ and on the estimated common factor $\hat{h}_t$. Second, there is a lack of coherency between the nonlinearities introduced in the LSTM and the linearity assumption in eq. (3.4). An alternative specification could allow each $h_{k,t}$ to be a nonlinear function of the same $h_t$ (instead of a linear affine function of $h_t$). Third, even if the rank of $H_t$ is close to one for any $t$, there is no reason for the reduced-rank constrained $h_t$ to belong in the range of $H_t$.

The assumption of a single factor determining the predictions at all horizons is very restrictive. Let us consider a linear process $y_t = \sum_{j=0}^{\infty} \theta_j \epsilon_{t-j}$,
where ($\epsilon_t$) is a strong white noise, $\theta_0=1$, and the optimal predictions, i.e. the conditional expectations. We have :

\begin{eqnarray*}
    E(Y_{t+1}|\underline{Y_t}) & = & \theta_1 \epsilon_t + \theta_2 \epsilon_{t-1} + \theta_3 \epsilon_{t-2} + \cdots\\
    E(Y_{t+2}|\underline{Y_t}) & = & \theta_2 \epsilon_t + \theta_3 \epsilon_{t-1} + \theta_4 \epsilon_{t-2} + \cdots
\end{eqnarray*}

These predictions are proportional if and only if there exists $\lambda$, $\lambda \neq 0$, such that  $\theta_{j+1} = \lambda \theta_j, \; \forall j \geq 1$, or equivalently $\theta_j = \lambda^{j-1} \theta_1, \, \forall j \geq 1$. Therefore, we have:

\begin{eqnarray*}
    y_{t} & = & \epsilon_t + \theta_1 L \sum_{j=1}^{\infty} (\lambda^{j-1} L^{j-1}) \epsilon_{t}   \\
    & = &  \epsilon_t +  \Frac{\theta_1 L}{1-\lambda L} \epsilon_{t},
\end{eqnarray*}

\nin or $(1-\lambda L) y_t = [1+ (\theta_1-\lambda) L] \epsilon_t$.

Therefore, in the linear dynamic framework, the assumption of one factor is equivalent to the assumption of an ARMA(1,1) representation of process $(y_t)$. This analysis is easily extended to nonlinear processes [\citet{GourierouxJasiak2026AffinePredictions}].

\section{Properties of the LSTM Filters}

The LSTM filter(s) transform the input series $(y_t)$ into state and output series through nonlinear transformations that depend on $(y_t)$, parameter $\theta$, and tuning parameters. Under mild regularity conditions, the filters at fixed $\theta$ and the joint filter LSTM-SGD transform a strictly stationary series $(y_t)$ into asymptotically stationary state and output series (see Appendix A.2).

However, the exact dynamic effects of these nonlinear dynamic filters are difficult to study theoretically due to the nonlinearities. The aim of this section is to examine them by simulations.
To facilitate the comparisons, we consider simple LSTM architectures with a limited number of memory states, starting values $h_0 = c_0 = 0$, one epoch, a batch size of 1, and the SGD algorithm without additional averaging. In this Section, we study the sub-optimality of LSTM predictions and introduce diagnostic tools that reveal their limitations.

\subsection{Design of input dynamics}

To choose the input dynamics, we have to consider the reason invoked for introducing the LSTM, i.e., including the long memory in the prediction formations. The long memory is present when the prediction depends on all lagged values of $(y_t)$ rather than on a finite number of lags, implying a fixed time window. In the linear time series literature, to create a dependence of the present value on the entire past, one can consider:

i) a moving average of order 1 (MA(1)) representation of a time series.

ii) a VAR(1) model applied to an extended series $(y_t, z_t)'$, where the process $(z_t)$ is unobserved. This approach is implicitly implemented in the LSTM.

Both examples of MA(1) and VAR(1) with a latent $(z_t)$ component in i) and ii) feature long memory of the prediction.

We want to explore if the long memory is captured by $c_t, h_t$, or both of them, and how. We use the following experimental stationary designs:

\medskip
Experiment 1: Strong white noise $y_t = \epsilon_t$,  with the following distributions: $\epsilon_t \sim N(0,1)$, and $\epsilon_t \sim C(0,1)$, where $C$ stands for the Cauchy distribution to see if the use of LSTM can create spurious long memory of the prediction.

\medskip
Experiment 2: MA(1) model: $y_t = \epsilon_t - 0.9 \epsilon_{t-1}, \; \epsilon_t \sim N(0,1)$, for which the optimal prediction is the conditional expectation $E[Y_{t+1}| \underline{y_t}] = -0.9(y_t + 0.9 y_{t-1} + 0.9^2 y_{t-2} + \cdots)$. This model features both long memory of the predictions.

\medskip
Experiment 3: Non-invertible MA(1) model $y_t = \epsilon_t - \epsilon_{t-1}, \; \epsilon_t \sim N(0,1)$, for which the optimal prediction at horizon 1 exists, but has no functional form:
$E[Y_{t+1}| \underline{y_t}] = lim_{H \rightarrow \infty} \frac{1}{H} \sum_{h=0}^{H-1} y_{t-h}$,

\medskip
Experiment 4: GARCH(1,1) model $y_t = \sigma_t \epsilon_t, \; \epsilon_t \sim N(0,1)$, where
$\sigma_t^2 = \gamma_0 + + \gamma_1 \epsilon_t^2 + \gamma_2 \sigma_{t-1}^2$, $ \gamma_i \geq 0, i=0,1,2$.

The GARCH(1,1) is stationary if $\gamma_1 +\gamma_2 <1$. When this sum is close to 1, the persistence in $y_t^2$ increases, while the persistence in $y_t$ is close to 0. In this experiment, we apply the LSTM to jointly predict $y_t$ and $y_t^2$ and to compare these predictions.

\subsection{The implementation and the diagnostic tools}

We proceed in the following steps to study the effects of filters $F_1$ and $F_1(\theta)$:

Step 1: Simulate a long series $y_t, \; t=1,...,T$.

\medskip
Step 2: Apply the online LSTM+SGD  filter to produce the series $z_t, h_t, c_t, \theta_t, \hat{y}_{t,t+1}$ 

\medskip
Step 3: Analyze the dynamic properties of these series: plot the ACF of $y_t$ and $y_t^2$, plot the evolution of the LSTM prediction error over time and plot the optimal prediction error over time and compare.

\medskip
Another LSTM filter that could be considered is the $F_1(\theta)$ filter with a fixed parameter $\theta$. The effect of this filter depends on the value of $\theta$. The steps are as follows:

\medskip

Step 1: Use the same series $y_t,\; t=1,...,T$, as for the LSTM+SGD.

\medskip

Step 2: Fix a large value  $T_0 <<T$ for the training period and the values $h_{T_0}, c_{T_0}, \theta_{T_0}$ obtained from the  LSTM+SGD filter at $T_0$ \footnote{Note that we do not reinitialize $h_{T_0}, c_{T_0}$ at zero.}

\medskip

Step 3: After date $T_0$ apply the LSTM with $\theta$ fixed at $\theta_{T_0}$, starting values $h_{T_0}, c_{T_0}$ for $h_t$ and $c_t$ to produce the series $z_t, h_t, c_t, \hat{y}_{t,t+1}$ for $t=T_0,...,T$.
\medskip

Step 4: Analyze the dynamic properties of these series.

These series depend on some controls that have been fixed.

In our experiment, the dimensions of $y_t$ and state variables are fixed with $n=m=p=1$. Then, the number of parameters is $dim(\theta) = 9+3+2=14$ [see, equation (2.8)]. For identification and consistency the training period has to be sufficiently long. Its length is fixed at $T_0=150$.

In order to illustrate the results, various diagnostic tools can be used. 
The root mean squared error (RMSE) is used to evaluate the accuracy of one-step-ahead forecasts and is defined as:
\[
\mathrm{RMSE}_n
=
\sqrt{\frac{1}{n}\sum_{t=1}^{n}
\left(y_{t+1}-\hat{y}_{t, t+1}\right)^2},
\]
where \(n\) is the number of forecasts, \(y_{t+1}\) is the observed value at time \(t+1\), and \(\hat{y}_{t, t+1}\) is the corresponding forecast based on information available at time \(t\).

Functionals are better diagnostic tools than scalar measures. Specifically, residual plots and their ACF are more informative about temporal dependence than $RMSE$ in the standard linear autoregressive models. We mainly use graphical displays of series such as
the joint series $y_t, \hat{y}_{t,t+1}$, the series of hidden states $h_t, c_t$, their ACF and PACF, and series of output weights and biases. Additional series are plotted in the online Appendix B.

\subsubsection{Exp 1: Noise input}

In this experiment, the LSTM prediction model includes the optimal prediction $E(Y_{t+1}|Y_t)=0$ obtained for $W_y=0, b_y =0$. Note that for  $W_y=0, b_y =0$, the remaining weight and bias parameters are not identifiable. Nevertheless, since the transitions are deterministic, their trained values, i.e. $\tilde{\theta}_t$, are uniquely defined for any given tuning parameters. However, the theoretical asymptotic behavior, including the consistency in such a non-identifiable framework, has not been examined in the theoretical literature.

Let us first consider the Gaussian noise. The series of realizations and predictions are displayed in Figure \ref{fig:gwn}, where we observe that the predictions are close to zero, i.e. the optimal prediction.

\medskip

\begin{figure}[H]
    \centering
        \setcounter{figure}{6}
    \caption{Gaussian White Noise: Realization and Online Prediction}
    \includegraphics[width=1\linewidth]{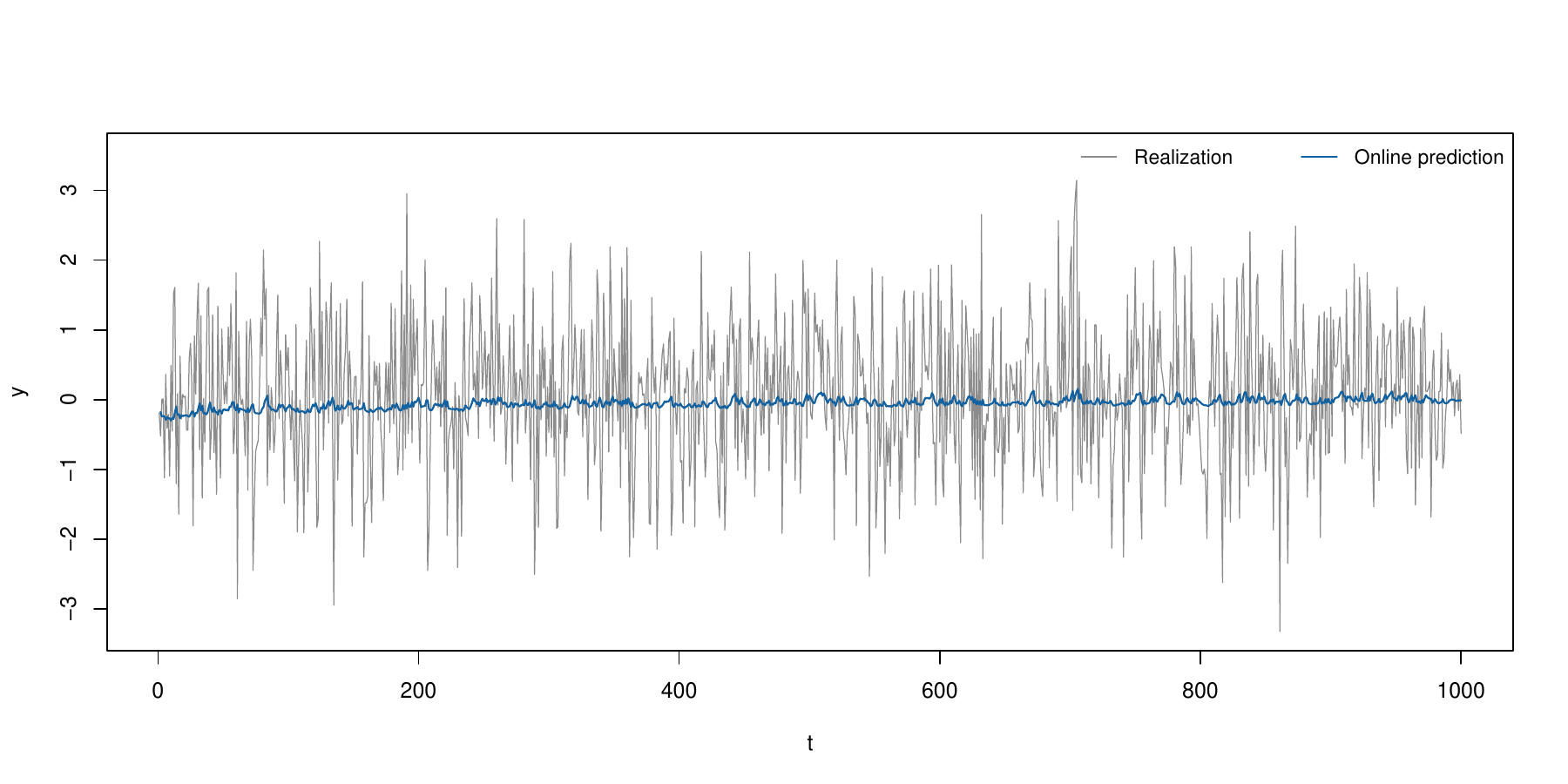}

    \label{fig:gwn}
\end{figure}

\medskip

\begin{figure}[H]
    \centering
            \caption{Gaussian White Noise: LSTM Parameter Evolutions}
    \includegraphics[width=1\linewidth]{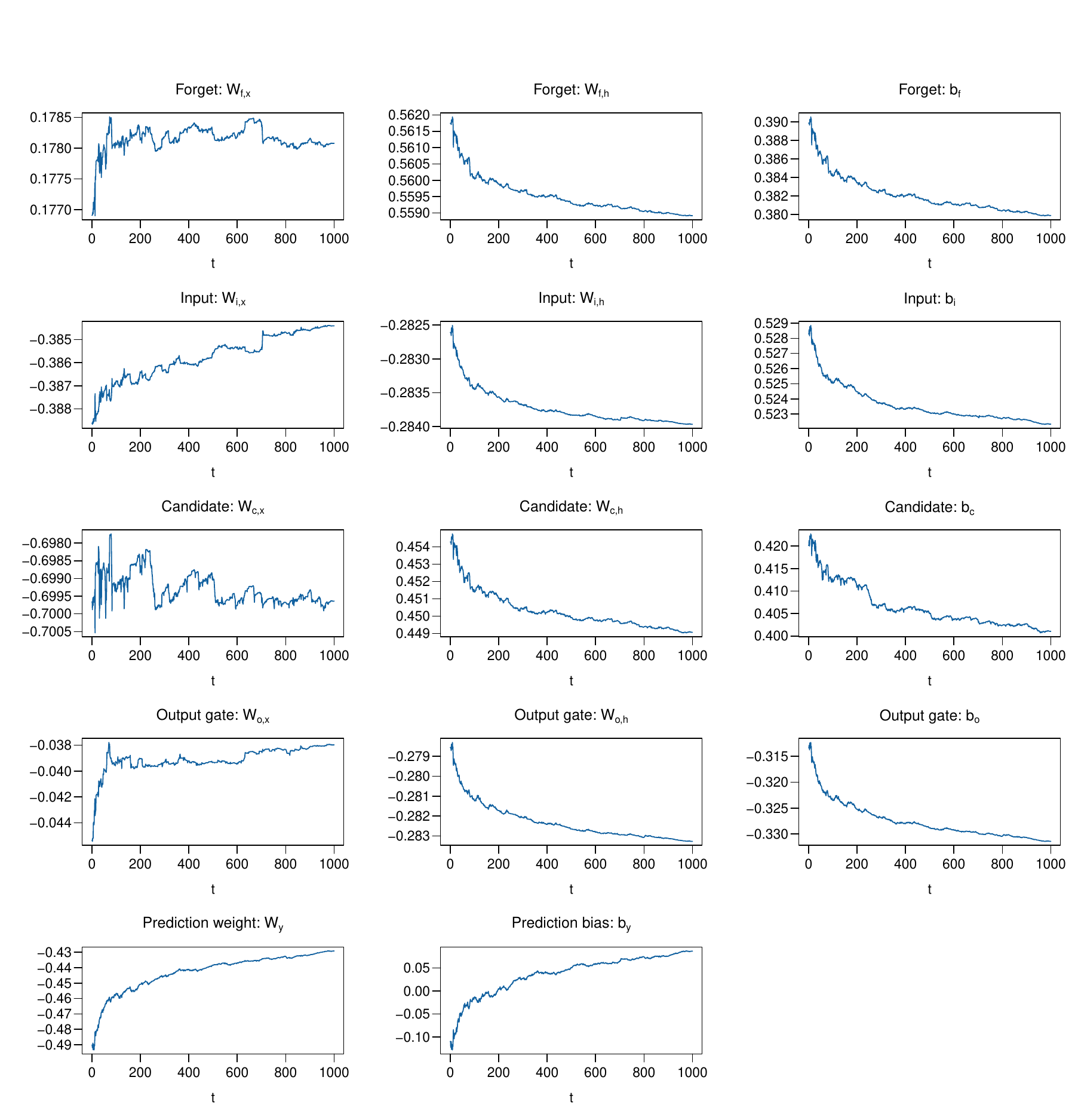}

    \label{fig:gwn_para}
\end{figure}

Therefore, it seems that the theoretical identification issue has been solved automatically by the choice of initial values in the SGD algorithm [see \citet{GourierouxMonfort2025AFS}]. This is confirmed in Figure \ref{fig:gwn_para}, showing the numerical convergence of the weight and bias parameters despite the identification issue. The effects of initial values depend on the parameters, whose estimated values become more stable after t=500.

\medskip
\medskip

The lack of optimality of online predictions can be revealed by checking empirically the conditions of 1) no bias of optimal predictions in prediction errors, and 2) the absence of error correlation with any function of $\underline{y}_{t-1}$. We know that the gradient in SGD algorithm is of the form 
$ \frac{\partial \hat{y}_{t-1,t} (\theta)}{ \partial \theta} (y_t -\hat{y}_{t-1,t})$. Therefore the online prediction errors satisfy by construction the condition of no serial correlation with $\frac{\partial \hat{y}_{t-1,t} (\theta)}{ \partial \theta}$ as the instrument. Following \citet{giacomini2006tests}, alternative instruments such as a constant equal to one, the indicator of positive $y_{t-1}$, or lagged prediction errors are used in Table \ref{tab:stat pred err}.

\medskip

\begin{table}[H]
    \centering
    \caption{Summary Statistics for Prediction Errors}
    \begin{tabular}{|c|c|}\hline
        Statistic & Prediction Errors\\\hline
        N   & 999\\\hline
        Mean & 0.1396\\\hline
         Standard Deviation& 1.0035\\\hline
         $mean(y_{t-1}|y_{t-1}\geq 0)$&0.8058\\\hline
         1st order autocorrelation& -0.3491\\\hline
         2nd order autocorrelation& -0.4667\\\hline
    \end{tabular}
    
    \label{tab:stat pred err}
\end{table}

\medskip

The standard deviation reported in Table 1 is close to the true standard deviation of the white noise. However, the reported mean and truncated mean 
are statistically significantly different from the true value of 0. We also observe significant autocorrelations of prediction errors. This means that even with this simple DGP, the online predictions are very sensitive to the choice of instruments potentially introduced in the gradient.

The evolutions of the hidden states $h_t, c_t$ are shown in Figures~\ref{fig:g_h_t series}--\ref{fig:g_c_t series} in the online Appendix B with their ACF in Figures~\ref{fig:g_h_t acf}--\ref{fig:g_c_t acf} \footnote{The ACF and PACF of $(y_t)$ are given in Figure b.1.}. For the Gaussian white-noise DGP, the hidden states $h_t, c_t$ display significant short-lag PACF and decaying ACF, with a pattern similar to that of an AR(1). This is likely a misleading effect of the initial values. 

The results are even more disappointing when the DGP is the Cauchy white noise. The series of realizations and predictions are significantly influenced by the extreme values, and the predictions are poor [see Figure~\ref{fig:cwn} and also \citet{zhong2026}]. The evolution of $h_t, c_t$ in Figures~\ref{fig:c_h_t series}--\ref{fig:c_c_t series} show trends and high volatility at the beginning up to about time t=400 and t=500, respectively, likely due to the effects of the starting values. After $t=500$, the means of the series are more stable, but the volatility increases. We observe more downturns in the $c_t$ series, which are unrelated with the extreme values observed in the input series. The ACF and PACF show unexpected "memory" in $c_t, h_t$.

\medskip

\subsubsection{Exp 2: MA(1) input}

\begin{figure}[H]
    \centering
    \caption{Scatter Plot of Optimal and Online Predictions}
    \includegraphics[width=\linewidth]{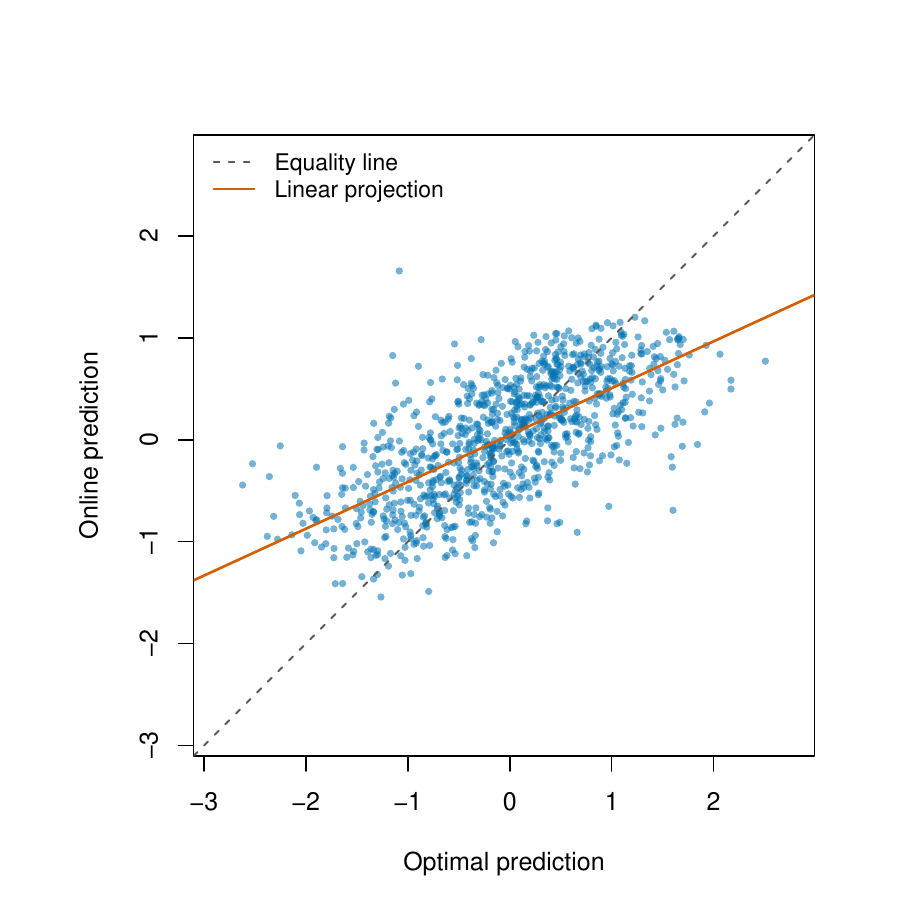}
    \caption*{Optimal predictions (horizontal) and online LSTM predictions (vertical). The dashed line indicates equality; the solid line is the sample linear projection of online predictions on optimal predictions, including an intercept.}
    \label{fig:ma1scatter1}
\end{figure}

The trajectories of predictions and realizations are given in Figure b.11. In this experiment, we do not expect the LSTM predictions to be optimal due to the nonlinear transformations $\sigma, \tanh$, and $\odot$ introduced in the filter. The optimal predictions can be computed analytically and compared with the LSTM predictions. They are plotted in Figure \ref{fig:ma1scatter1} as a scatter plot of optimal predictions measured on the x-axis and the sub-optimal predictions on the y-axis.

\medskip

\medskip

The estimated LSTM root mean squared error is equal to 1.1390 and can be compared with the estimated optimal one equal to 0.9660. This comparison is averaged over the dates, i.e., environments. Figure \ref{fig:ma1scatter1} is more informative on how the LSTM prediction varies with respect to the optimal prediction. The dependence is almost linear with larger (resp. smaller) values for negative (resp. positive) optimal predictions. This is illustrated further in Figures~\ref{fig:ma1onlinevstrue}--\ref{fig:ma1optimalvstrue}
that show the predictions and the realizations. We get similar patterns. The regression lines cross the axes close to the origin (0,0), indicating that both types of predictions are globally unbiased. Since the realizations $y_{t}$ and their optimal predictions $E[Y_{t+1}|Y_t]$ are positively correlated, the slope of this regression line is positive. It is higher for the online predictions, revealing their lower accuracy.  

\subsubsection{Exp. 2: MA(1) Input; Predictions of $y$ and $y^2$ }

\begin{figure}[H]
    \centering
            \caption{Online Prediction and Realization of $y_t^2$ (1 units, 1 epochs)}
    \includegraphics[width=1\linewidth]{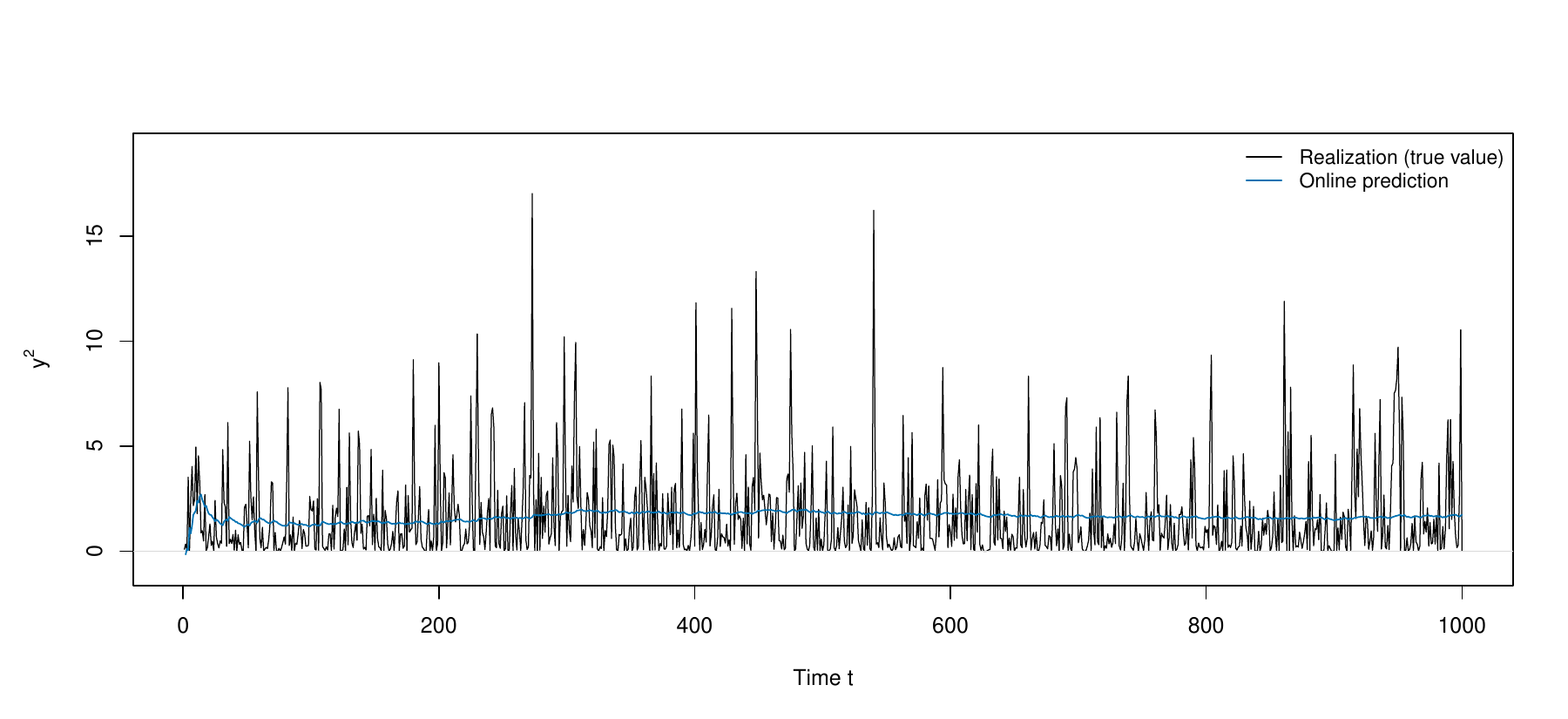}

    \label{fig:y2prediction}
\end{figure}

Let us now examine the predictions of $y_t$ and $y_t^2$ at horizon 1 performed in parallel with separate $h_t, c_t$ for each of these series.
The predictions of $y_t^2$ are shown in Figure \ref{fig:y2prediction}

\medskip

Let us now discuss the "long memory effects" in $y_t$ and $y_t^2$. The evolutions of the hidden states $h_t, c_t$ are given in Figures~\ref{fig:ma1_h_t series}--\ref{fig:ma1_c_t series} for $y_t$ and Figures~\ref{fig:y2_h_t series}--\ref{fig:y2_c_t series} for $y_t^2$, and their ACF/PACF in 
Figures~\ref{fig:ma1_h_t acf}--\ref{fig:ma1_c_t acf} for $y_t$ and Figures~\ref{fig:y2_h_t acf}--\ref{fig:y2_c_t acf} for $y_t^2$, in
the online Appendix. The evolutions of hidden states for $y_t$ and $y_t^2$ are quite different, although they have common peaks in the $c_t$ and $h_t$ series corresponding to $y^2$. The $c_t$ and $h_t$  are almost uncorrelated for $y_t$, whereas for $y_t^2$ they display exponentially decreasing autocorrelations.

The experiment is repeated in a bivariate setup using both  $y_t$ and $y_t^2$ as inputs, and a common $h_t, c_t$.
The evolutions of these state variables are given in Figures 11 and 12 and can be compared with the evolutions of the states obtained from parallel estimation. Let us compare the hidden states computed for $y_t$ only (Figures b.14-b.15). Whereas at first their trajectories are very different, they become similar after about $t=300$. Obviously, this visual inspection is insufficient and could be improved by a Principal Component Analysis of the series $c_t, h_t$ (see Section 4.3). 

\medskip
The bivariate hidden- and cell-state ACFs and PACFs are shown in Figures~\ref{fig:Bivariate_h_t acf}--\ref{fig:Bivariate_c_t acf}.
\medskip

\begin{figure}[H]
    \centering
        \caption{Hidden State $h_t$ Evaluation for MA(1) Bivariate  Series}
    \includegraphics[width=1\linewidth]{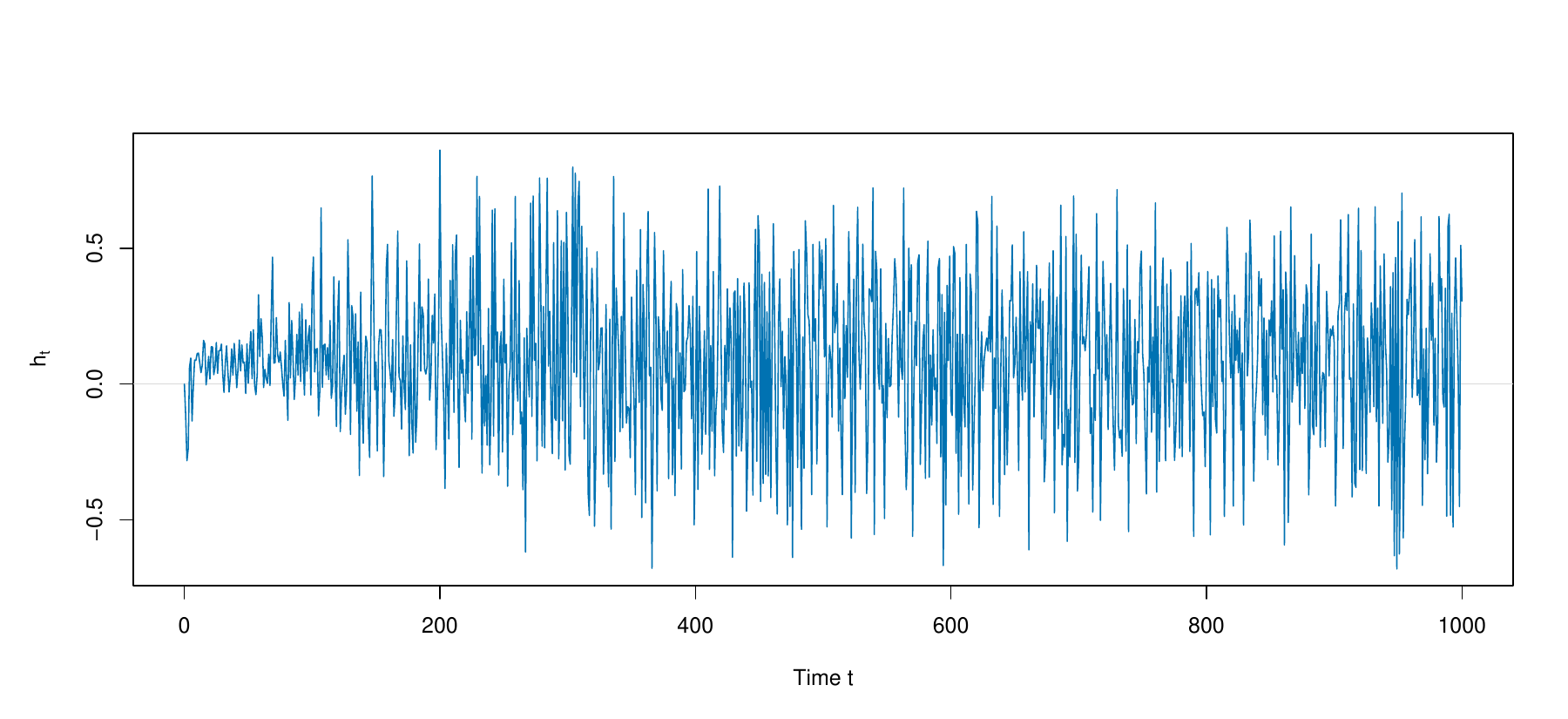}

    \label{fig:Bivariate_h_t series}
\end{figure}

\begin{figure}[H]
    \centering
        \caption{Cell State $c_t$ Evaluation for MA(1) Bivariate Series}
    \includegraphics[width=1\linewidth]{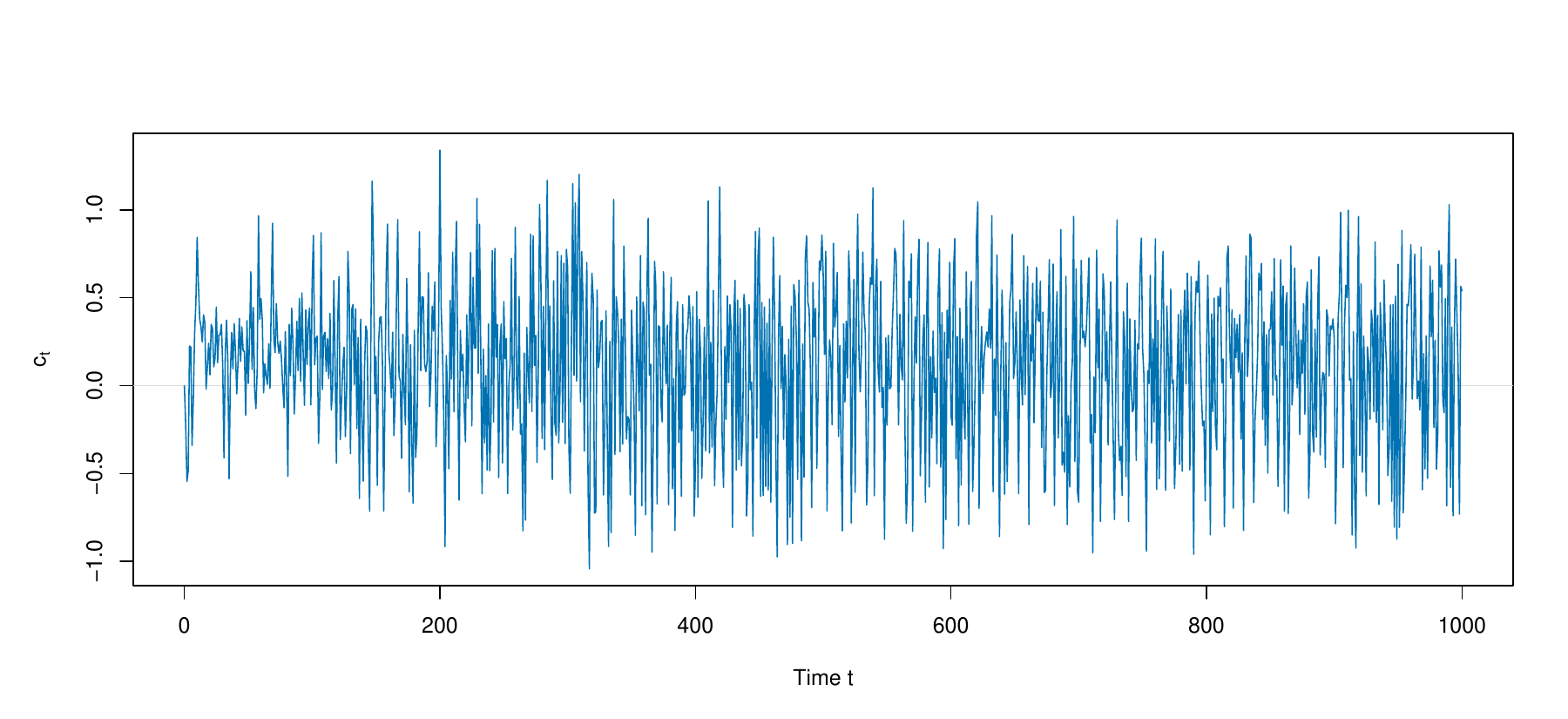}

    \label{fig:Bivariate_c_t series}
\end{figure}

The LSTM prediction models used in parallel or jointly cannot ensure that the nonlinear inequality constraints hold between pointwise predictions of the series and its transformations. For example, we may want to estimate the short-term volatility as:

$$\widehat{Vol}_{t,t+1} = \widehat{(y^2)}_{t,t+1} - (\hat{y}_{t,t+1})^2,$$

\nin but this quantity is not always positive. This is illustrated in Figure 13, based on one common pair of state variables.

\medskip

\begin{figure}[H]
    \centering
    \caption{Conditional Volatility}
    \includegraphics[width=\textwidth]{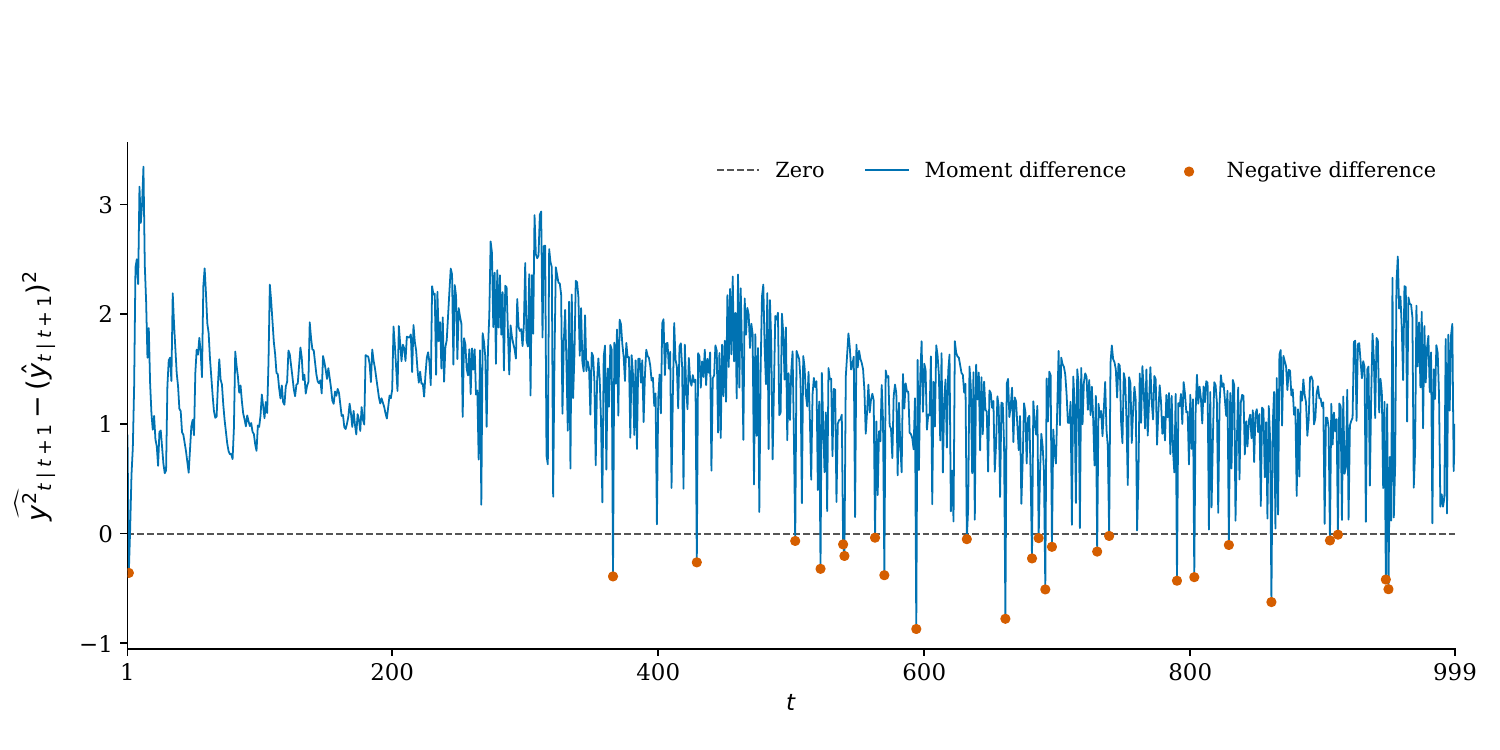}
    
    \caption*{Trajectories of $(\widehat{y^2})_{t,t+1}-(\hat y_{t,t+1})^2$ for the bivariate MA(1) model (case iii), with one common pair $h_t, c_t$. The dashed line denotes zero; orange points indicate negative volatility.}
    \label{fig:ma1_bivariate_moment_difference}
\end{figure}

\subsubsection{Exp 3: Non-invertible MA(1) input}

The aim of this experiment is to detect a potential spurious effect of non-invertibility of the MA model on the predictions of the LSTM+SGD algorithm.

\begin{figure}[H]
    \centering
    \caption{ACF (left) and PACF (right) of the hidden state $h_t$ (top) and cell state $c_t$ (bottom) for the non-invertible MA(1) online LSTM.}
    \includegraphics[width=0.95\textwidth]{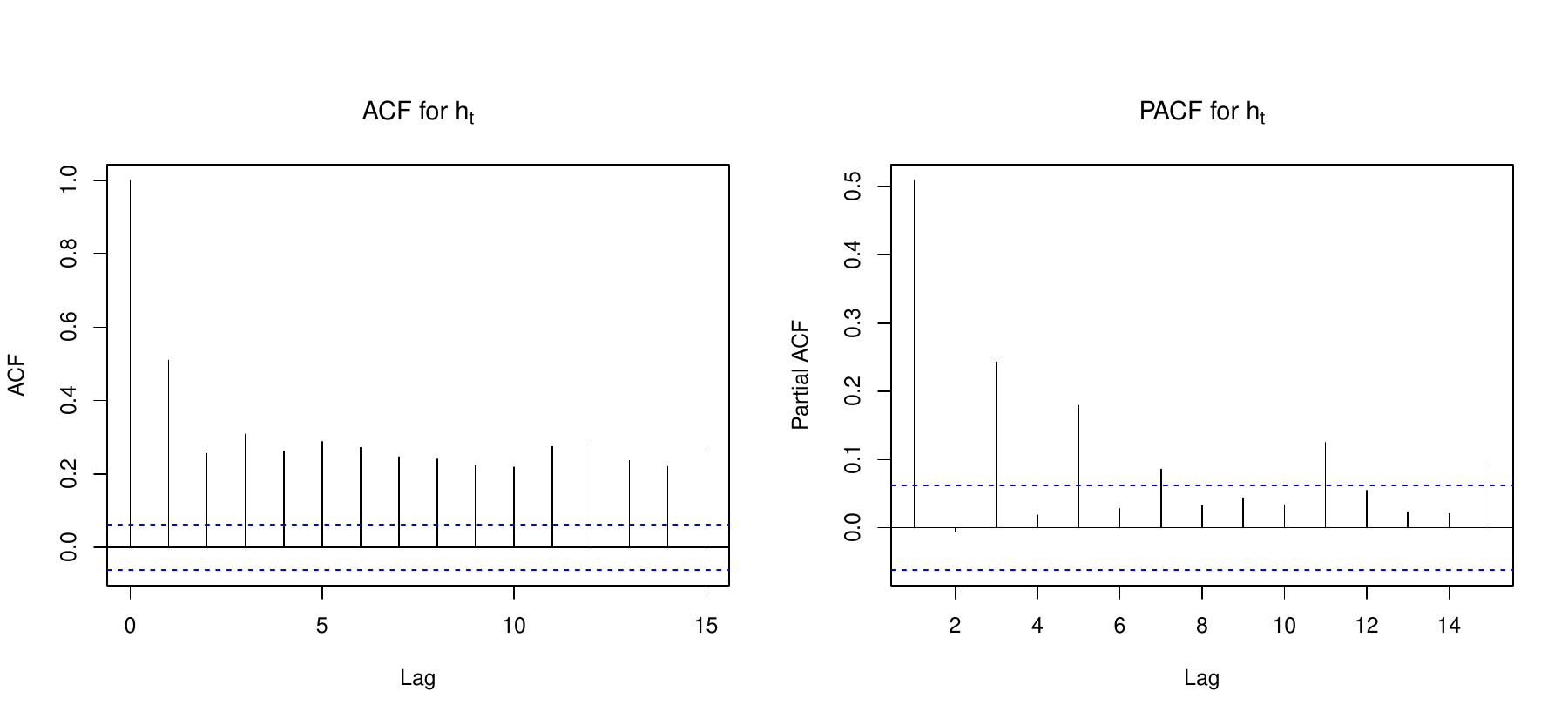}
    \par\medskip
    \includegraphics[width=0.95\textwidth]{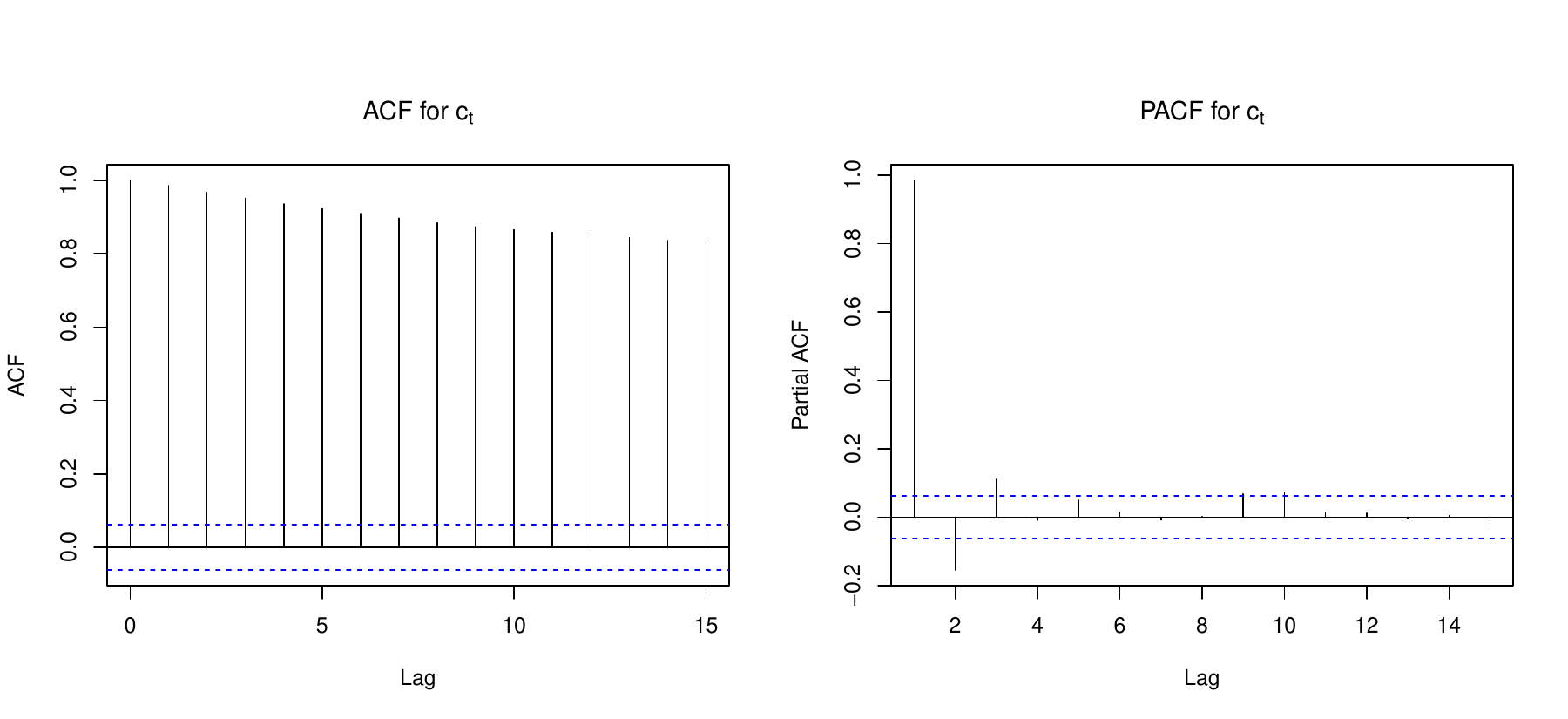}

    \label{fig:ma1_theta1_states_acf_pacf}
\end{figure}

We generate the Gaussian MA(1) input according to:
\[
y_t=\varepsilon_t+\varepsilon_{t-1},\qquad t=1,\ldots,1000,
\]
where $\varepsilon_0,\ldots,\varepsilon_{1000}$ are independent $N(0,1)$ variables. The trajectories of realizations and online predictions are given in Figure b.11. The sample ACF and PACF are shown in Figure 14. All correlograms use a maximum lag of 15 and the 95\% white-noise confidence bands. Let us now examine the dynamic properties of the hidden states $c_t, h_t$ [see Figure \ref{fig:ma1_theta1_states_acf_pacf}]. They reveal that  $c_t, h_t$ show both long memory and long-range dependence close to a unit root for $c_t$ \footnote{This may explain why some machine learning experts consider $c_t$ rather than $h_t$ as a memory state.} in this framework, where the input series has short 
memory

\subsubsection{Exp 4: GARCH(1,1) input}

In the previous experiments, the DGP corresponds to a linear dynamic model, whereas the LSTM is designed to accommodate smooth regime changes by using gates as well as more complex hidden nonlinear features by using back-propagation. Let us now consider the filter behaviour when the DGP has nonlinear dynamics.

The data are generated from the GARCH(1,1) process
\[
y_t=\sigma_t \epsilon_t,\qquad \epsilon_t\overset{\mathrm{iid}}{\sim}N(0,1),
\]
\[
\sigma_t^2=\gamma_0+\gamma_1 y_{t-1}^2+\gamma_2 \sigma_{t-1}^2
=0.1+0.3y_{t-1}^2+0.6\sigma_{t-1}^2.
\]
The stationarity condition $\gamma_1+\gamma_2=0.3+0.6=0.9<1$ is satisfied with a value close to 1 to create volatility clustering. The unconditional variance is
\[
\operatorname{Var}(y_t)=\frac{\gamma_0}{1-\gamma_1-\gamma_2}
=\frac{0.1}{1-0.3-0.6}=1.
\]

This experiment can be performed to see the effect of the choice of input. In our case, we can apply the LSTM to input $y_t$, or $y_t^2$ separately, or to the bivariate vector $(y_t, y_t^2)$ [see Section 4.2.3 for a similar study with the  linear MA(1) DGP]. These correspond to different information sets $\underline{y_t}$, $\underline{y_t^2}$, $(\underline{y_t}, \underline{y_t^2})$, respectively.

More precisely, we consider th following three applications of the LSTM+SGD filter:

i) applied to $(y_t)$ as input with a single pair ($h_t, c_t$) and output $\hat{y}_{t,t+1}$

ii) applied to $(y_t^2)$ as input with a single pair ($h_t, c_t$) and output $(\hat{y}^2)_{t,t+1}$

The training loss is the squared prediction error for $y_{t+1}^2$:
\[
\ell_{t+1}^{(ii)}(\theta)
=\left[y_{t+1}^2-(\widehat{y^2})_{t,t+1}(\theta)\right]^2.
\]

iii) applied to $(y_t, y_t^2)$ as input with univariate ($h_t, c_t$) and outputs $(\hat{y}, \hat{y}^2)_{t,t+1}$

The training loss is the weighted mean of the squared errors for the two output channels:
\begin{align*}
\ell_{t+1}^{(iii)}(\theta)
&=\left[y_{t+1}-\hat y_{t,t+1}(\theta)\right]^2\\
&\quad+\pi\left[y_{t+1}^2-(\widehat{y^2})_{t,t+1}(\theta)\right]^2.
\end{align*}

\nin where $\pi$ is a weight used to compensate the different units in the multitask objective function. Here $(\widehat{y^2})_{t,t+1}$ denotes the direct prediction of $y_{t+1}^2$.

\begin{figure}[H]
    \centering
        \caption{Realizations and Online Predictions of $y_t$ from Applications i) and iii)}
    \includegraphics[width=\textwidth]{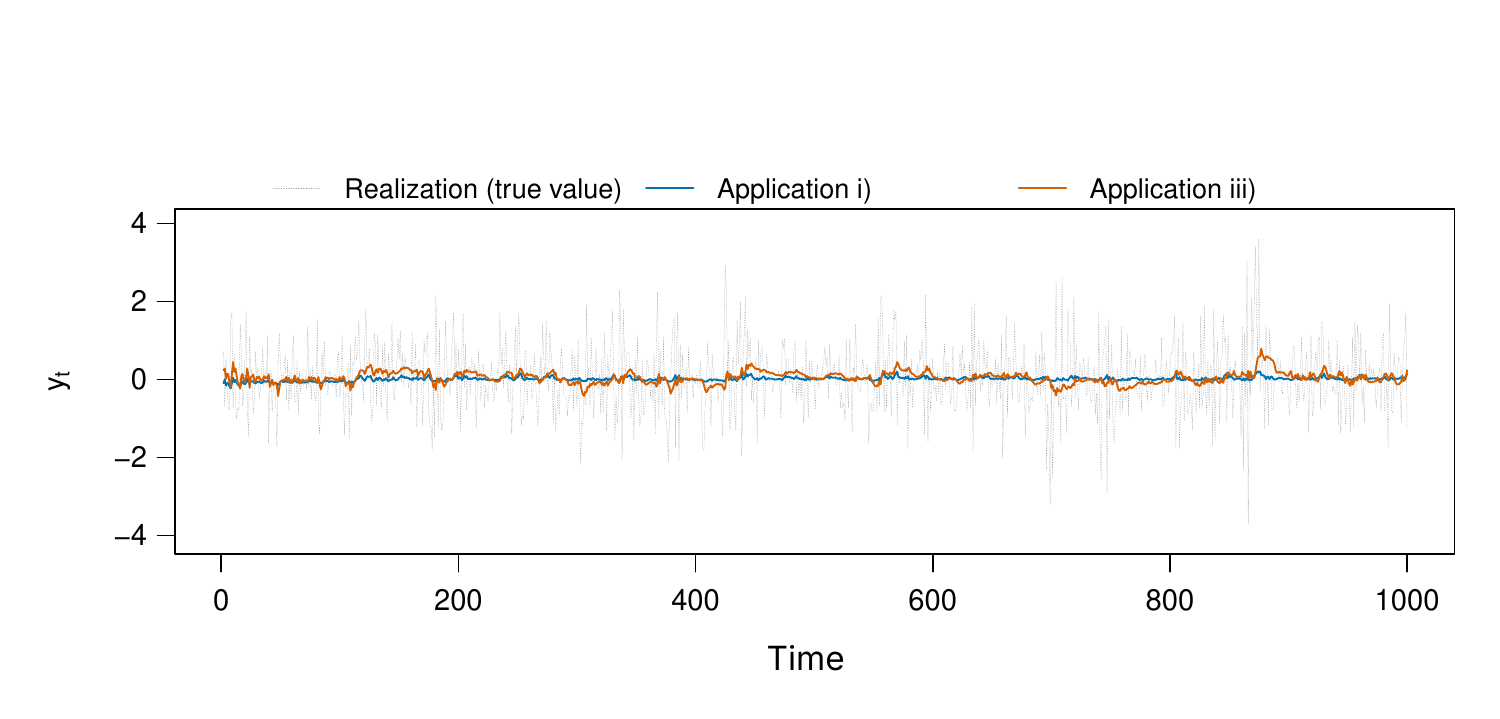}

    \label{fig:garch11_y_predictions_i_iii}
\end{figure}

\begin{figure}[H]
    \centering
        \caption{Realizations and Online Predictions of $y_t^2$ from Applications ii) and iii)}
    \includegraphics[width=\textwidth]{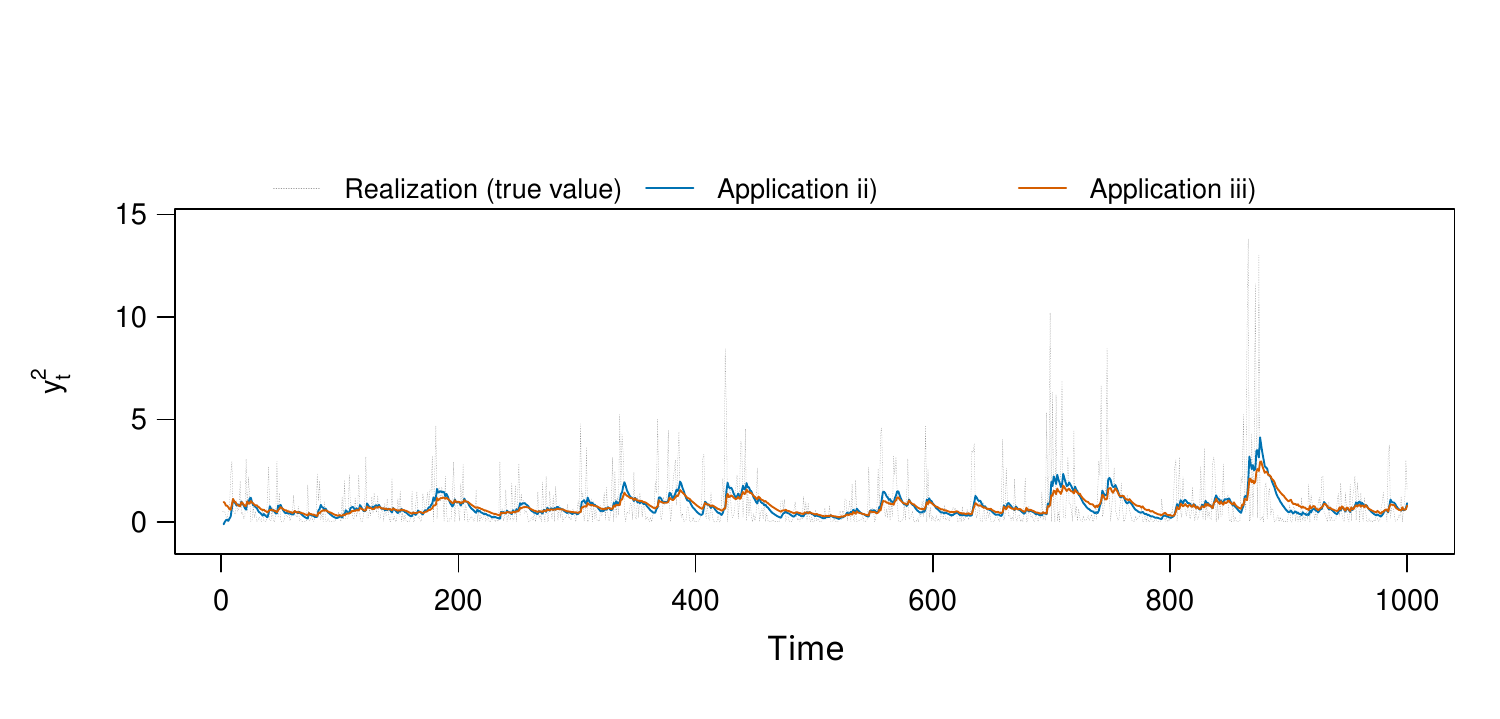}

    \label{fig:garch11_y_squared_predictions_ii_iii}
\end{figure}

Then, we compare these predictions in Figures~\ref{fig:garch11_y_predictions_i_iii}  and~\ref{fig:garch11_y_squared_predictions_ii_iii}. 

\medskip

\medskip

\medskip

\begin{figure}[H]
    \centering
        \caption{ACF and PACF of $h_t$ and $c_t$ for application i).}
    \includegraphics[width=0.8\textwidth,height=0.8\textheight,keepaspectratio]{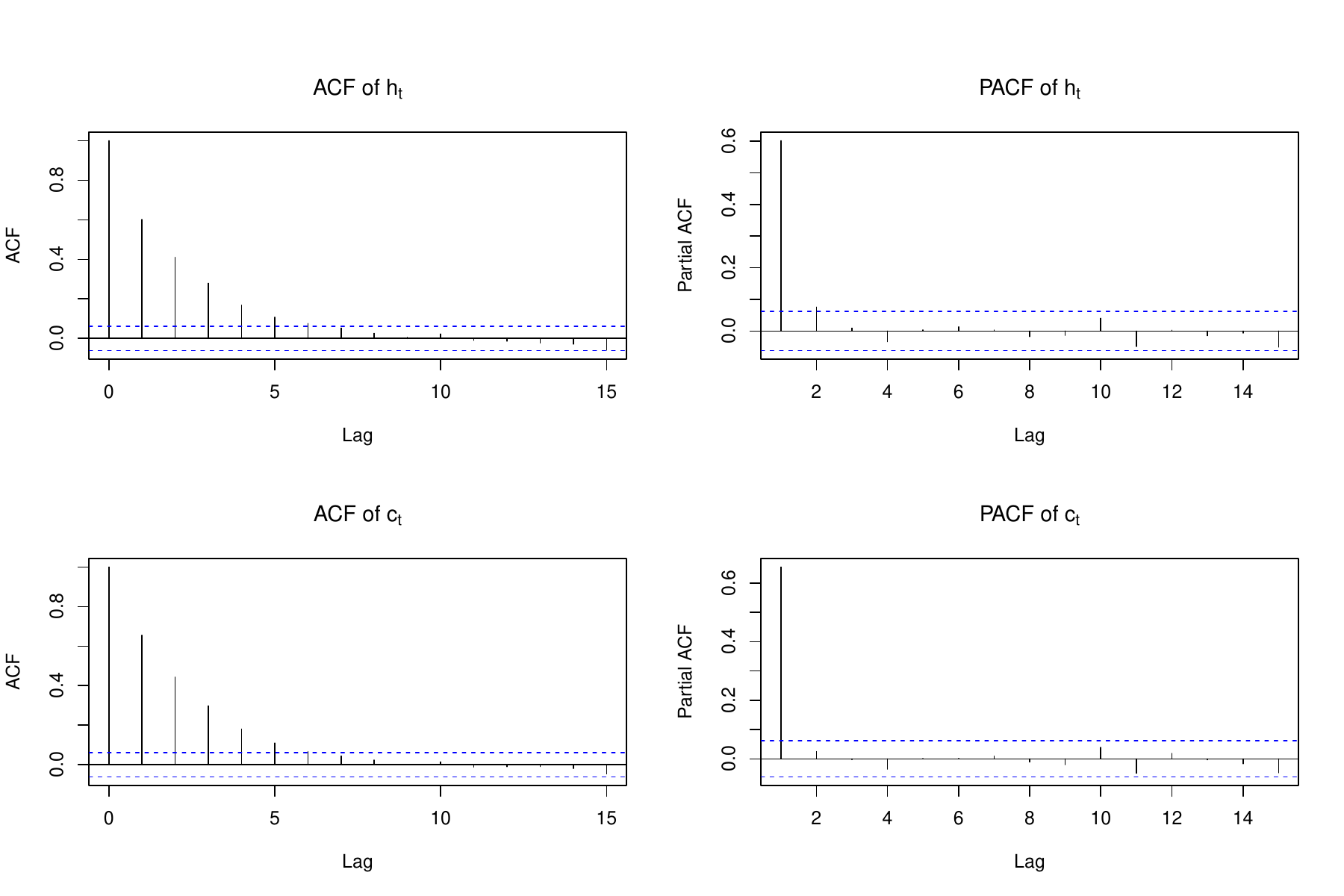}

    \label{fig:garch11_app_i_state_acf_pacf}
\end{figure}

\begin{figure}[H]
    \centering
        \caption{ACF and PACF of $h_t$ and $c_t$ for application ii).}
    \includegraphics[width=0.8\textwidth,height=0.8\textheight,keepaspectratio]{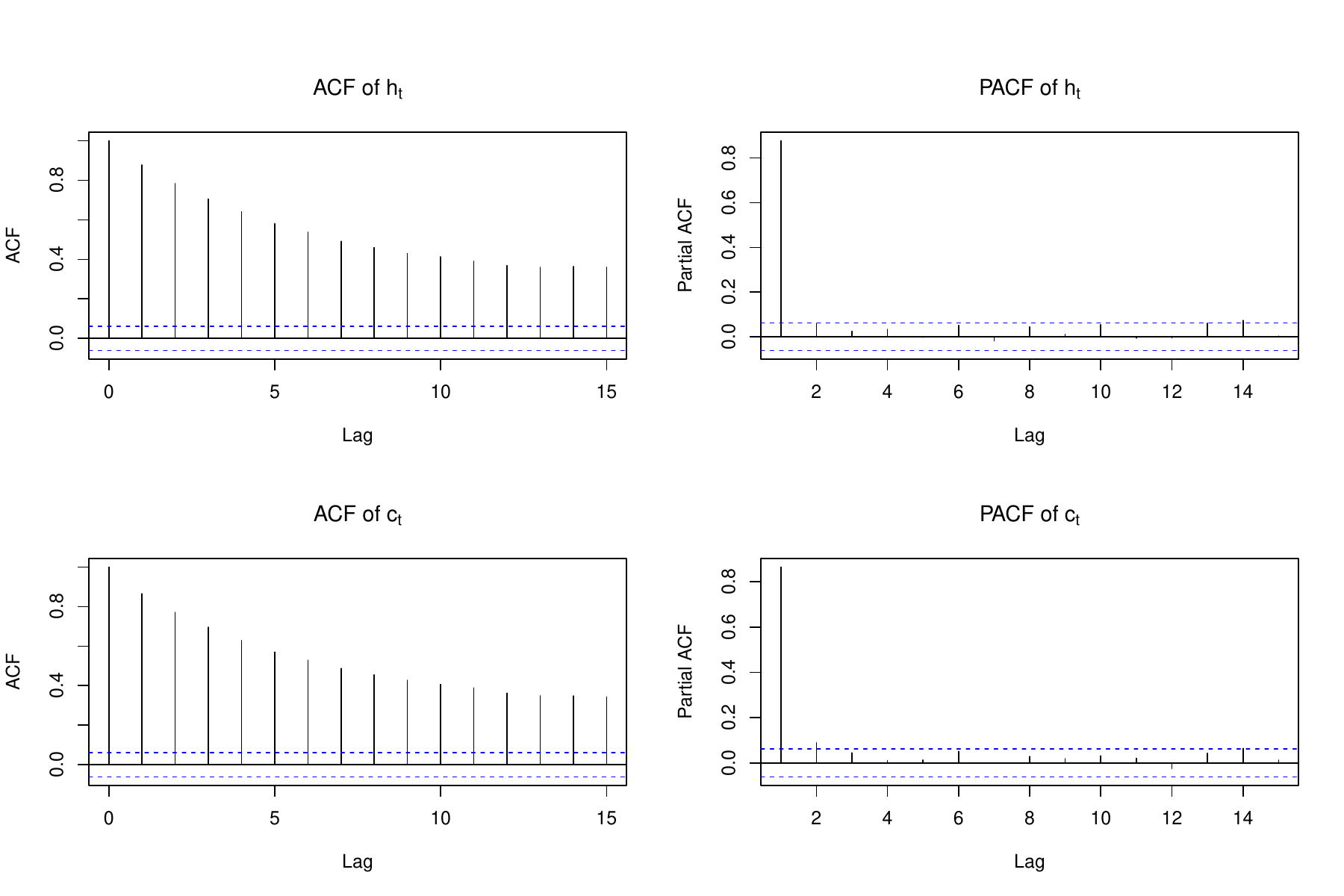}

    \label{fig:garch11_app_ii_state_acf_pacf}
\end{figure}

\begin{figure}[H]
    \centering
        \caption{ACF and PACF of $h_t$ and $c_t$ for application iii).}
    \includegraphics[width=0.8\textwidth,height=0.8\textheight,keepaspectratio]{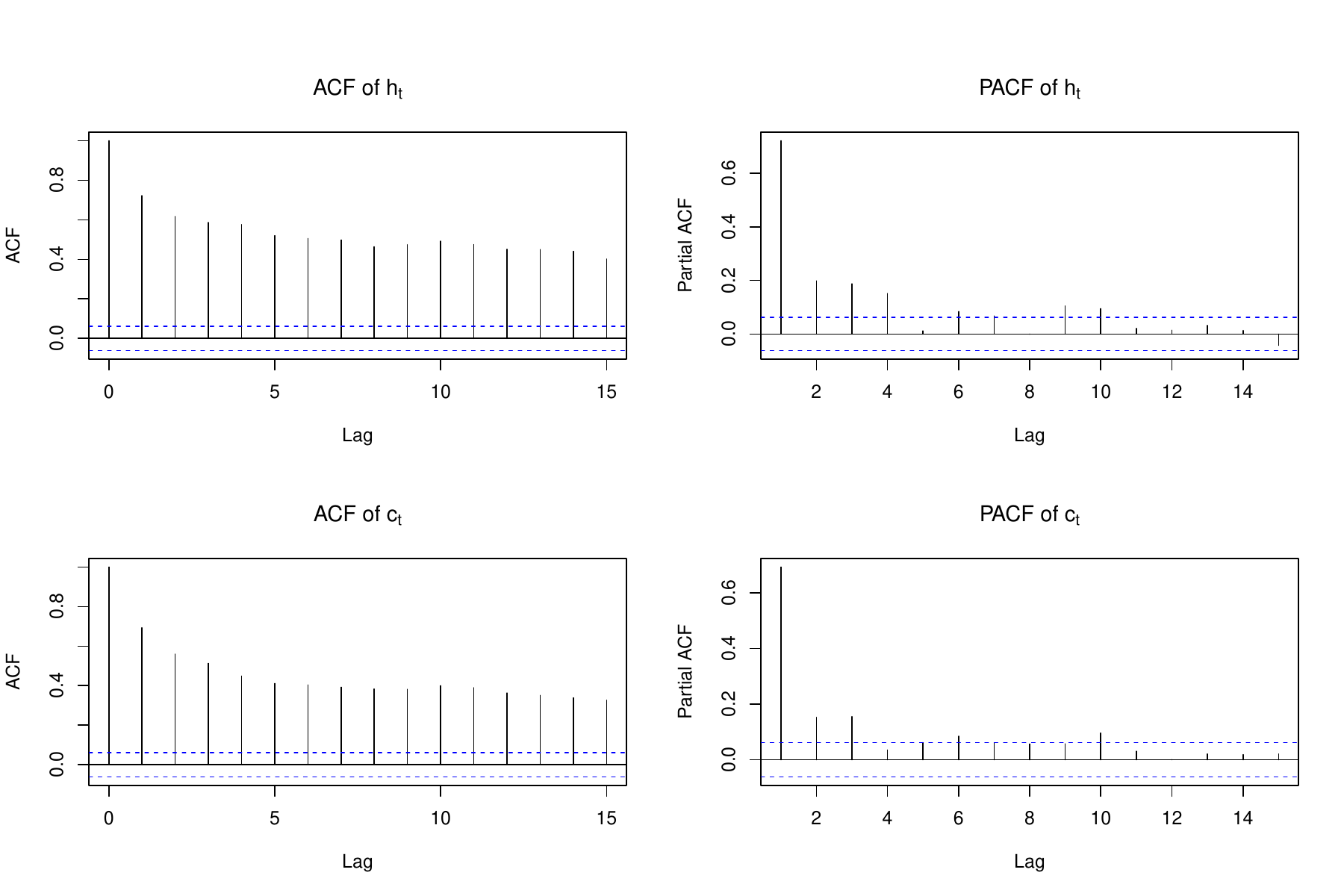}

    \label{fig:garch11_app_iii_state_acf_pacf}
\end{figure}

It is also interesting to see how the long run memory is transferred through the hidden states. Figures~\ref{fig:garch11_app_i_state_acf_pacf},~\ref{fig:garch11_app_ii_state_acf_pacf}, and~\ref{fig:garch11_app_iii_state_acf_pacf} show the univariate ACFs and PACFs of $h_t$ and $c_t$ for cases i), ii), and iii), respectively.

\medskip

\medskip

\medskip

\medskip

The dynamic of the hidden states depend on the type of prediction considered in the LSTM implementation. When $y_t^2$ is introduced as an input, the ACF displays slow decay similar to a close to a unit root dynamics of both $h_t, c_t$.

Let us now consider the estimation of conditional volatility in case iii) of joint bivariate analysis. We observe that the risk of negative volatility is significantly reduced compared to the MA(1) DGP of Section 4.2.3.

We conclude that the LSTM performance is not universal but depends on the input, the quantity to predict, and the DGP.

\begin{figure}[H]
    \centering
        \caption{Online Conditional Volatility for Application iii), $\widehat{E}(Y_{t+1}^2\mid\underline{Y_t})-\widehat{E}(Y_{t+1}\mid\underline{Y_t})^2$}
    \includegraphics[width=0.8\textwidth,height=0.8\textheight,keepaspectratio]{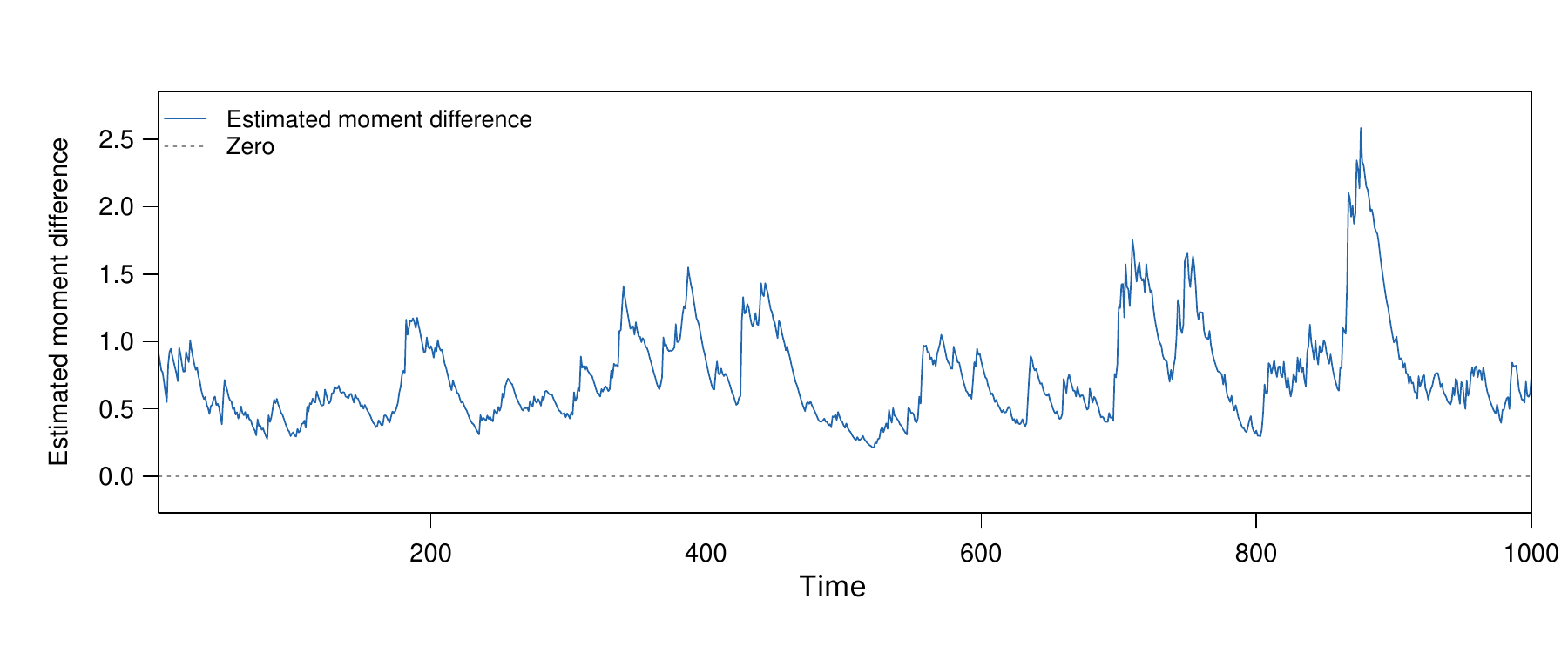}

    \label{fig:garch11_app_iii_moment_difference}
\end{figure}

\medskip

\subsection{Estimated Term Structure of Predictions}

The aim of this section is to compare the direct (local projection) and recursive (structural) approaches in LSTM for estimation of the term structure of predictions. For a numerical illustration, we use the invertible MA(1) process of experiment 2:
\[
y_t=\varepsilon_t-0.9\varepsilon_{t-1},\qquad
\varepsilon_t\stackrel{\mathrm{iid}}{\sim}N(0,1),\qquad t=1,\ldots,1000,
\]

\nin and consider $K=10$ horizons.

\subsubsection{Local projections}

% BEGIN direct-horizon MA(1) predictions and RMSE (2026-10-16)

Ten separate LSTM models directly predict $y_{t+k}$ from observations through time $t$, for $k=1,\ldots,10$. Each model uses the input $y_t$, one hidden unit (i.e., one pair $h_{k,t}, c_{k,t}$) and $h_{k,0}=c_{k,0}=0$.  The learning rate is $\delta_{k,t}=\alpha_k (t+1)^{-\beta_k}$, where $t=1,\ldots,1000-k$ indexes completed updates. For $k=1$, we use the earlier specification $(\alpha_1,\beta_1)=(1.1,0.7)$. For $k=2,\ldots,10$, $(\alpha,\beta)$ is selected by the lowest online RMSE over the available dates $t=k+1,\ldots,1000$.

\begin{figure}[H]
    \centering
    \caption{Direct Online Predictions}
    \includegraphics[width=\textwidth,height=0.78\textheight,keepaspectratio]{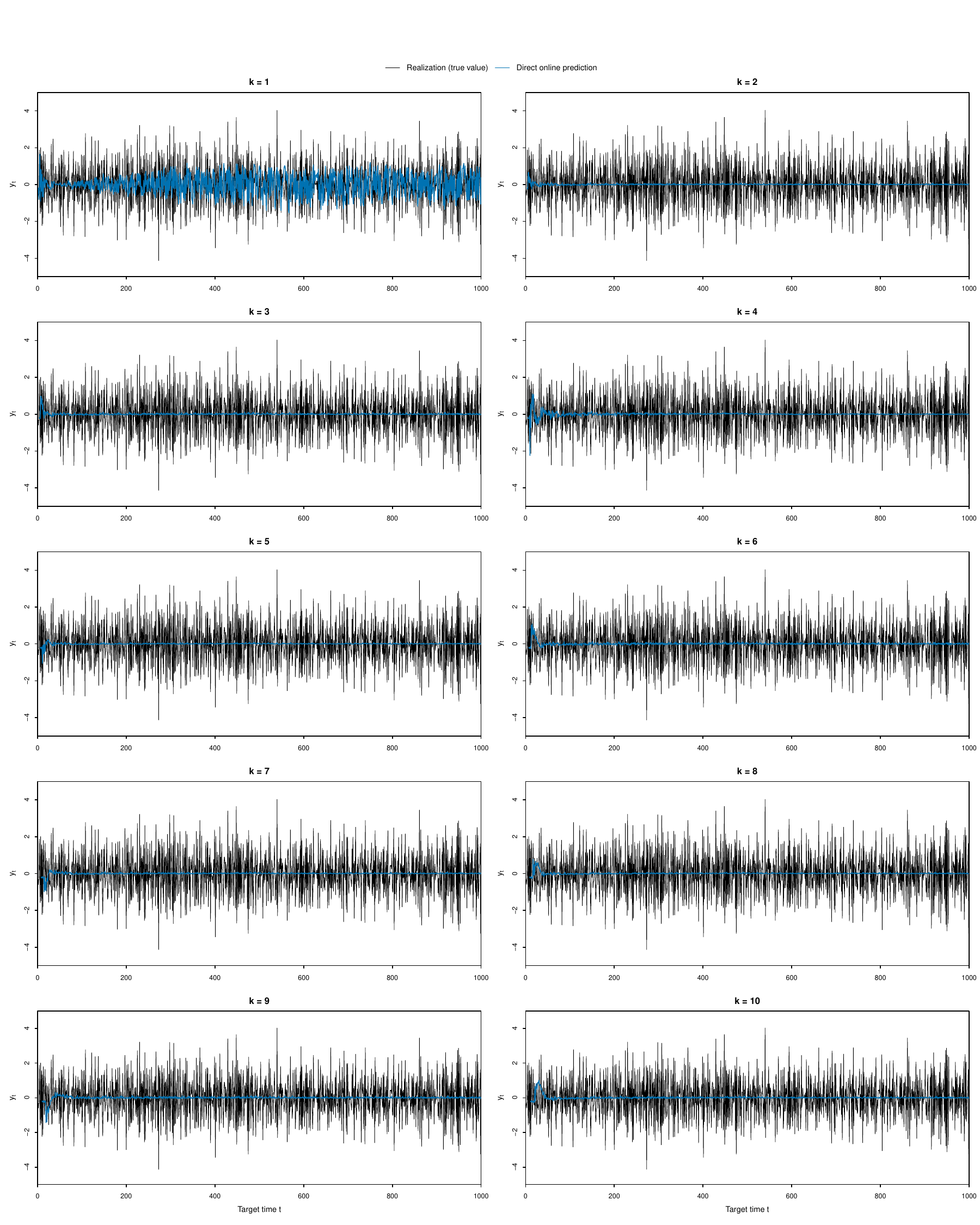}
    \caption*{Realizations (black) and direct online LSTM predictions (blue) for the invertible MA(1) process, at horizons $k=1,\ldots,10$. Each panel uses a separate model with one pair of hidden units. Predictions are aligned with their target times; the first $k$ observations have no reported predictions.}
    \label{fig:ma1_direct_horizons}
\end{figure}

Figure~\ref{fig:ma1_direct_horizons} displays the realizations and predictions. Each forecast computed at time $t-k$ is plotted jointly with the corresponding realization $y_t$.

Table~\ref{tab:ma1_direct_horizon_rmse} compares the prediction errors over the interval $t=11,\ldots,1000$, where the prediction horizon is evaluated from the same 990 realizations:
\[
\operatorname{RMSE}_k=\left\{\frac{1}{990}\sum_{t=11}^{1000}
\left(y_t-\widehat y_{t-k, t}\right)^2\right\}^{1/2}.
\]
The associated optimal RMSE's are equal to 1 for $k=1$ and to $\sqrt{1.81} = 1.34$ for $k \geq 2$.

\begin{table}[H]
    \centering
    \caption{RMSE of the Reported Direct Online LSTM Predictions for the Invertible MA(1) Process}
    \label{tab:ma1_direct_horizon_rmse}
    \begin{tabular}{crrr}
        \hline
        Horizon $k$ & $\alpha$ & $\beta$ & RMSE \\
        \hline
        1 & 1.1 & 0.7 & 1.12783 \\
        2 & 0.5 & 0.9 & 1.28992 \\
        3 & 0.5 & 0.9 & 1.28796 \\
        4 & 1.9 & 0.9 & 1.28931 \\
        5 & 0.5 & 0.9 & 1.29018 \\
        6 & 0.5 & 0.9 & 1.28991 \\
        7 & 0.5 & 0.9 & 1.29065 \\
        8 & 0.5 & 0.9 & 1.29027 \\
        9 & 0.5 & 0.9 & 1.29182 \\
        10 & 0.5 & 0.9 & 1.29400 \\
        \hline
    \end{tabular}

    \medskip
    \begin{minipage}{0.95\textwidth}
    \small 
    \end{minipage}
\end{table}
% END direct-horizon MA(1) predictions and RMSE

Compared to the term structure of RMSE corresponding to the optimal pattern, as expected, we observe an almost constant value for $k \geq 2$ and a positive difference between the values at horizons 1 and 2. Apparently, this stability does not depend on the learning rate (set differently at horizon 4 for comparison) but the size of the increase in RMSE between $k=1$ and $k=2$ does.

\begin{figure}[H]
    \centering
    \caption{Hidden-state Evolutions for Direct Online Estimation}
    \includegraphics[width=\textwidth,height=0.78\textheight,keepaspectratio]{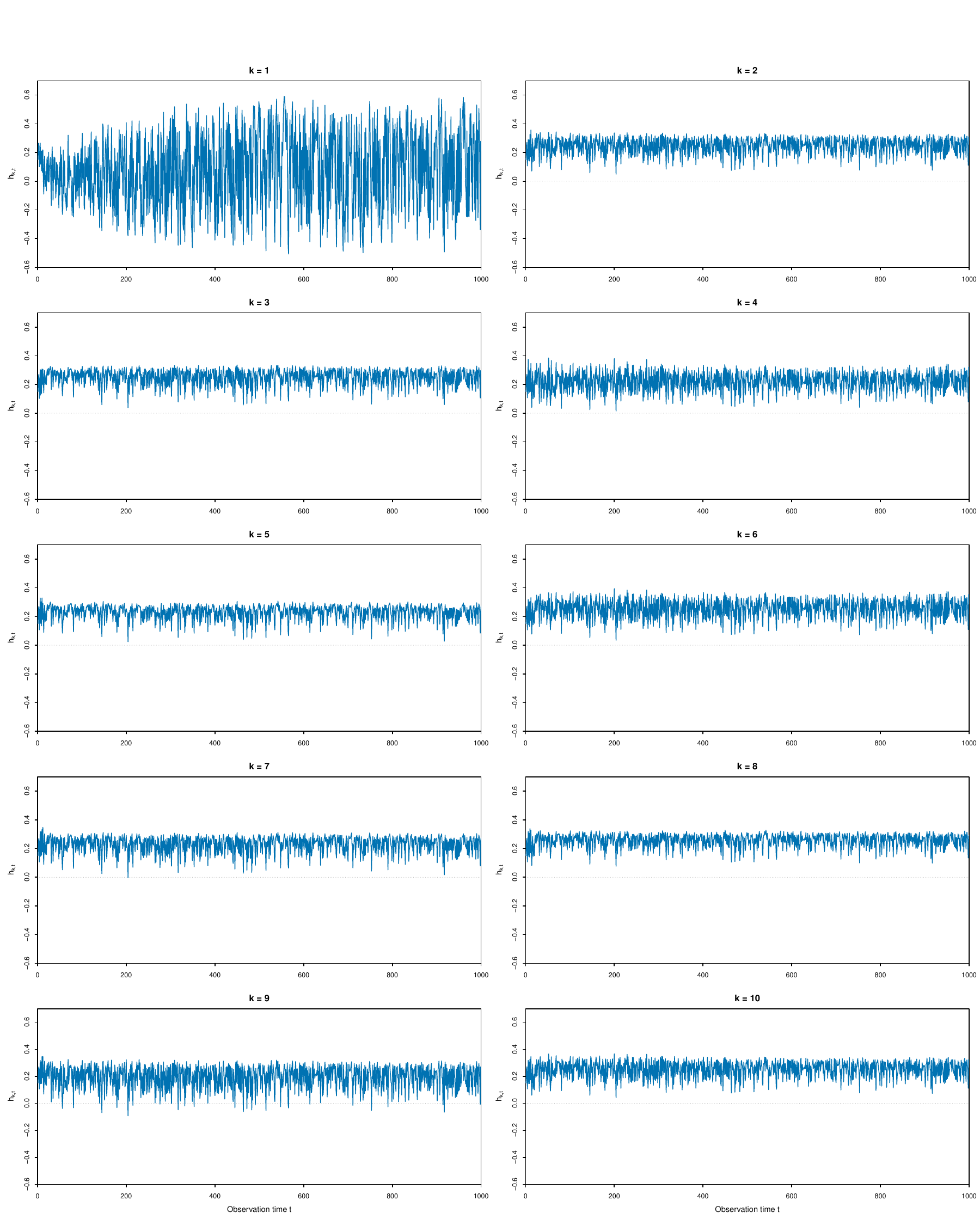}
    \caption*{Hidden-state evolutions $h_{k,t}$ for $k=1,\ldots,10$ from local projections}
    \label{fig:ma1_horizon_hidden_states}
\end{figure}

Figure~\ref{fig:ma1_horizon_hidden_states} shows the evolutions of the hidden states $h_{k,t}$ for $k=1,\ldots,10$.

\medskip
\medskip

\begin{figure}[H]
    \centering
    \caption{Cell-state Evolutions of Direct Online Estimation}
    \includegraphics[width=\textwidth,height=0.78\textheight,keepaspectratio]{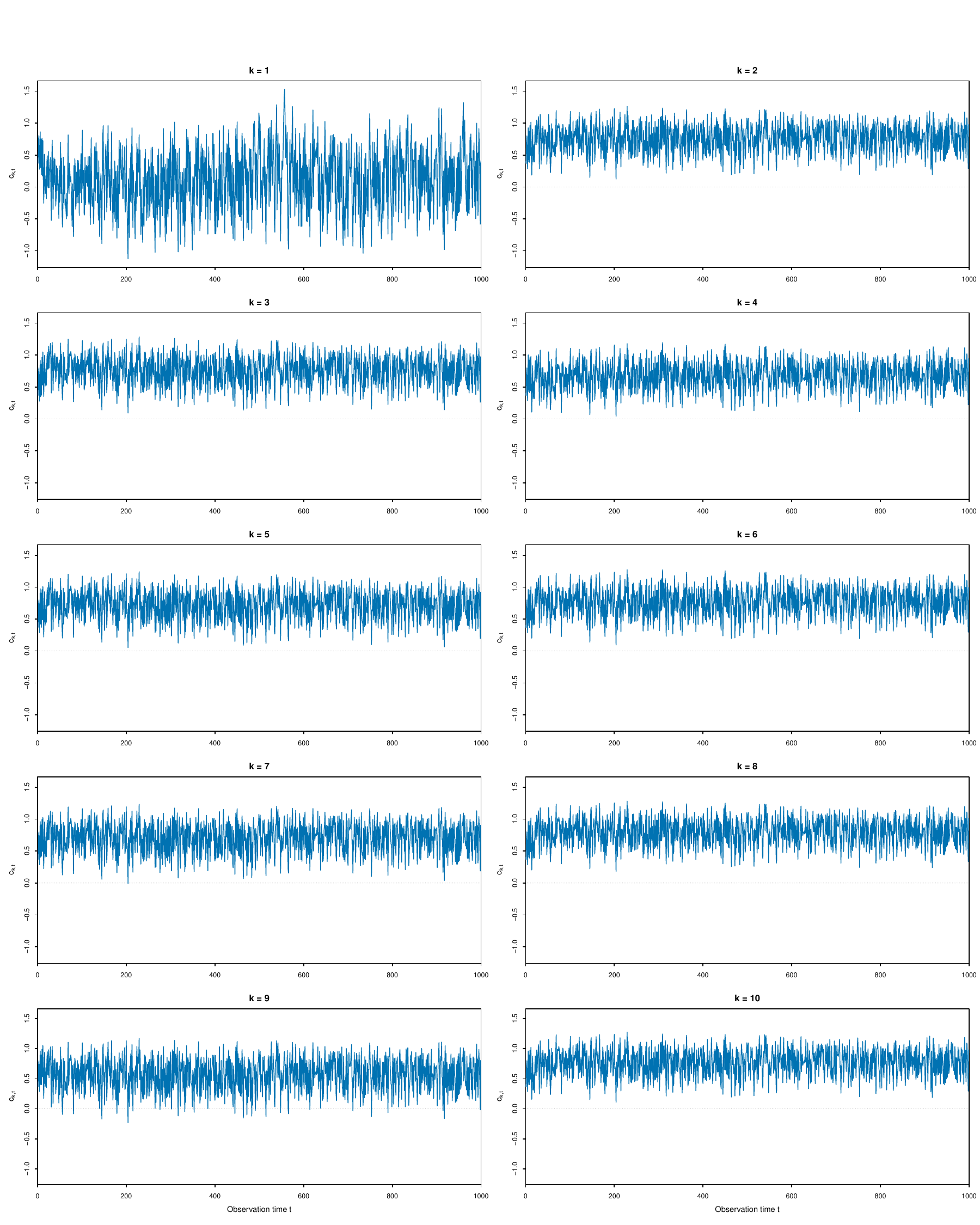}
    \caption*{Cell-state evolutions $c_{k,t}$ for $k=1,\ldots,10$, with local projections}
    \label{fig:ma1_horizon_cell_states}
\end{figure}

Figure~\ref{fig:ma1_horizon_cell_states} shows the corresponding cell-state evolutions $c_{k,t}$ for $k=1,\ldots,10$.

\medskip
\medskip

\clearpage

Visual inspection reveals similar patterns of $(h_t, c_t)$ for $k \geq 2$ that could result from reduced rank of $H_t$ (resp. for $c_t$). This possibility of reduced rank will be systematically discussed in the context of the two other architectures.
These evolutions are reported after date $t=11$ for comparison. Then, the first reported date for the states correspond to $h_{k,11} =0.2, c_{k,11}=0.5$ that differ from $h_0=c_0=0$.

\subsubsection{Multitask Approach}

Let us now consider a parallel implementation with a multitask approach assuming a common $h_t, c_t$. The multitask objective function is equiweighted. The MSE objective functions of Section 4.3.1 could also be considered as an alternative weighting (1,0,...,0), (0,1,0...,0),....(0,0,...,0,1), respectively. These eleven implementations can be done in parallel. The realizations and online predictions are given in Figure b.29. Due to the constraint of a common unit, (i.e. pair $h_t, c_t$)  the RMSE are larger than in predictions with separate units, based on larger numbers of parameters. The evolutions of the hidden states show more persistence in $c_t$ than $h_t$, with different volatility episodes (see also the unit root effect in their ACF's in Figures b.30, b.31). These unexpected evolutions of $h_t, c_t$ is likely due to the constraint of a single pair ($h_t, c_t$), whereas it seems from Section 4.3.1 (and also later 4.3.3) that $H_t$ is likely of a reduced rank equal to 2. In particular, $c_t$ has almost reached its maximum admissible value of 2 at some point in time. It would be misleading to try to interpret these evolutions as combinations of the evolutions of the ten pairs $h_{k,t}, c_{k,t}$ of Section 4.3.1 with switching regimes. There is no reason to expect a link between parameters $\theta_k^*, k=1,...,10$ and $\theta^*$.

\begin{table}[H]
    \centering
    \caption{RMSE of the joint online LSTM predictions for the invertible MA(1) process. All horizons use the same learning-rate parameters $(\alpha,\beta)=(1.1,0.2)$ and are evaluated on the common 990 times $t=11,\ldots,1000$.}
    \label{tab:ma1_joint_horizon_rmse}
    \begin{tabular}{cr}
        \hline
        Horizon $k$ &  RMSE \\
        \hline
        1 & 1.20996 \\
        2 & 1.29329 \\
        3 & 1.29075 \\
        4 & 1.29244 \\
        5 & 1.29211 \\
        6 & 1.29470 \\
        7 & 1.29321 \\
        8 & 1.29383 \\
        9 & 1.29291 \\
        10 & 1.29303 \\
        \hline
    \end{tabular}
\end{table}

\medskip

\medskip
\begin{figure}[H]
    \centering
        \caption{Evolution of the Multitask Approach's $h_t$ and $c_t$}
    \includegraphics[width=\textwidth]{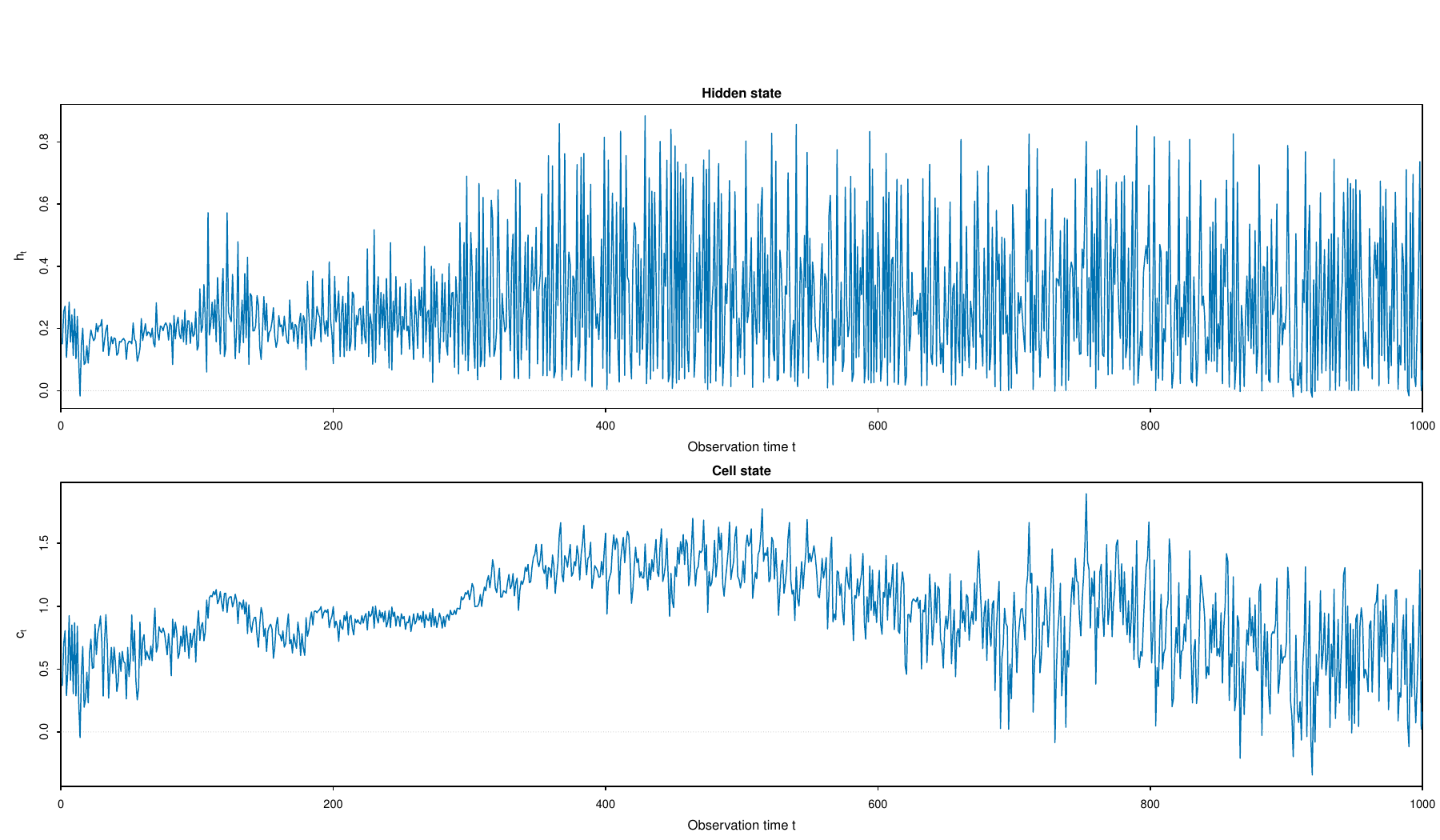}

    \label{fig:ma1_joint_hidden_cell_evolutions}
\end{figure}
\medskip

\subsubsection{Recursive Approach}

The analysis is now performed recursively with the predictions and realizations displayed in Figure b.32 and the RMSE reported in Table 4.

\begin{table}[H]
    \centering
    \caption{Recursive online forecast RMSE on the common 990 target dates $11,\ldots,1000$. The shared learning rate $\eta_n=2.1(n+1)^{-0.9}$ is selected by one-step RMSE on the same realization.}
    \label{tab:ma1_recursive_horizon_rmse}
    \begin{tabular}{crrr}
        \hline
        Horizon $k$ & $\alpha$ & $\beta$ & RMSE \\
        \hline
        1 & 2.1 & 0.9 & 1.10421 \\
        2 & 2.1 & 0.9 & 1.30008 \\
        3 & 2.1 & 0.9 & 1.29501 \\
        4 & 2.1 & 0.9 & 1.29196 \\
        5 & 2.1 & 0.9 & 1.29783 \\
        6 & 2.1 & 0.9 & 1.29082 \\
        7 & 2.1 & 0.9 & 1.29616 \\
        8 & 2.1 & 0.9 & 1.30139 \\
        9 & 2.1 & 0.9 & 1.29601 \\
        10 & 2.1 & 0.9 & 1.29062 \\
        \hline
    \end{tabular}
\end{table}

The Principal Component Analysis (PCA) of the ten series of predictions shows only two significant eigenvalues. That implies that the space generated by the predictions at different horizons is of dimension 2. It is almost entirely determined by the first two horizons. A visual inspection of Figure b.32 suggests that the predictions are almost equal at horizons $k \geq 2$, which is larger than the dimension of the affine space generated by the optimal predictions, which is equal to 1.

The analysis of Table 4 reveals a jump of RMSE from 1.10 to 1.30 between horizons 1 and 2, respectively. For the optimal predictions in the MA(1) model, this jump would be from 1 to $\sqrt{1.81} = 1.34$. The RMSE values in Table 4 are closer to the optimal values than in the local projections approach [see Table 2].

The finite dimension effect is also visible in the hidden states that become almost identical after horizon 4 for $h_t$ (see, Figure 26) and after horizon 3 for $c_t$ (see Figure 27).

\begin{figure}[H]
    \centering
       \caption{Recursive hidden states $h_{t+k-1\mid t}$ for $k=1,\ldots,10$.}
    \includegraphics[width=\textwidth,height=0.78\textheight,keepaspectratio]{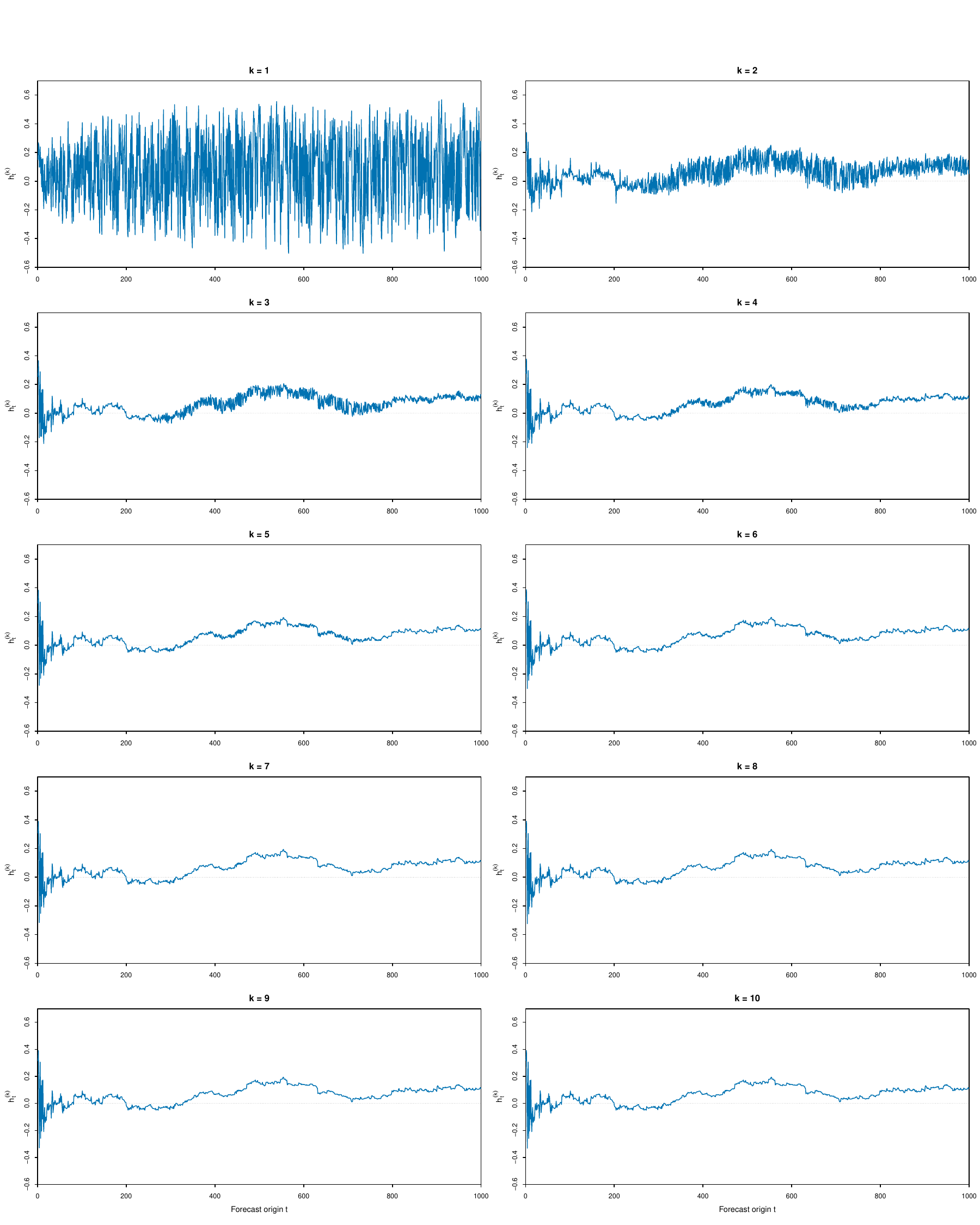}
 
    \label{fig:ma1_recursive_hidden_states}
\end{figure}

\clearpage
\begin{figure}[H]
    \centering
        \caption{Recursive cell states $c_{t+k-1\mid t}$ for $k=1,\ldots,10$}
    \includegraphics[width=\textwidth,height=0.78\textheight,keepaspectratio]{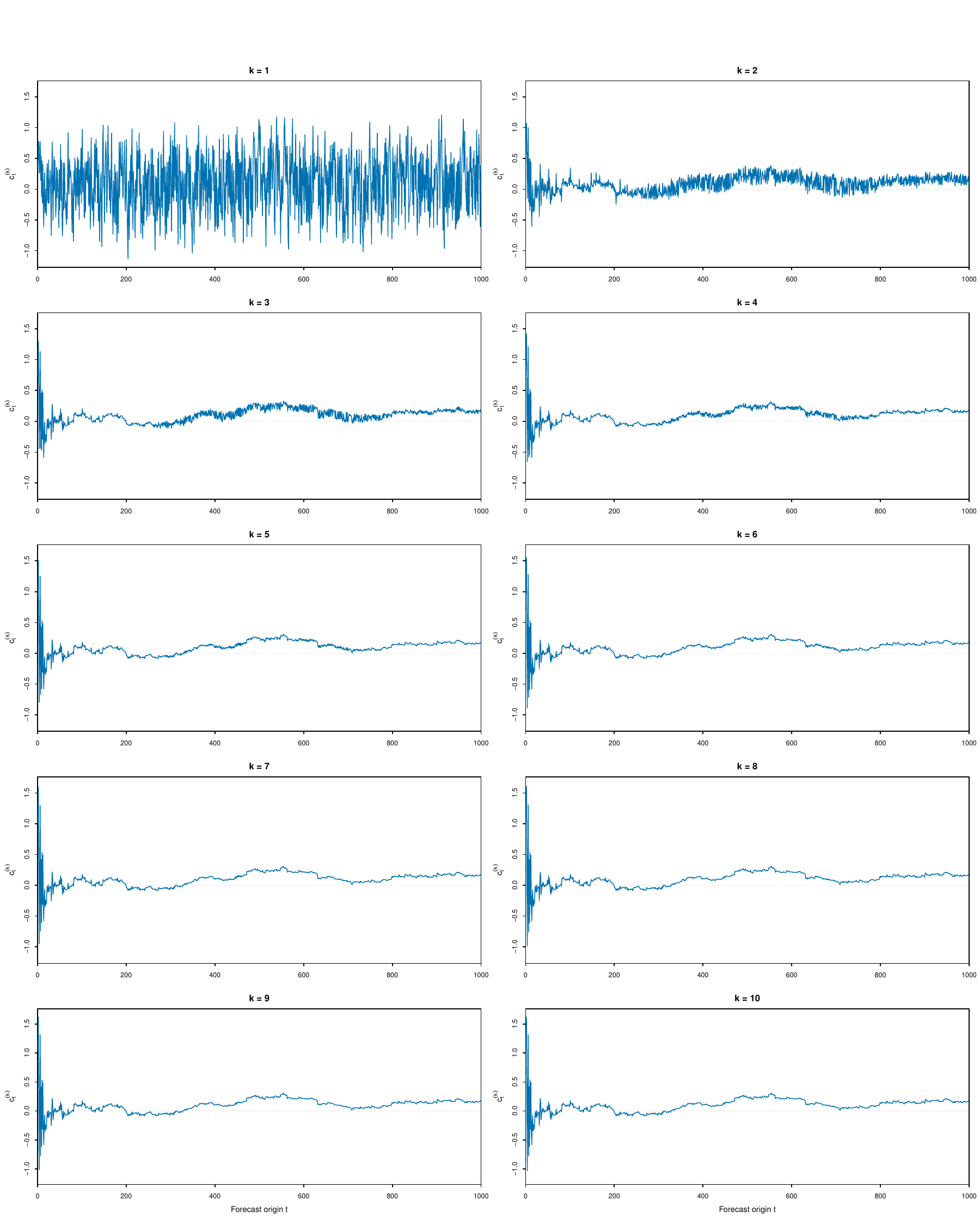}

    \label{fig:ma1_recursive_cell_states}
\end{figure}
\clearpage
% END fixed-parameter recursive MA(1) results

\section{Concluding Remarks}

There has been a large development of neural network prediction  models such as the RNN, LSTM, GRU for application to univariate and multivariate time series, due in particular to the availability of software such as PyTorch and TensorFlow.  In the literature, their performance has been analyzed by simulations and compared to the performance of other prediction models, such as random forest, Boltzmann machine, Vector Autoregressive models, state-space models, and ARCH-type models. The conclusions of these comparisons are unclear, ranging from the universal use of the LSTM prediction models [\citet{RosenbaumZhang2024Universality}, \citet{Sprecher1993UniversalMapping}] to much better performance of the RNN or TKAN prediction models [\citet{genet2025recurrent}, \citet{chen2026recurrent}], or to the difficulty to use the basic NN prediction models for more structural prediction problems such as deriving the term structure of predictions, or the predictive distributions [\citet{GourierouxMonfort2025AFS}] among the recent papers.

The aim of this paper was to study the predictions from  LSTM models applied to processes commonly used in economics and finance with an emphasis on the term structure of predictions.
In practice, this can be done by Monte-Carlo simulation methods, but it is necessary to follow the online updating of the estimates of the hyperparameters, of the hidden states, predictions, MSE, and their dynamics along the training process as diagnostic tools. This results in multiple plots shown in our paper and is only feasible if the number of parameters and horizons is not too large.
It is important to distinguish the respective roles of a parametric set of admissible prediction functions, the performance measures, i.e., the objective functions used for training, and the learning algorithms of SGD type.

In particular, the selected prediction set can be too restrictive to detect specific nonlinear characteristics of the time series of interest, such as the conditional heteroscedasticity or the speculative bubbles [\citet{zhong2026}]. More importantly, the same parametric set of prediction functions cannot be adequate for different objectives. It is difficult to have a parametric prediction model that would work for a large spectrum of DGP's and problems of interest. This approach results in sub-optimal predictions, and their sub-optimality is difficult to measure and depends on the environment.

\bigskip

\nocite{*}
\bibliographystyle{apalike-first-initial}
\bibliography{references}

\bigskip

\section*{Appendices}

\section*{Appendix A}

\subsection*{A Recursive LSTM +SGD with common LSTM unit}

This version  is obtained by applying automatic multistep forecasting strategies [see e.g. \citet{BenTaiebHyndman2012RecursiveDirect}, Section 2.1, \citet{GreenEtAl2025Stratify}]. More precisely, in contrast to the recursive approach described in Section 3.1, where the LSTM parameters $\theta_k$ at step $k$ were assumed functionally independent and optimized separately, we now assume that $\theta_k = \tilde{\theta}$ is independent of horizon $k$. Then, Figure~\ref{fig:recursive_scheme} is modified with the same previous $\tilde{\theta}_{t-1}$ and still the $h_{k, t-1}, \; k=1,...,K$. Then,  $\tilde{\theta}_{t}$
is deduced by optimizing a multi-task objective function taking into account the different horizons.

This approach leads to final layers of the type:

$$\hat{y}_{t,t+k} = \hat{w}_{y,t} h_{k,t} + \hat{b}_{y,t}, \;\; k=1,...,K.$$

\nin They differ from the equations corresponding to the ASTM with a single factor (see, (3.2)) that are:

$$\hat{y}_{t,t+k} = \hat{w}_{y,k,t} h_{t} + \hat{b}_{y,k,t}, \;\; k=1,...,K.$$

\nin In fact, the multi-task versions of the direct and recursive LSTM +SGD approaches do not share the hidden layers in the same way. They share the parameters and state variables in the direct approach, and only the parameters in the recursive approach.

\bigskip

\begin{center}
\section*{Appendix B}
\end{center}

% Automatic figure numbering for Appendix B.
\setcounter{figure}{0}
\renewcommand{\thefigure}{b.\arabic{figure}}
\renewcommand{\theHfigure}{appendixB.\arabic{figure}}

B.1 Gaussian and Cauchy white noises

\begin{figure}[H]
    \centering
        \caption{ACF and PACF of Gaussian White Noise}
    \includegraphics[width=1\linewidth]{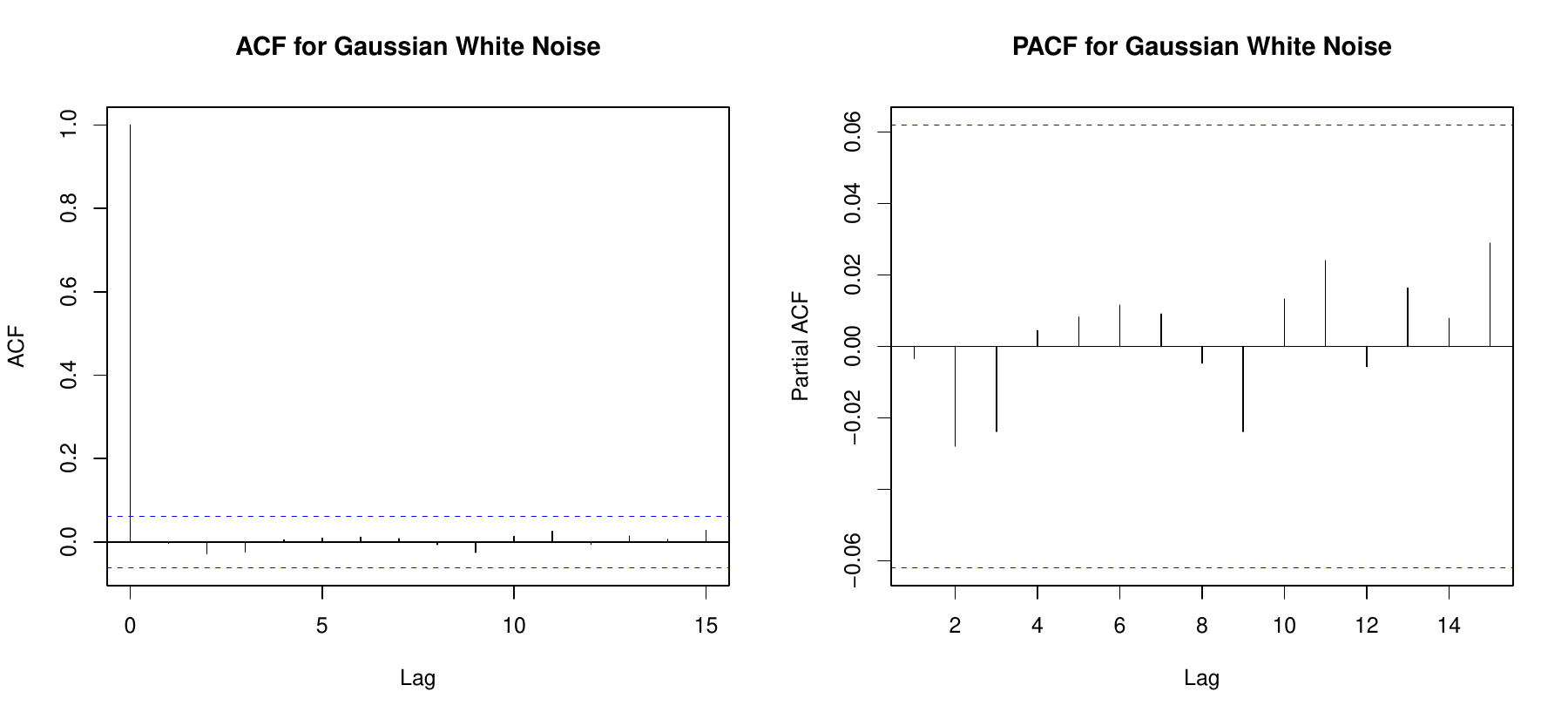}

    \label{fig:gwn_acf}
\end{figure}

\begin{figure}[H]
    \centering
        \caption{Hidden State $h_t$ Evolution for Gaussian White Noise Series}
    \includegraphics[width=1\linewidth]{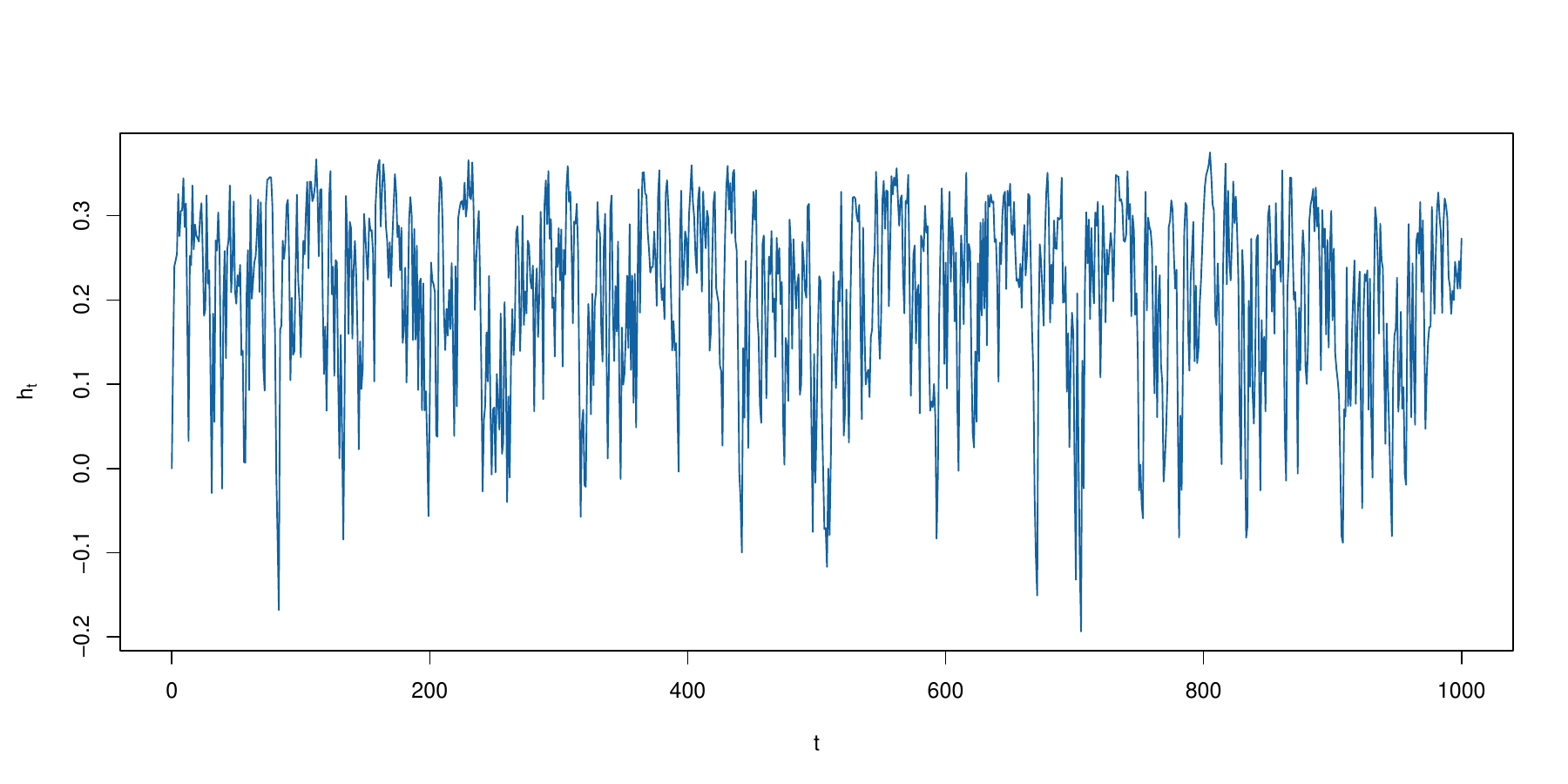}

    \label{fig:g_h_t series}
\end{figure}

\begin{figure}[H]
    \centering
        \caption{Cell State $c_t$ Evolution for Gaussian White Noise Series}
    \includegraphics[width=1\linewidth]{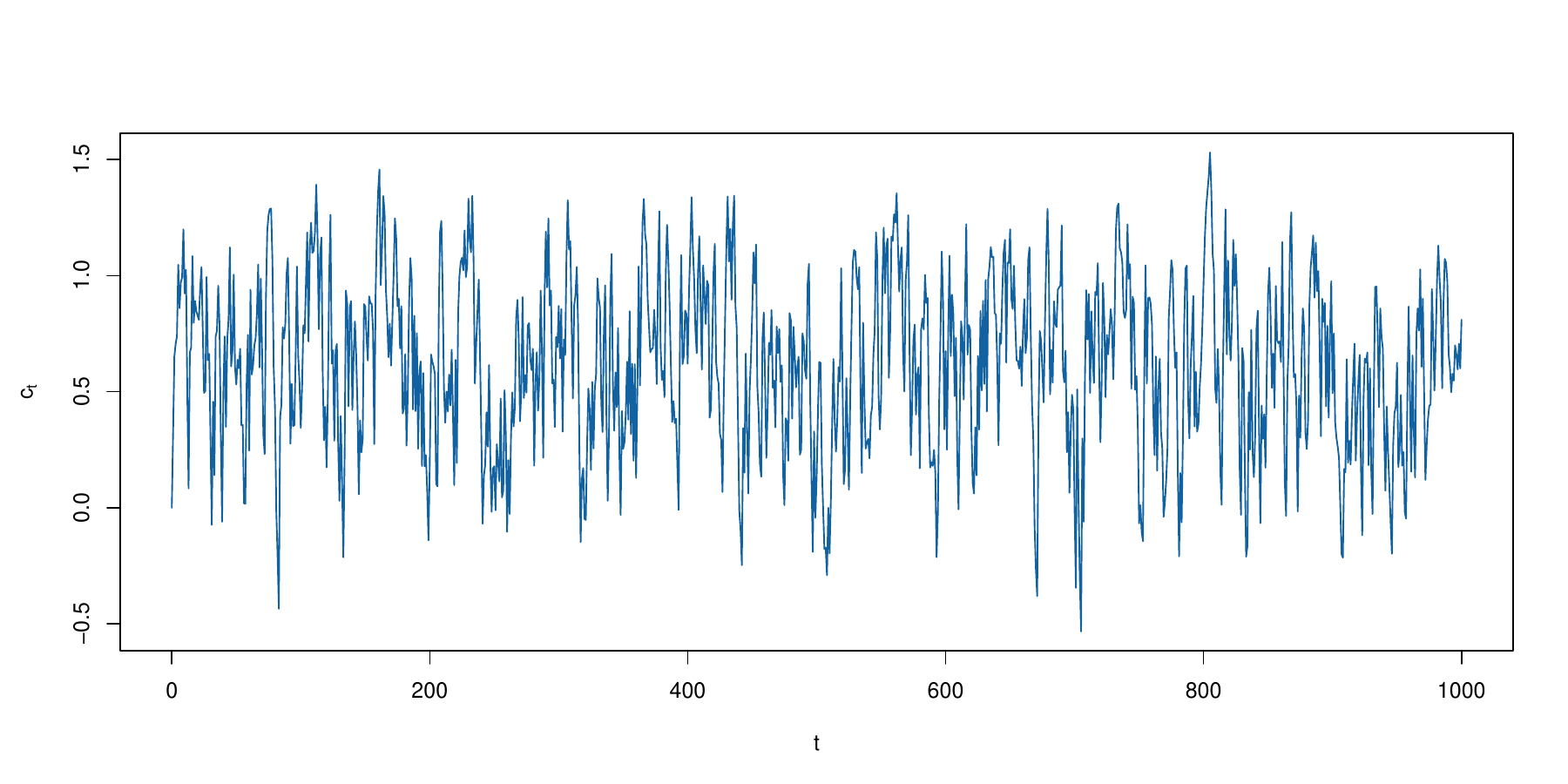}

    \label{fig:g_c_t series}
\end{figure}

\begin{figure}[H]
    \centering
        \caption{ACF and PACF of Hidden State $h_t$ for Gaussian White Noise Series}
    \includegraphics[width=1\linewidth]{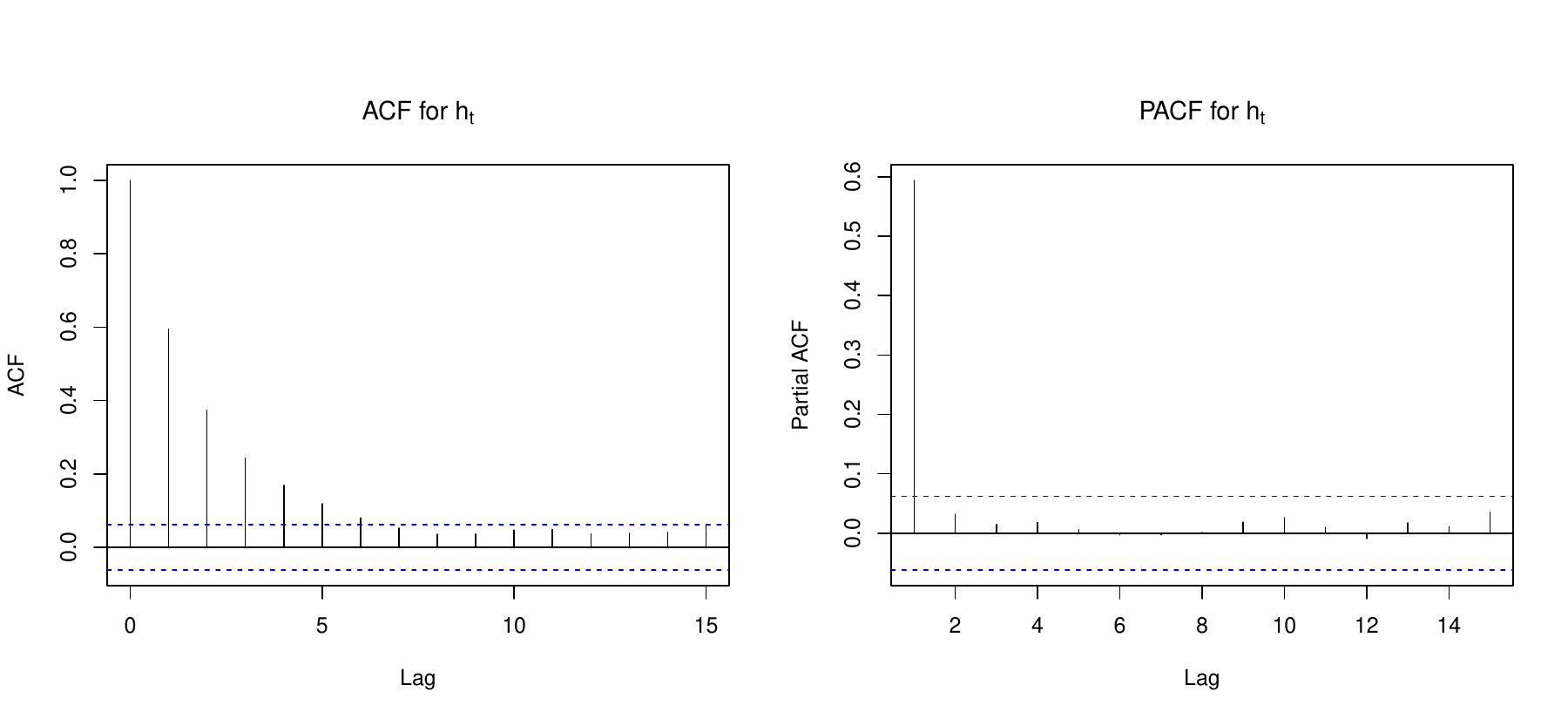}

    \label{fig:g_h_t acf}
\end{figure}

\begin{figure}[H]
    \centering
        \caption{ACF and PACF of Cell State $c_t$ for Gaussian White Noise Series}
    \includegraphics[width=1\linewidth]{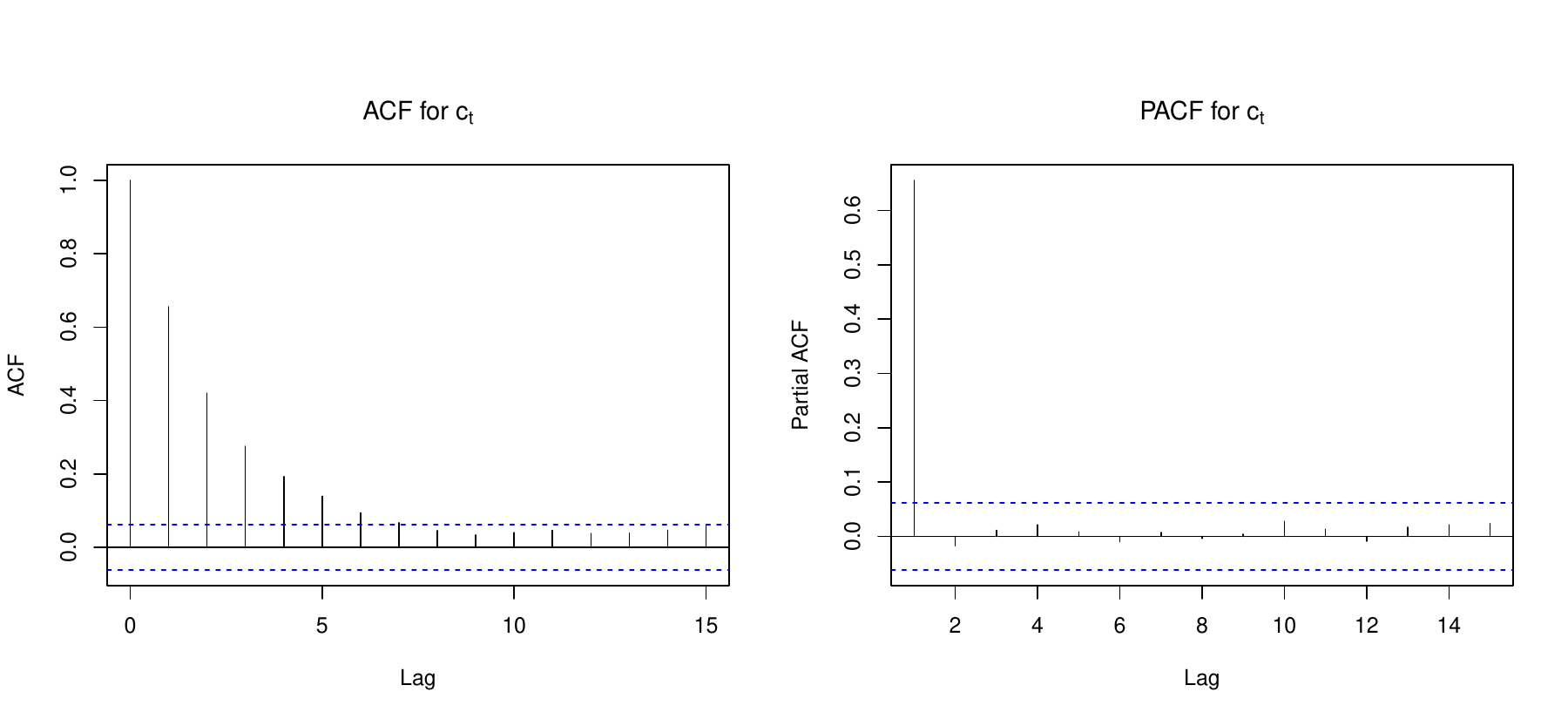}

    \label{fig:g_c_t acf}
\end{figure}

\begin{figure}[H]
    \centering
        \caption{Realizations and Predictions: Cauchy White Noise}
    \includegraphics[width=1\linewidth]{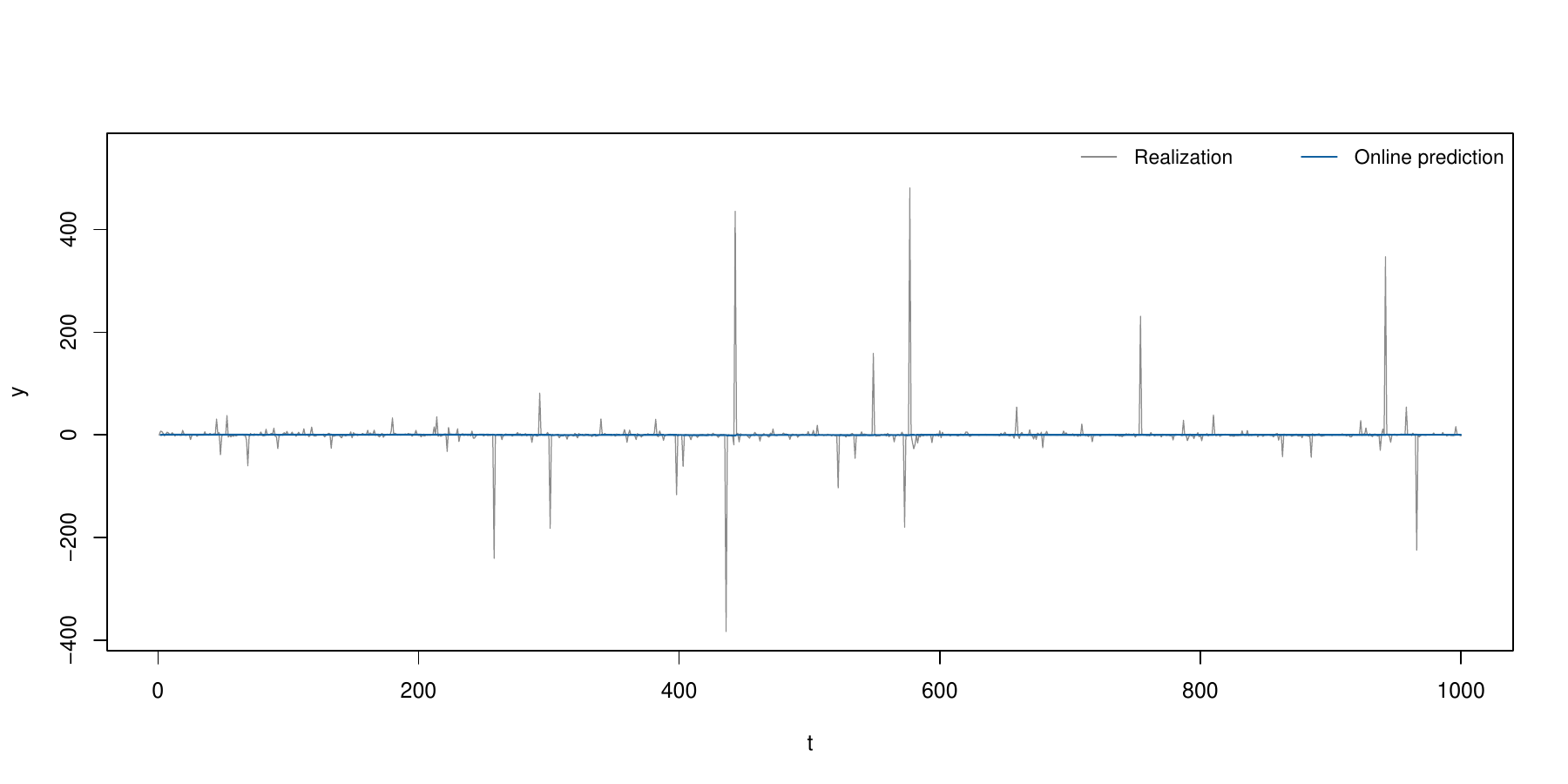}

    \label{fig:cwn}
\end{figure}

\begin{figure}[H]
    \centering
        \caption{Hidden State $h_t$ Evaluation for Cauchy White Noise Series}
    \includegraphics[width=1\linewidth]{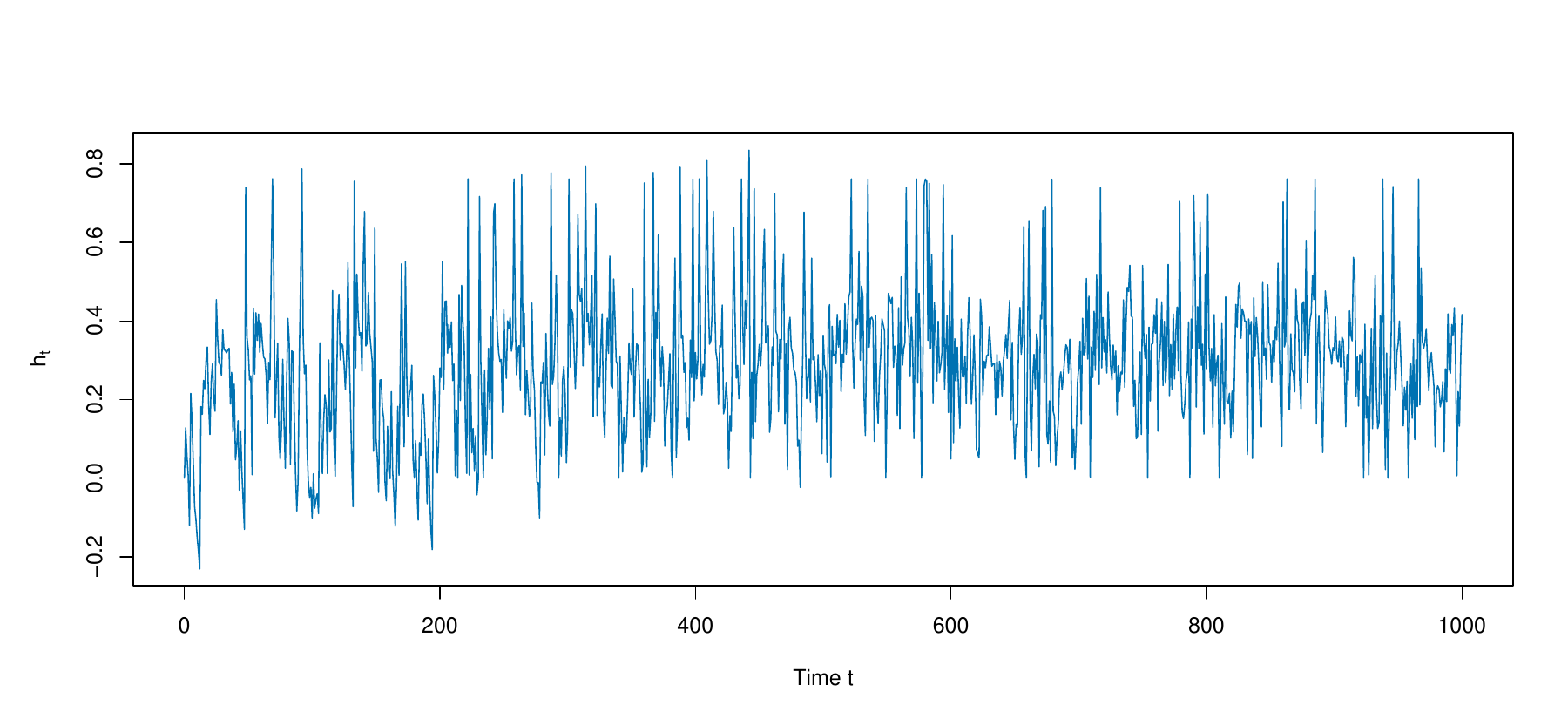}

    \label{fig:c_h_t series}
\end{figure}

\begin{figure}[H]
    \centering
        \caption{Cell State $c_t$ Evolution for Cauchy White Noise Series}
    \includegraphics[width=1\linewidth]{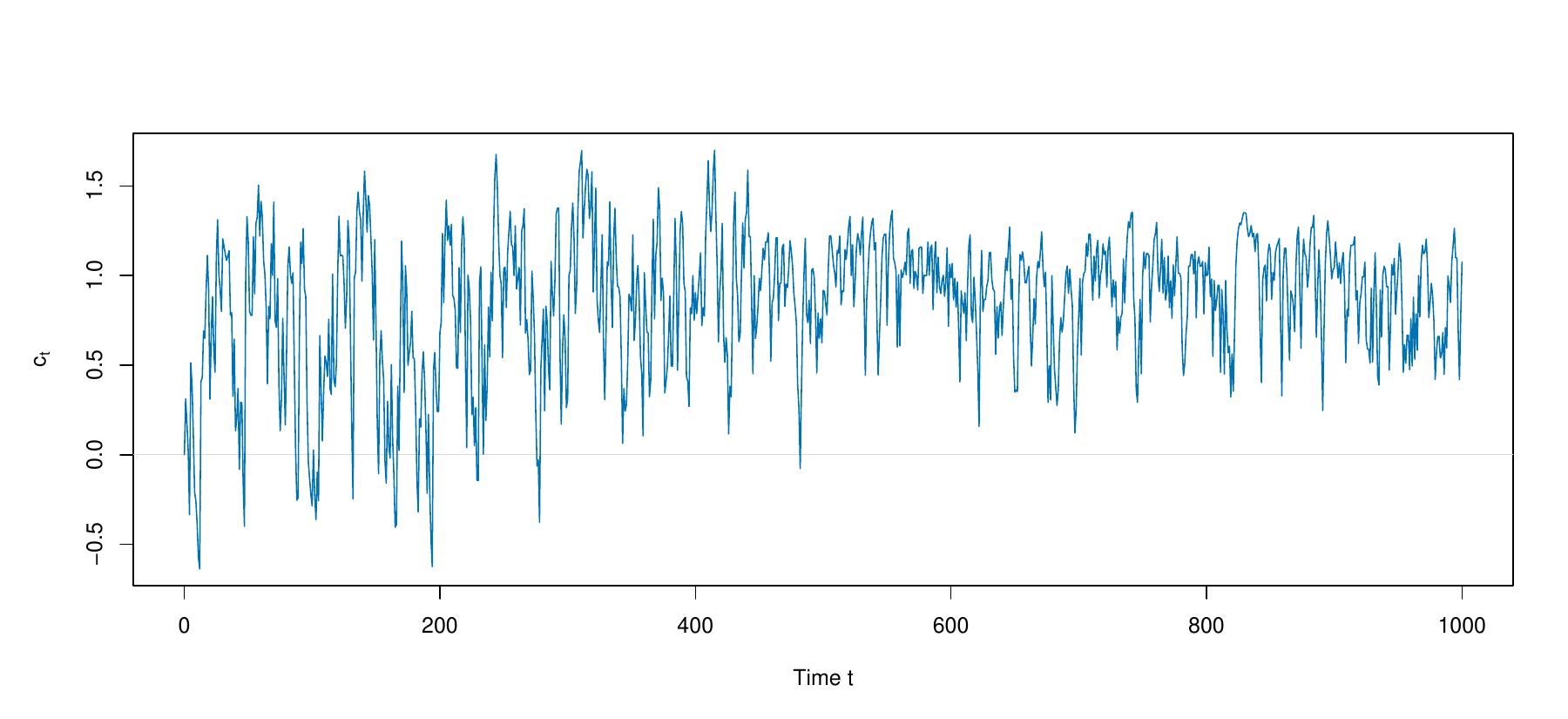}

    \label{fig:c_c_t series}
\end{figure}

\begin{figure}[H]
    \centering
        \caption{ACF and PACF of Hidden State $h_t$ for Cauchy Simulation Series}
    \includegraphics[width=1\linewidth]{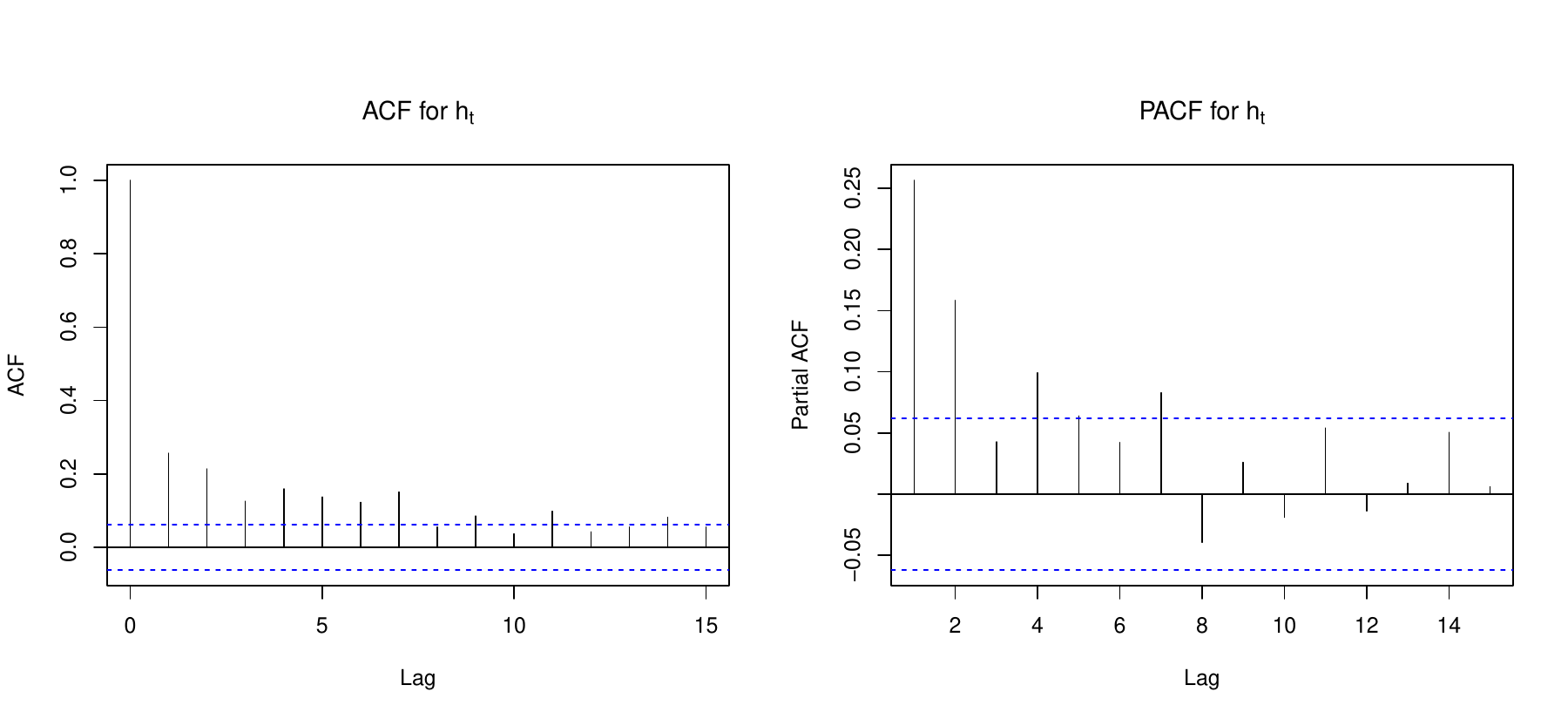}

    \label{fig:c_h_t acf}
\end{figure}

\begin{figure}[H]
    \centering
        \caption{ACF and PACF of Cell State $c_t$ for Cauchy White Noise Series}
    \includegraphics[width=1\linewidth]{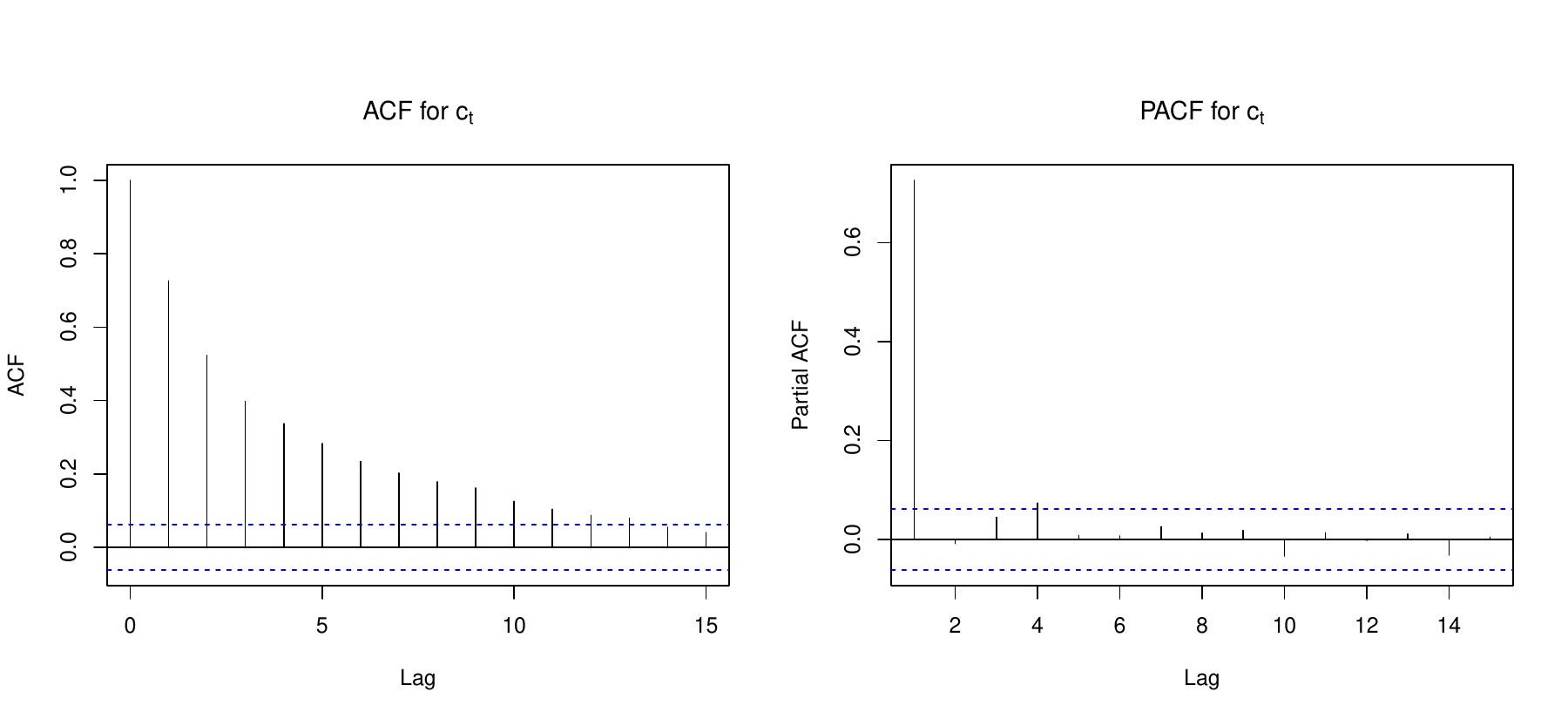}

    \label{fig:c_c_t acf}
\end{figure}

\clearpage
\nin B.2 MA(1) experiments

\begin{figure}[H]
    \centering
        \caption{Realizations and Predictions For MA(1) Input (1 unit, 1 epoch)}
    \includegraphics[width=1\linewidth]{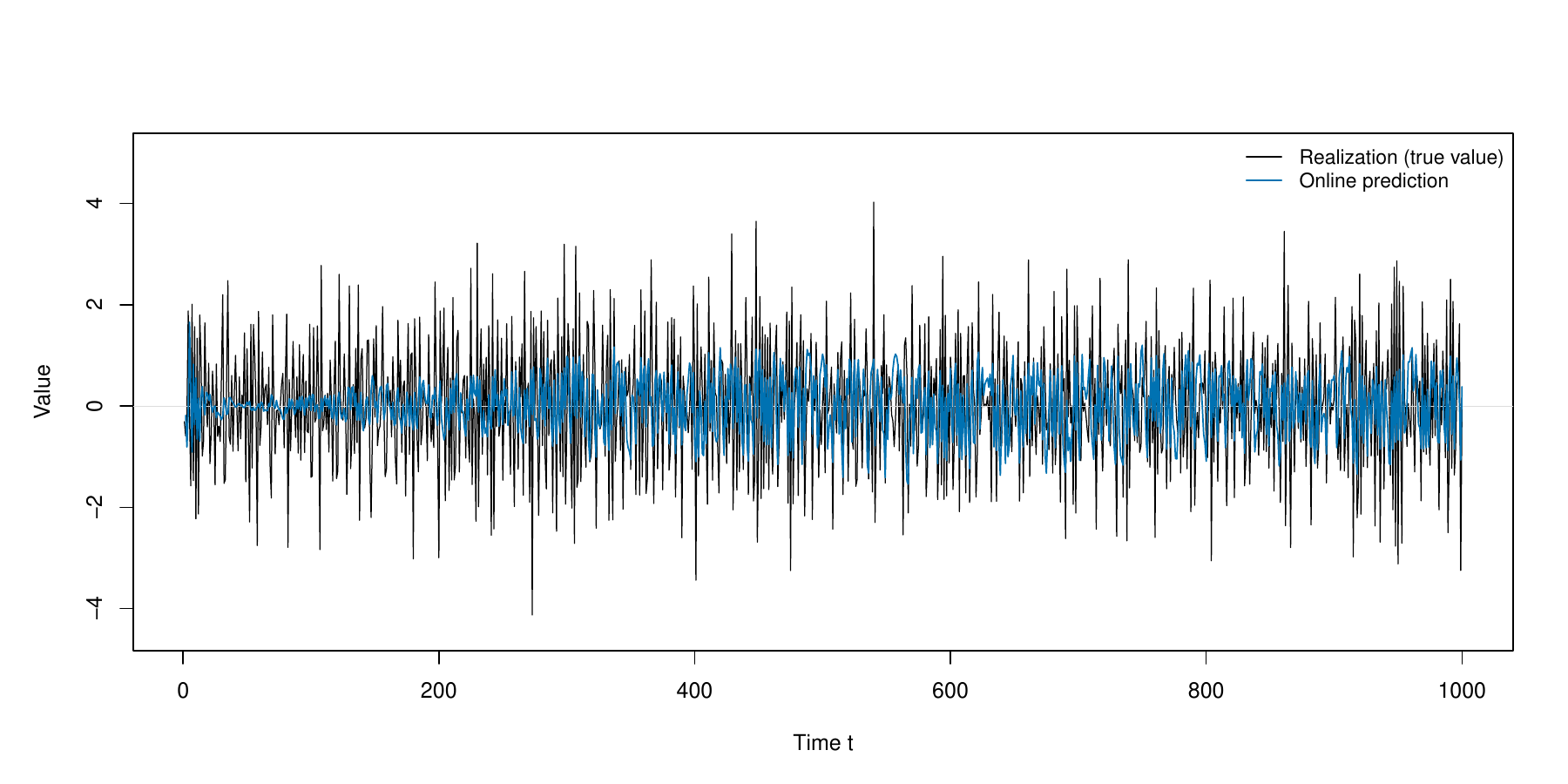}

    \label{fig:ma1u1}
\end{figure}

\begin{figure}[H]
    \centering
    \caption{Scatter Plot of Online Predictions and Realizations}
    \includegraphics[width=\linewidth]{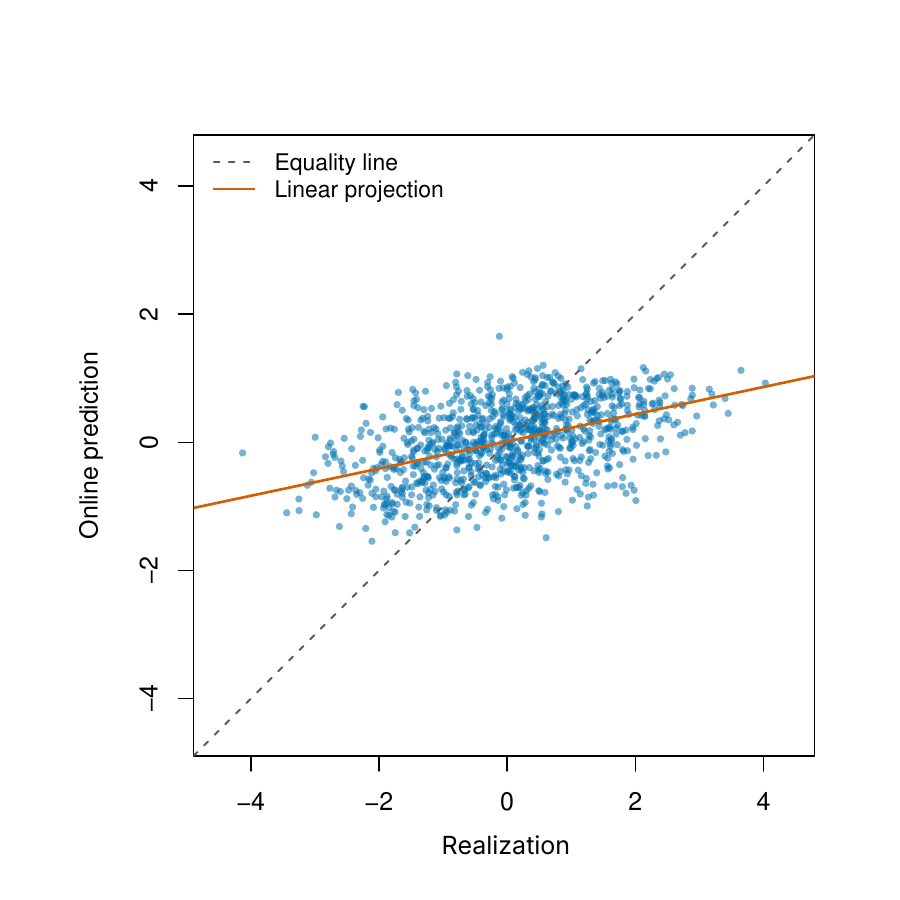}
    \caption*{MA(1) Realizations (horizontal) and online LSTM predictions (vertical). The dashed line indicates equality; the solid line is the sample linear projection of predictions on realizations, including an intercept.}
    \label{fig:ma1onlinevstrue}
\end{figure}

\begin{figure}[H]
    \centering
        \caption{Scatter Plot of Optimal Predictions and Realizations}
    \includegraphics[width=\linewidth]{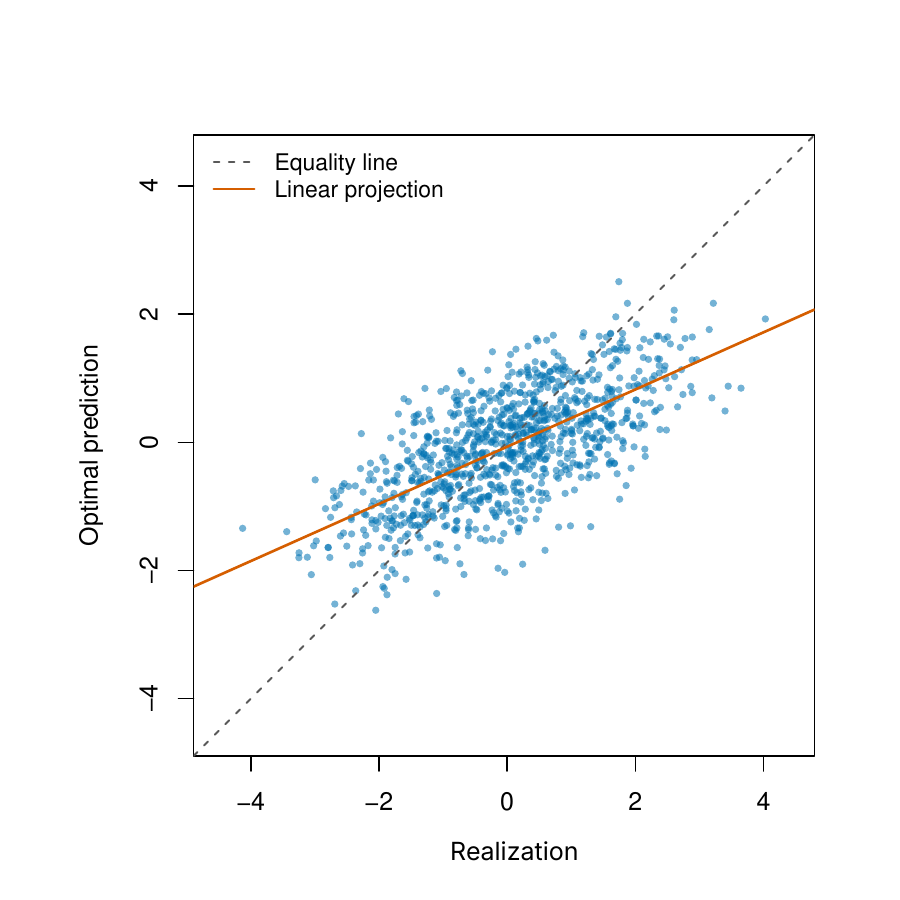}
    \caption*{MA(1) Realizations (horizontal) and optimal predictions (vertical). The optimal forecast is the conditional mean given finite observed history. The dashed line indicates equality; the solid line is the sample linear projection of predictions, including an intercept.}
    \label{fig:ma1optimalvstrue}
\end{figure}

\begin{figure}[H]
    \centering
        \caption{Hidden State $h_t$ Evolution of MA(1) Simulated $y_t$ Series}
    \includegraphics[width=1\linewidth]{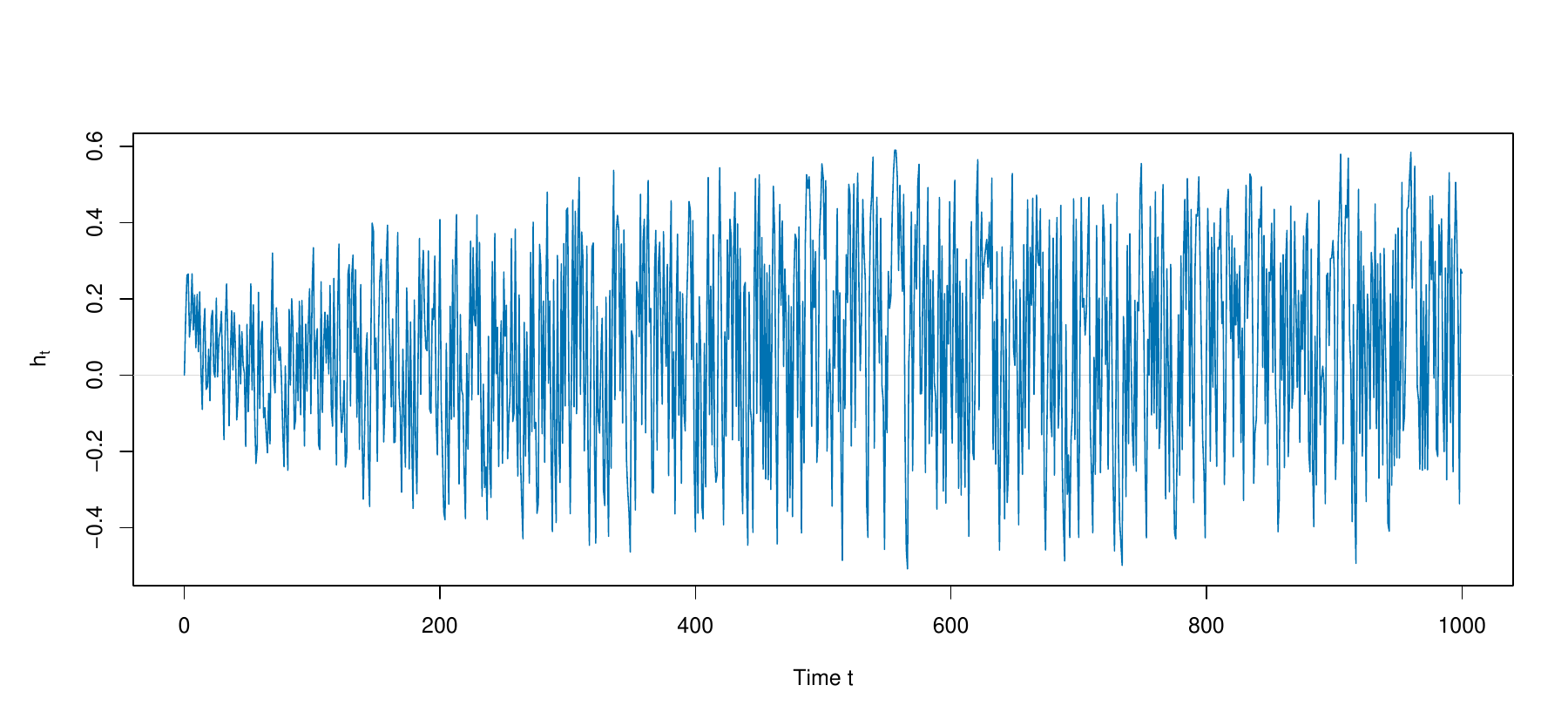}

    \label{fig:ma1_h_t series}
\end{figure}

\begin{figure}[H]
    \centering
        \caption{Cell State $c_t$ Evolution of MA(1) Simulated $y_t$ Series}
    \includegraphics[width=1\linewidth]{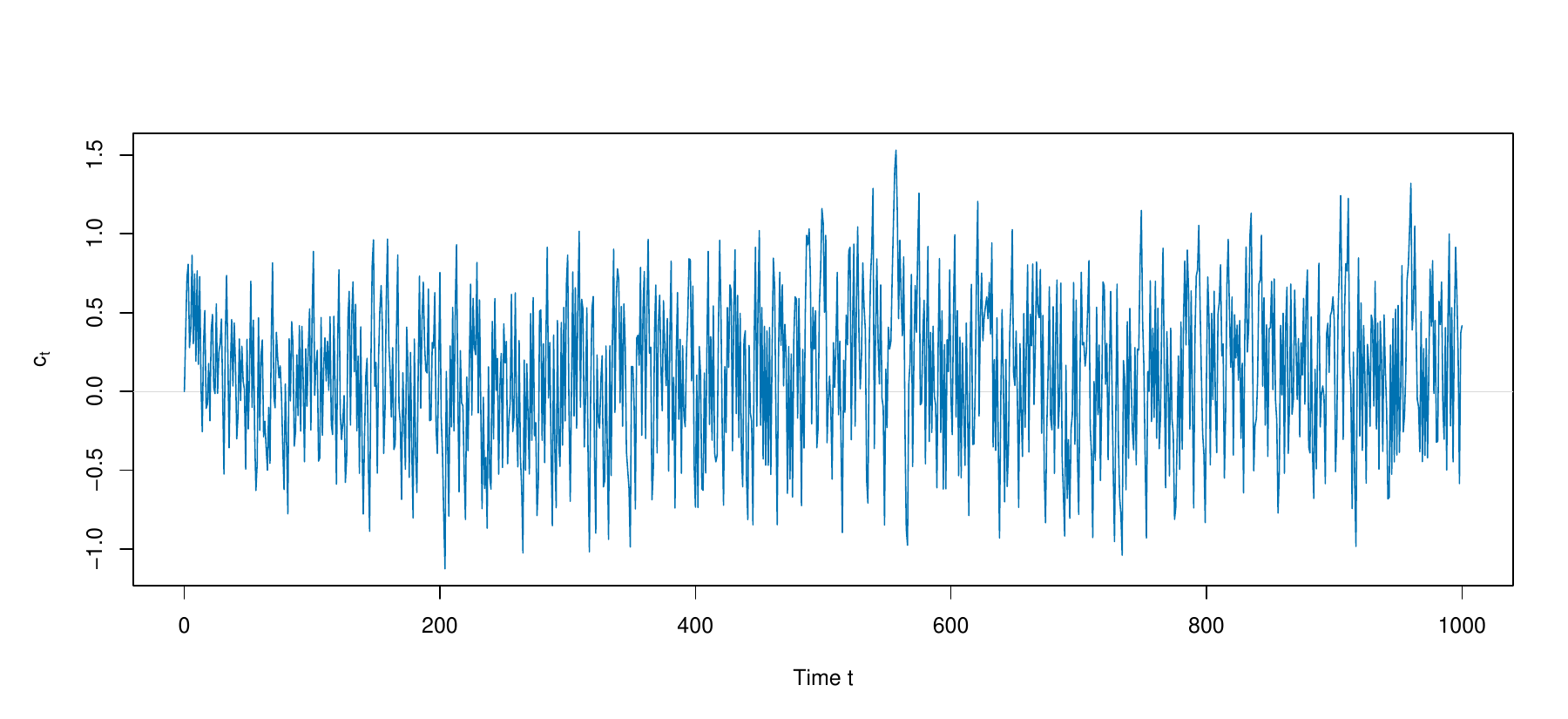}

    \label{fig:ma1_c_t series}
\end{figure}

\begin{figure}[H]
    \centering
        \caption{ACF and PACF of Hidden State $h_t$ for MA(1) Simulated $y_t$ Series}
    \includegraphics[width=1\linewidth]{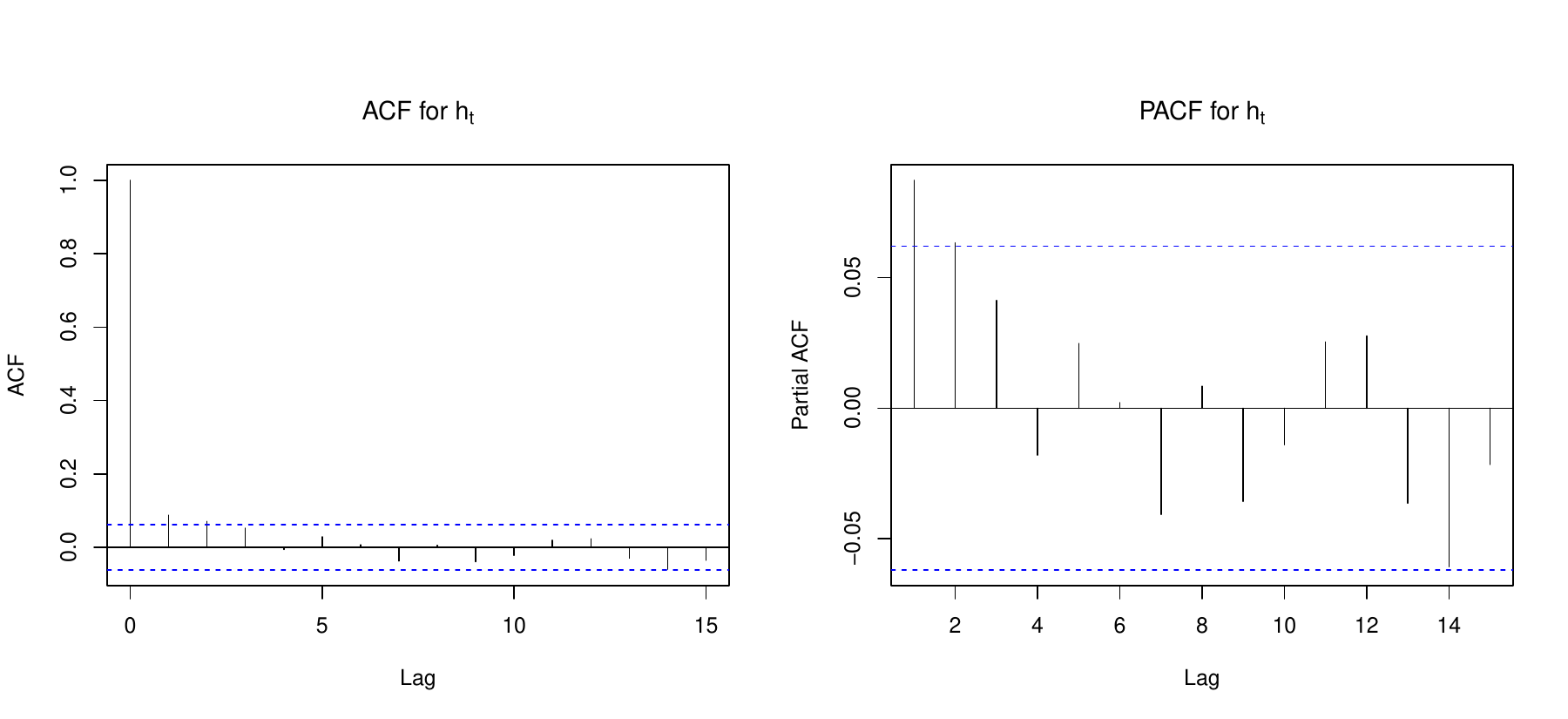}

    \label{fig:ma1_h_t acf}
\end{figure}

\begin{figure}[H]
    \centering
        \caption{ACF and PACF of Cell State $c_t$ for MA(1) Simulated $y_t$ Series}
    \includegraphics[width=1\linewidth]{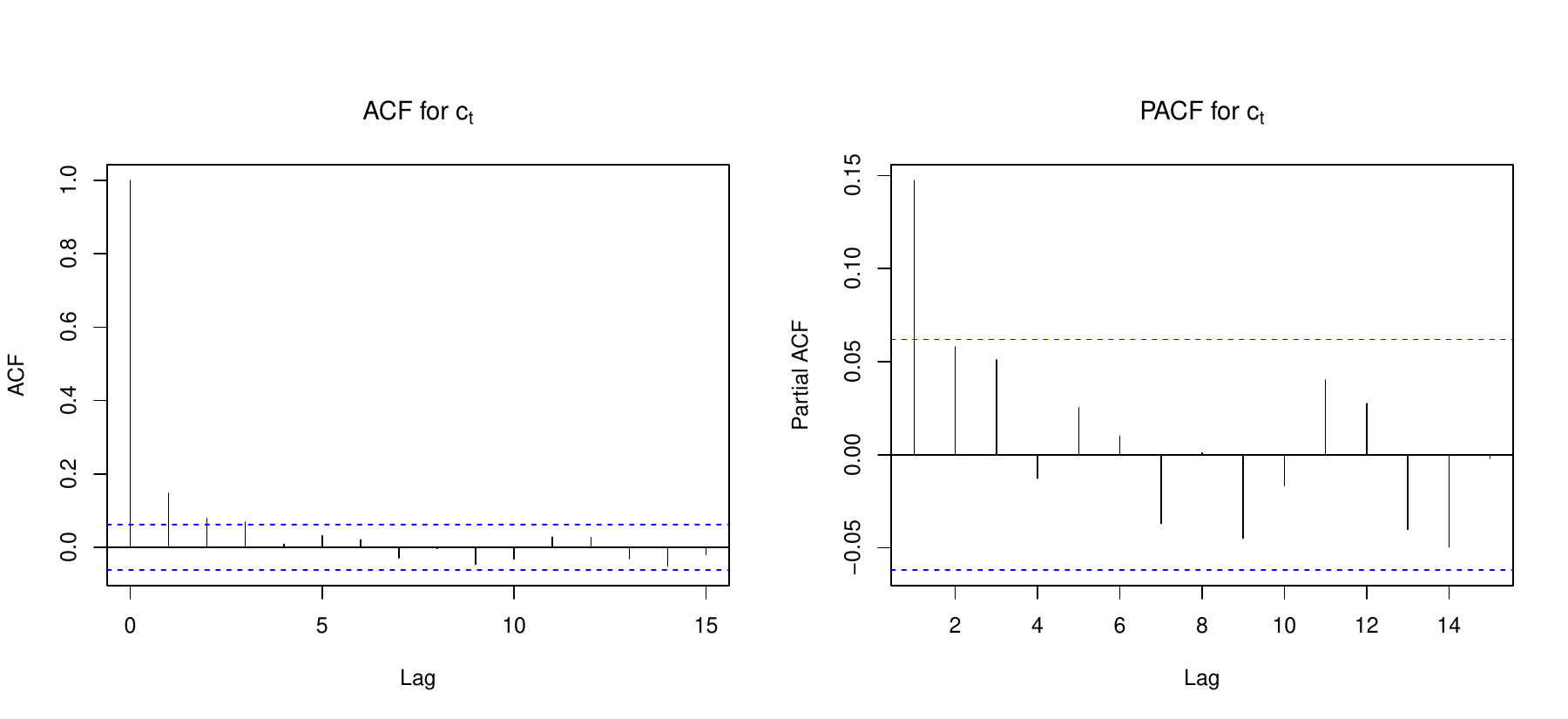}

    \label{fig:ma1_c_t acf}
\end{figure}

\begin{figure}[H]
    \centering
        \caption{Hidden State $h_t$ Evaluation for MA(1) $y_t^2$  Series}
    \includegraphics[width=1\linewidth]{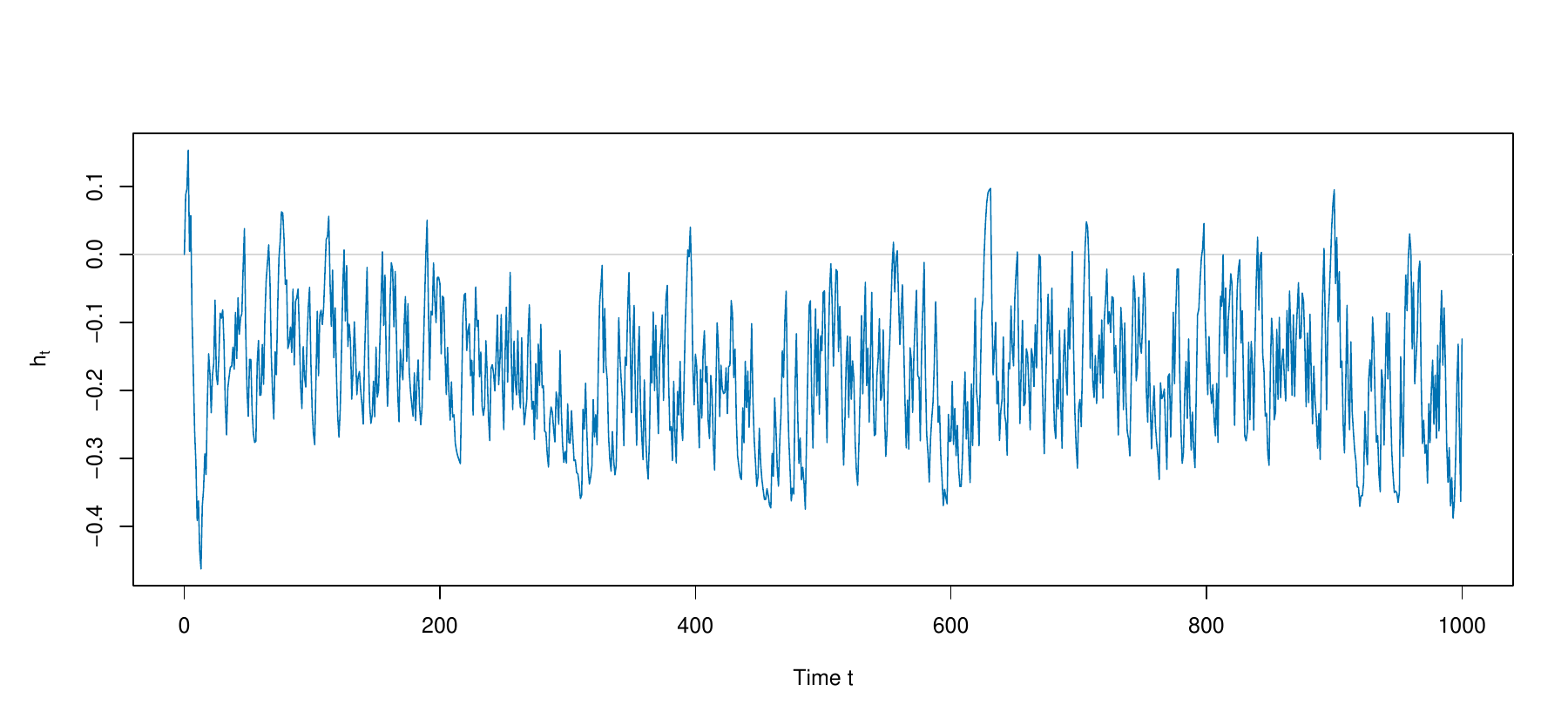}

    \label{fig:y2_h_t series}
\end{figure}

\begin{figure}[H]
    \centering
        \caption{Cell State $c_t$ Evaluation for MA(1) $y_t^2$ Series}
    \includegraphics[width=1\linewidth]{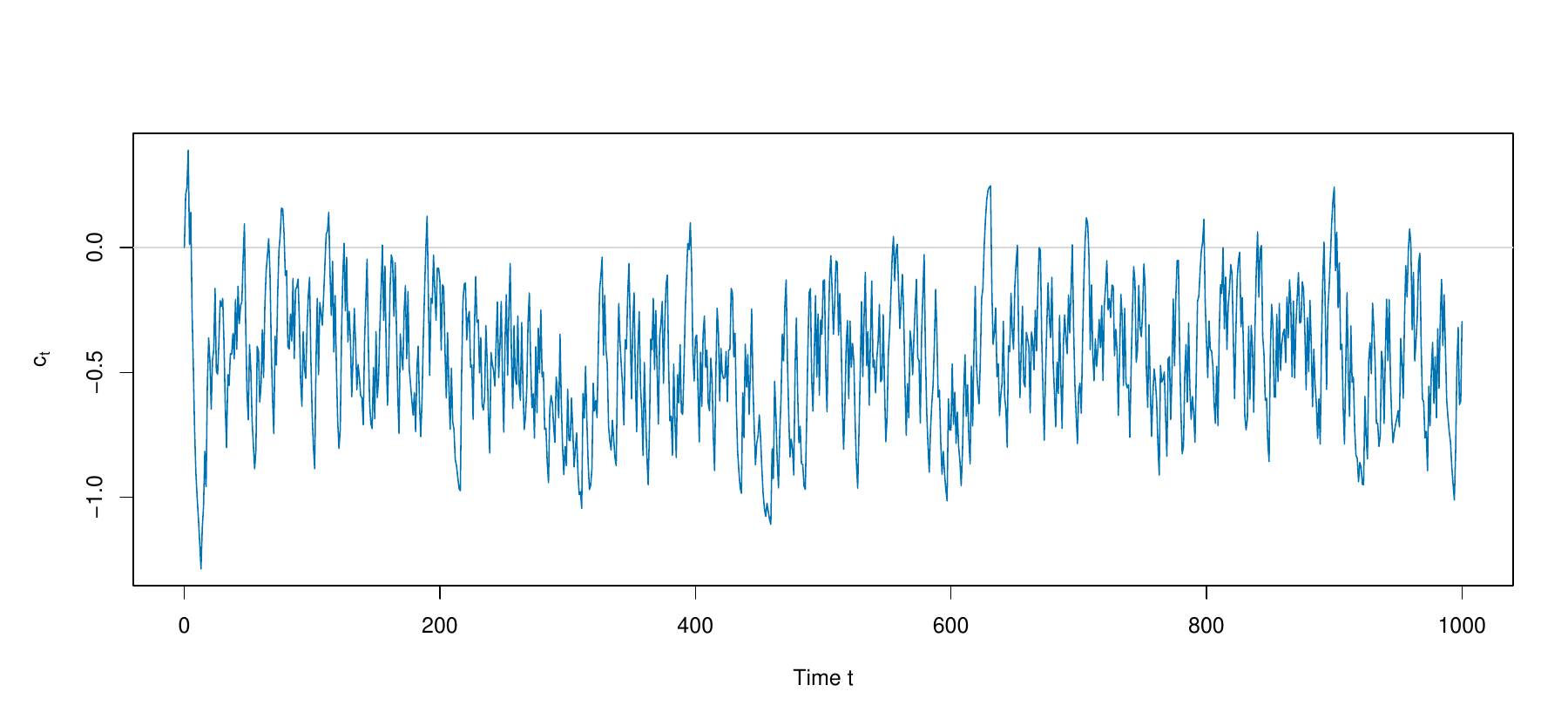}

    \label{fig:y2_c_t series}
\end{figure}

\begin{figure}[H]
    \centering
        \caption{ACF and PACF of Hidden State $h_t$ for MA(1) $y_t^2$ Series}
    \includegraphics[width=1\linewidth]{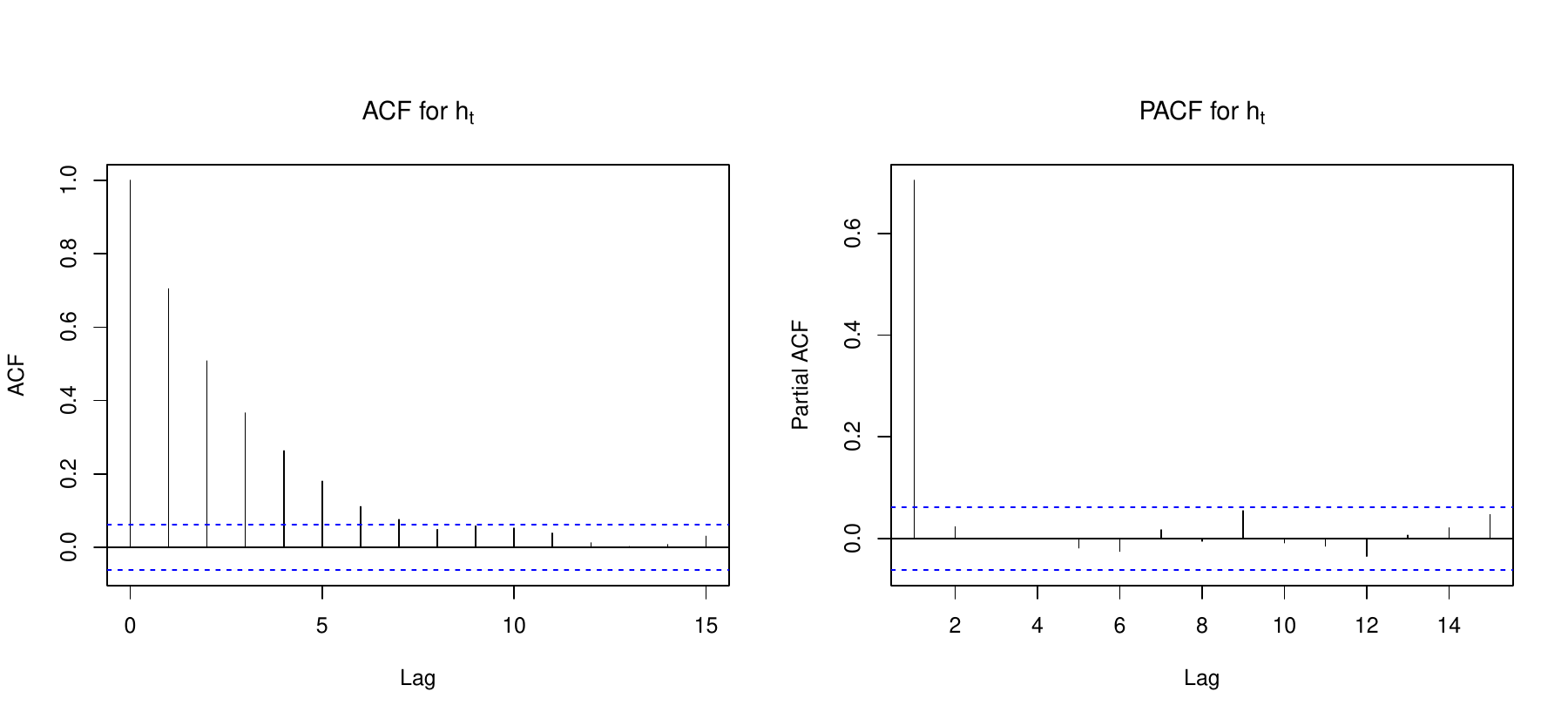}

    \label{fig:y2_h_t acf}
\end{figure}

\begin{figure}[H]
    \centering
        \caption{ACF and PACF of Cell State $c_t$ for MA(1) $y_t^2$ Series}
    \includegraphics[width=1\linewidth]{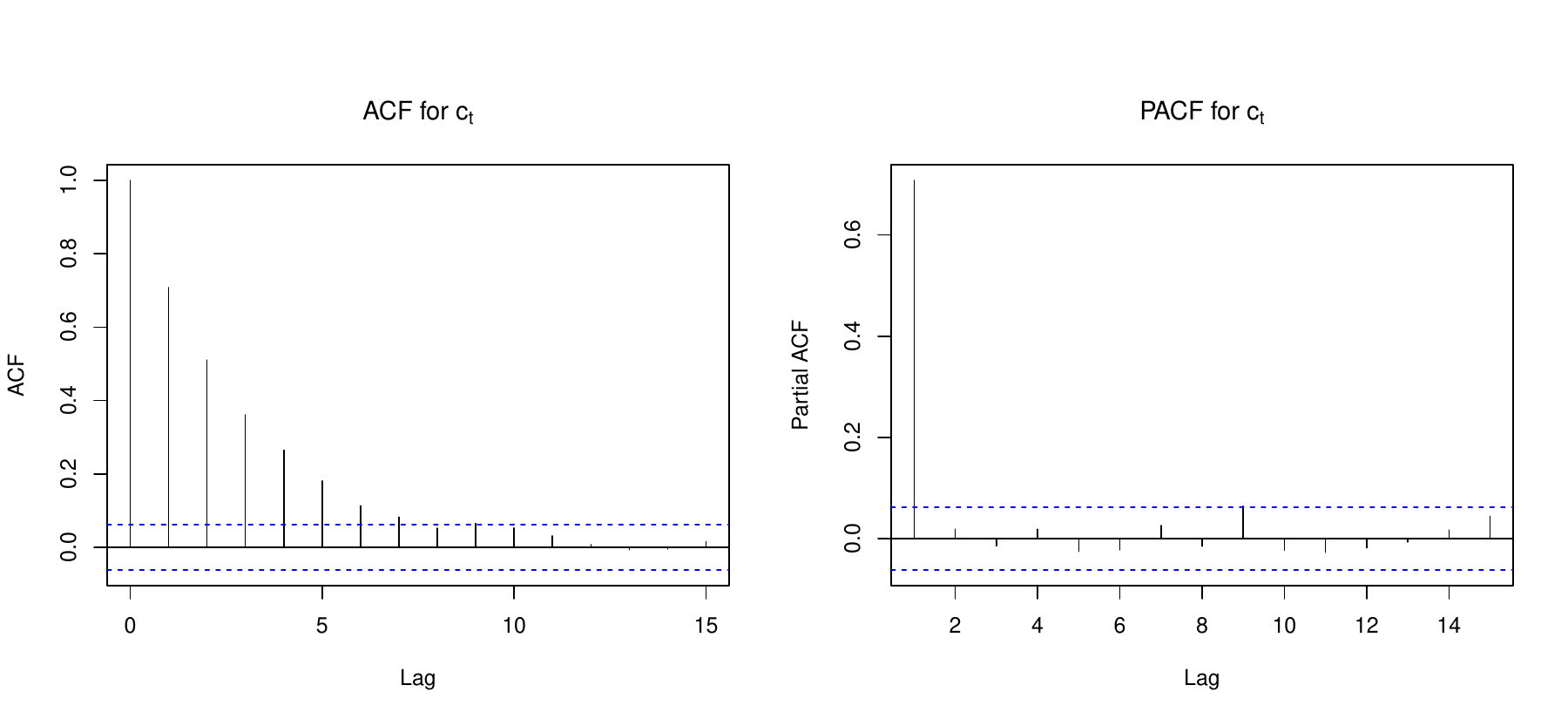}

    \label{fig:y2_c_t acf}
\end{figure}

\begin{figure}[H]
    \centering
        \caption{Bivariate Prediction and Realizations of $y_t$}
    \includegraphics[width=1\linewidth]{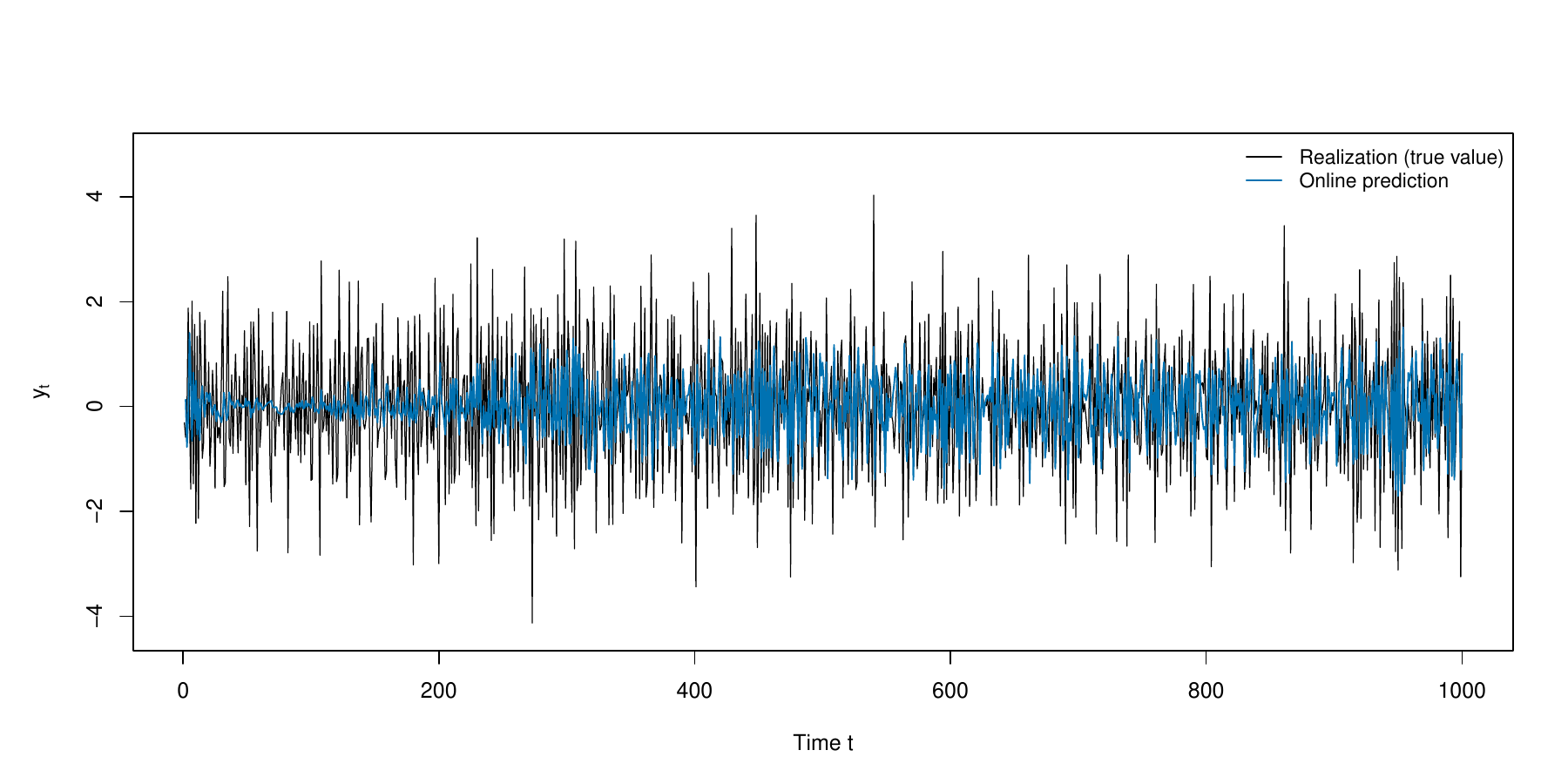}

    \label{fig:bi_y}
\end{figure}

\begin{figure}[H]
    \centering
        \caption{Bivariate Prediction and Realizations of $y_t^2$ }
    \includegraphics[width=1\linewidth]{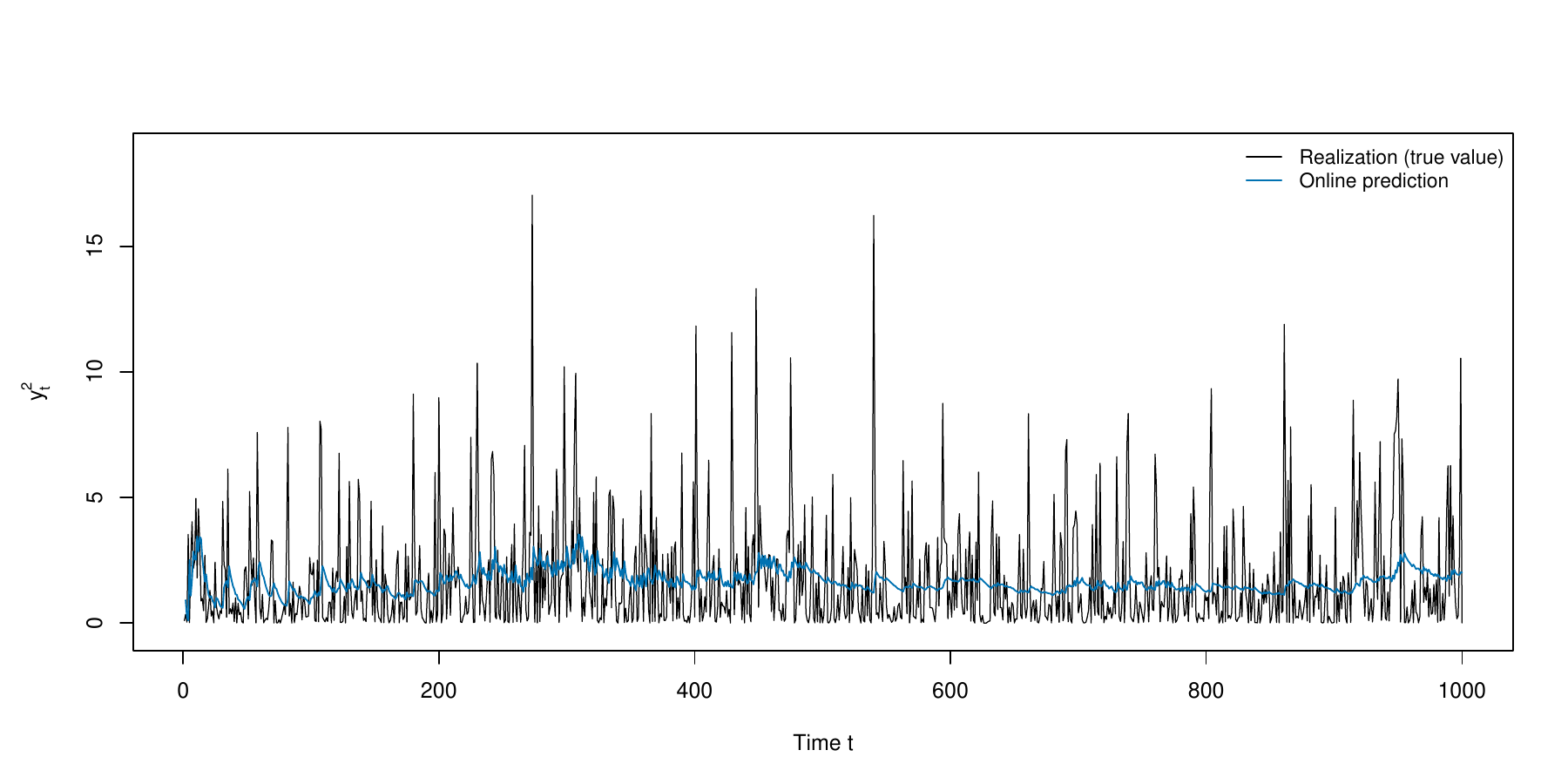}

    \label{fig:bi_y_squared}
\end{figure}

\begin{figure}[H]
    \centering
        \caption{ACF and PACF of Hidden State $h_t$ for MA(1) Bivariate Series}
    \includegraphics[width=1\linewidth]{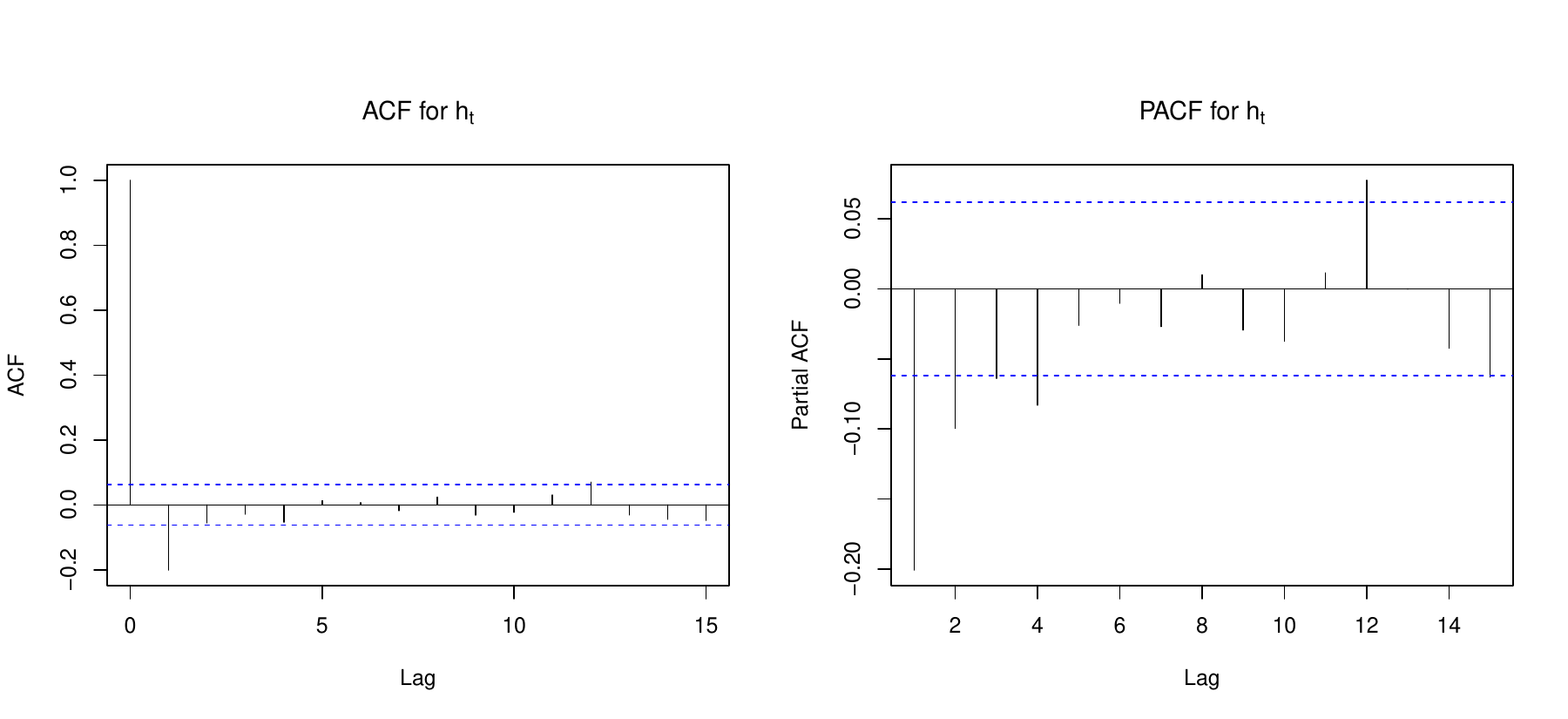}

    \label{fig:Bivariate_h_t acf}
\end{figure}

\begin{figure}[H]
    \centering
        \caption{ACF and PACF of Cell State $c_t$ for MA(1) Bivariate Series}
    \includegraphics[width=1\linewidth]{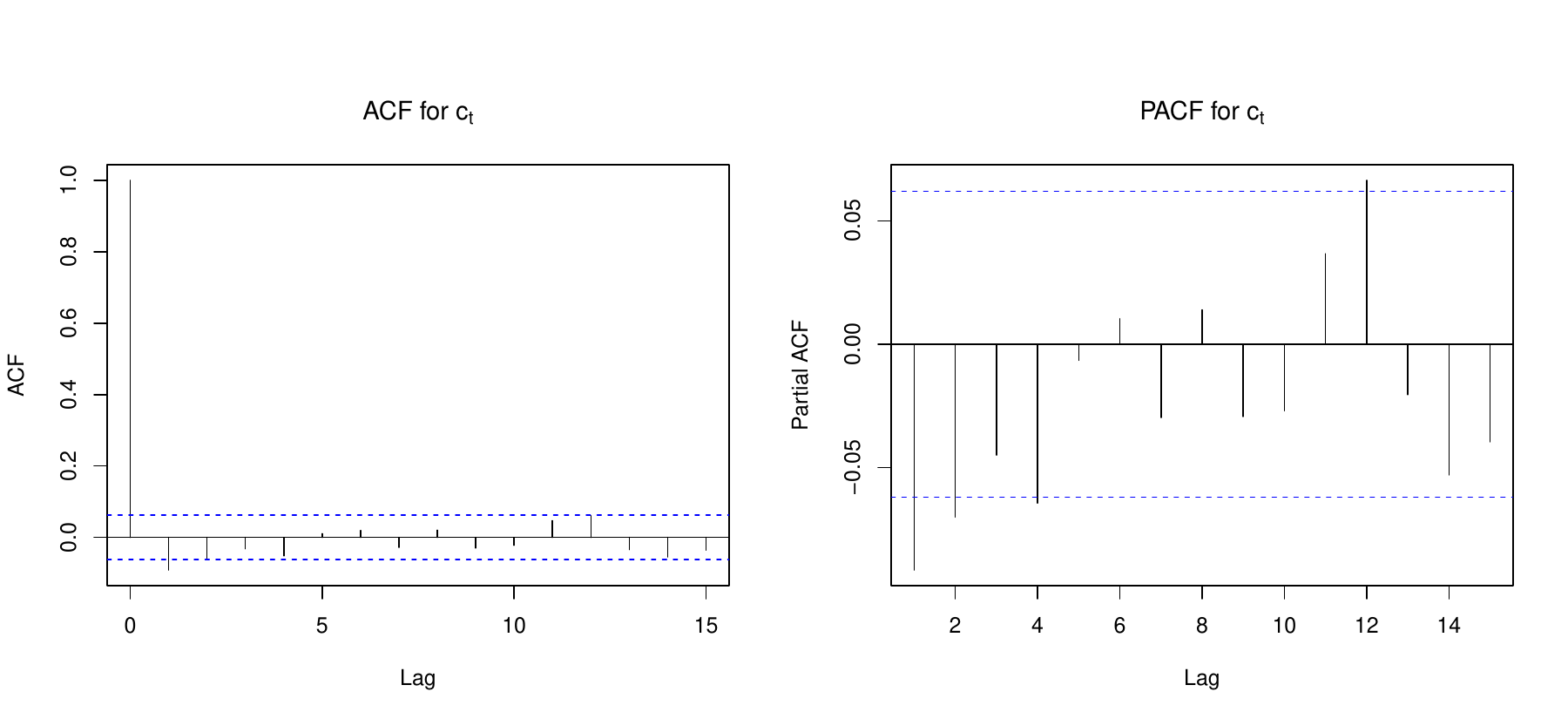}

    \label{fig:Bivariate_c_t acf}
\end{figure}

\clearpage
\textbf{MA(1) Non-invertible Experiment}

\begin{figure}[H]
    \centering
        \caption{Realizations and online one-step predictions of the non-invertible MA(1) series}
    \includegraphics[width=\textwidth]{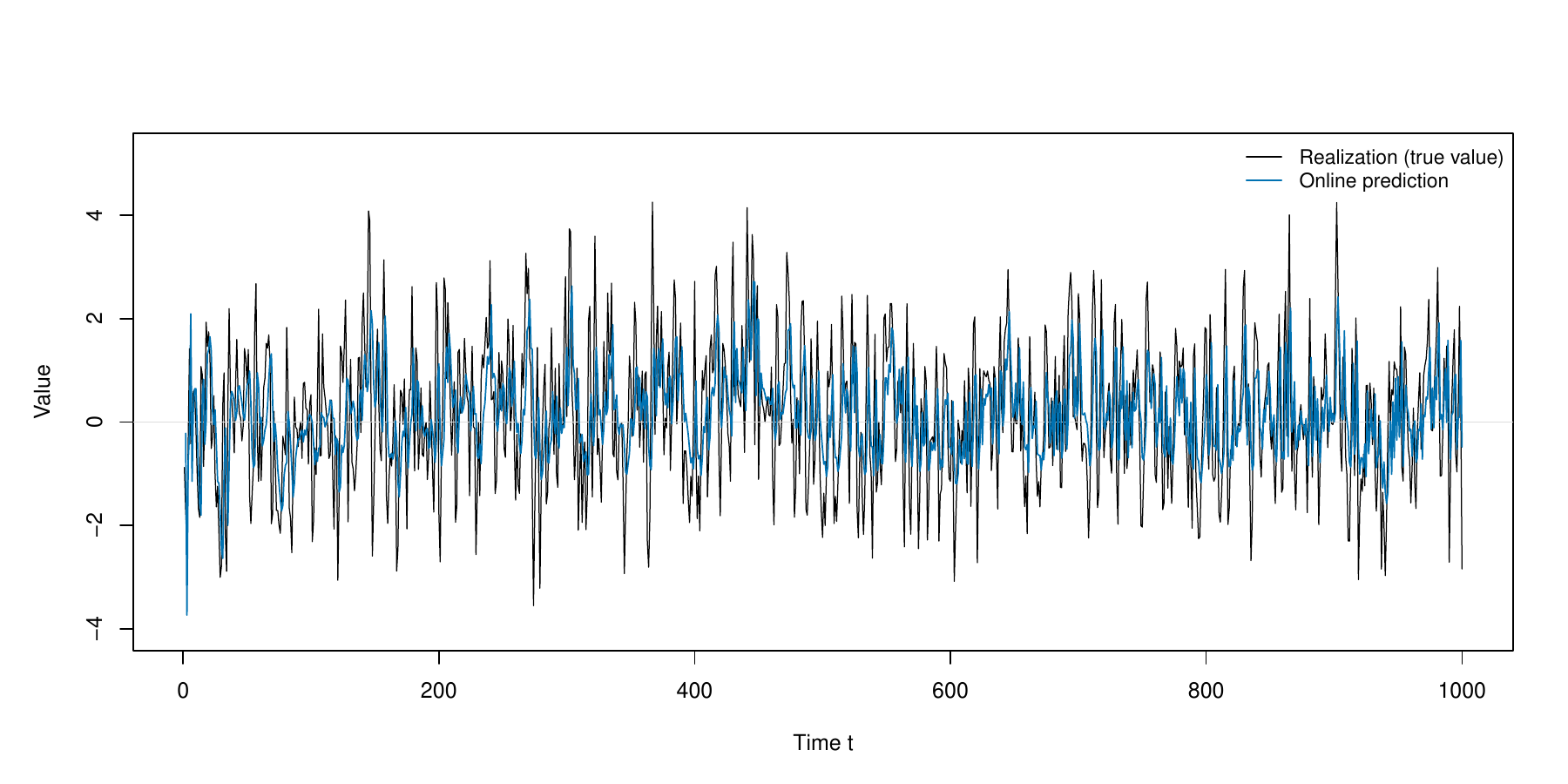}

    \label{fig:ma1_theta1_prediction}
\end{figure}

\begin{figure}[H]
    \centering
        \caption{Sample ACF  and PACF  of the generated MA(1) series}
    \includegraphics[width=\textwidth]{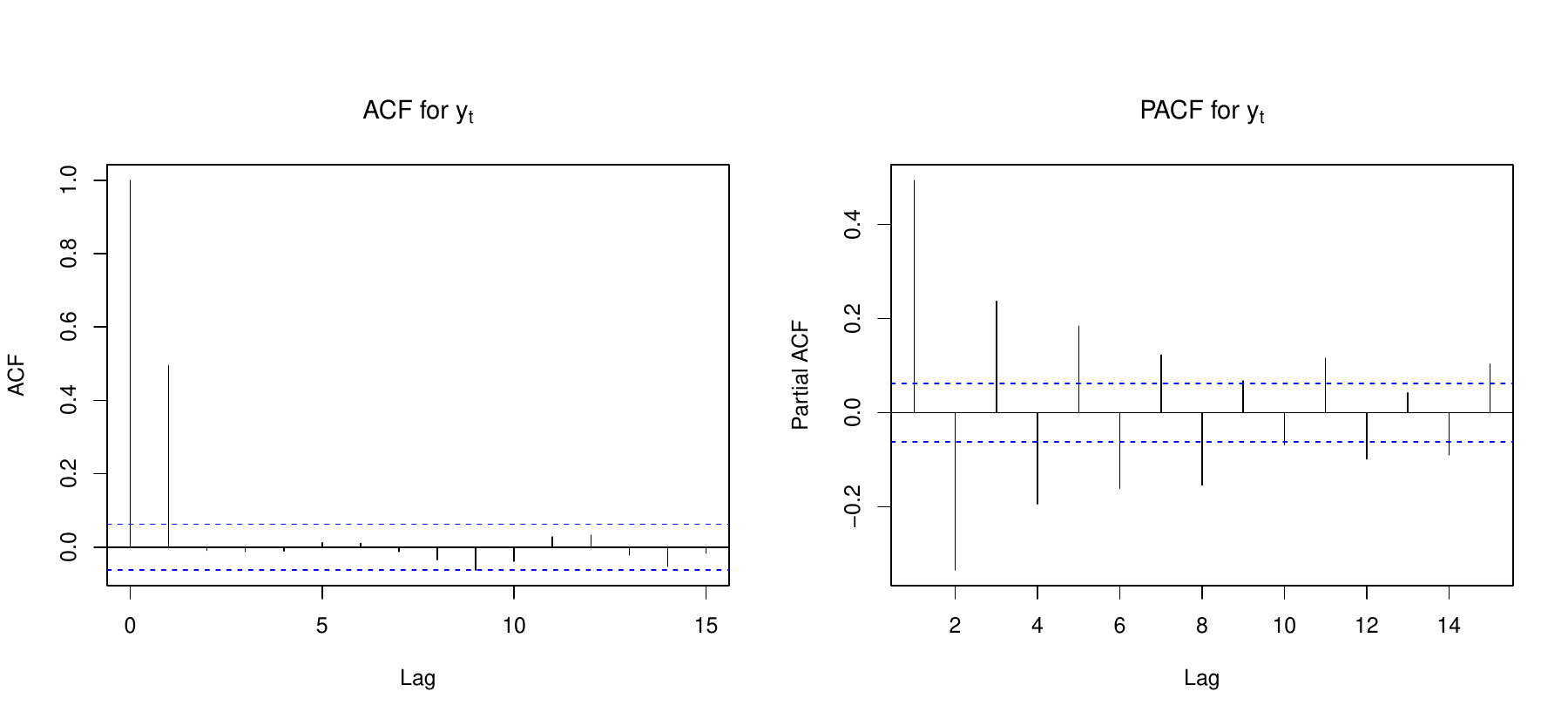}

    \label{fig:ma1_theta1_series_acf_pacf}
\end{figure}

\begin{figure}[H]
    \centering
        \caption{ACF and PACF of the Online Predictions for the Non-invertible MA(1) Series}
    \includegraphics[width=\textwidth]{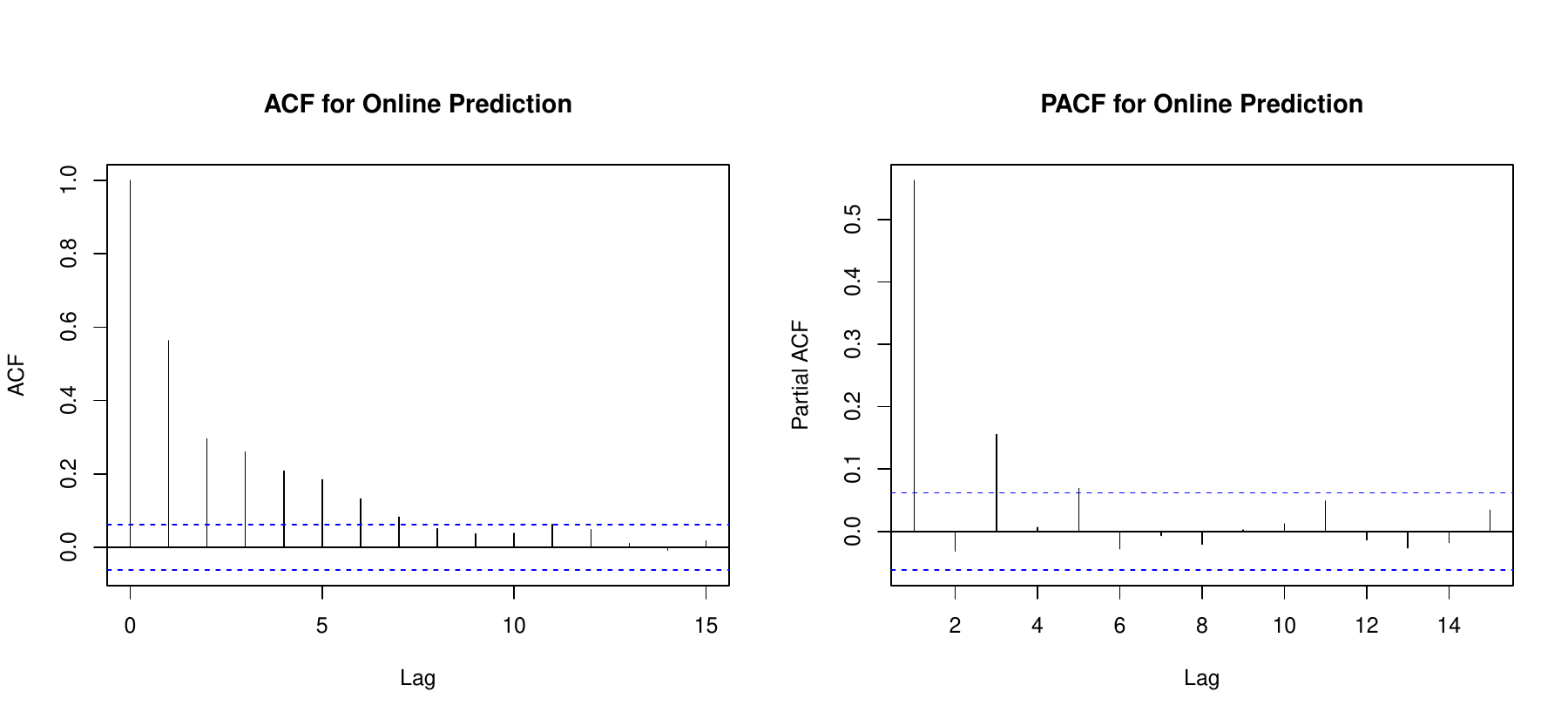}

    \label{fig:ma1_theta1_prediction_acf_pacf}
\end{figure}

\begin{figure}[H]
    \centering
    \caption{Realizations and Joint Online Prediction of the Multitask Approach}
    \includegraphics[width=\textwidth,height=0.78\textheight,keepaspectratio]{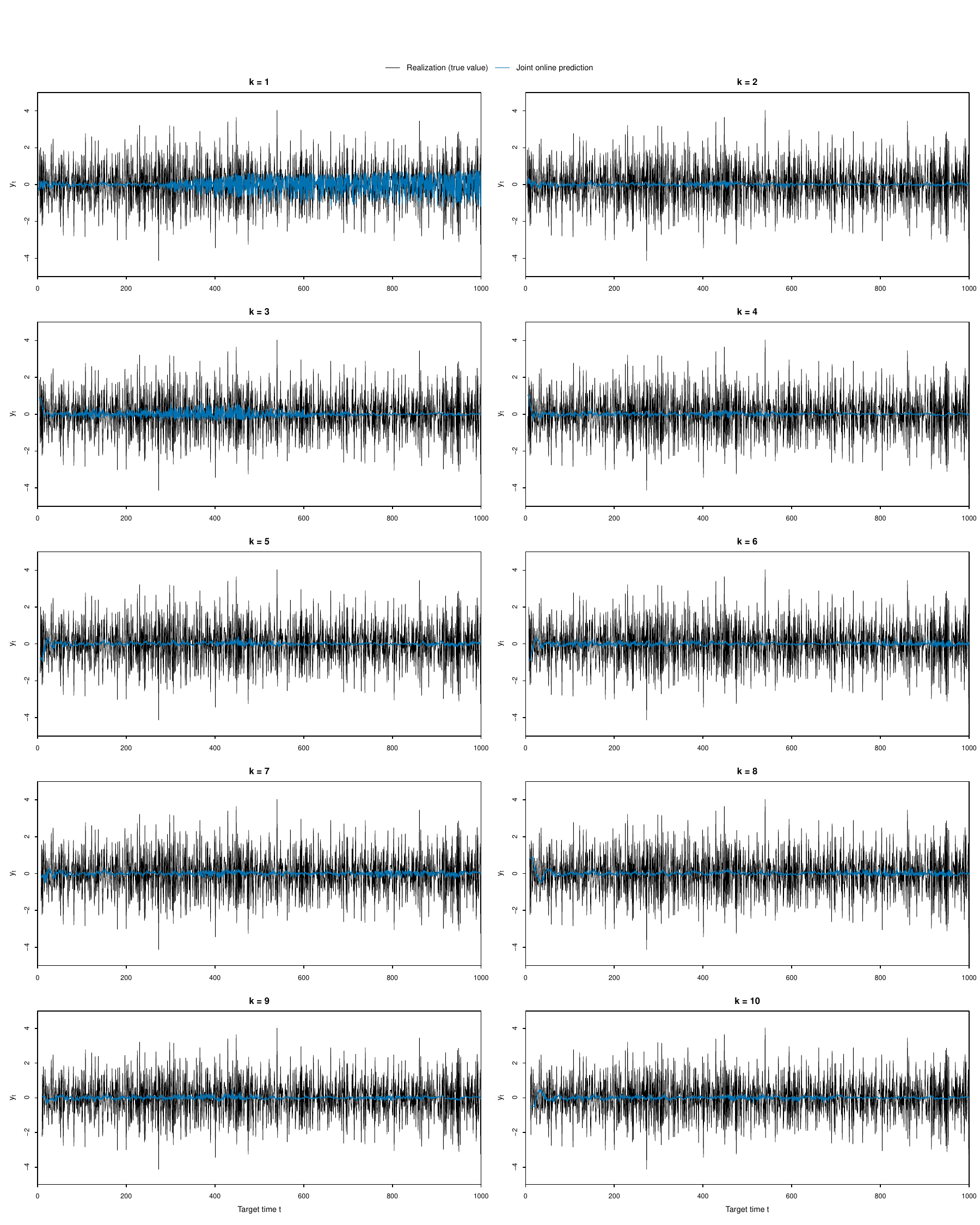}
    \caption*{Realizations (black) and joint online LSTM predictions (blue) of the invertible MA(1) process for horizons $k=1,\ldots,10$. The model has one shared hidden unit and $(\alpha,\beta)=(1.1,0.2)$. All available predictions are shown, for $t=k+1,\ldots,1000$.}
    \label{fig:ma1_joint_horizon_predictions}
\end{figure}
\medskip

\clearpage
\begin{figure}[H]
    \centering
        \caption{ACF and PACF of the Multitask Approach's $h_t$.}
    \includegraphics[width=\textwidth]{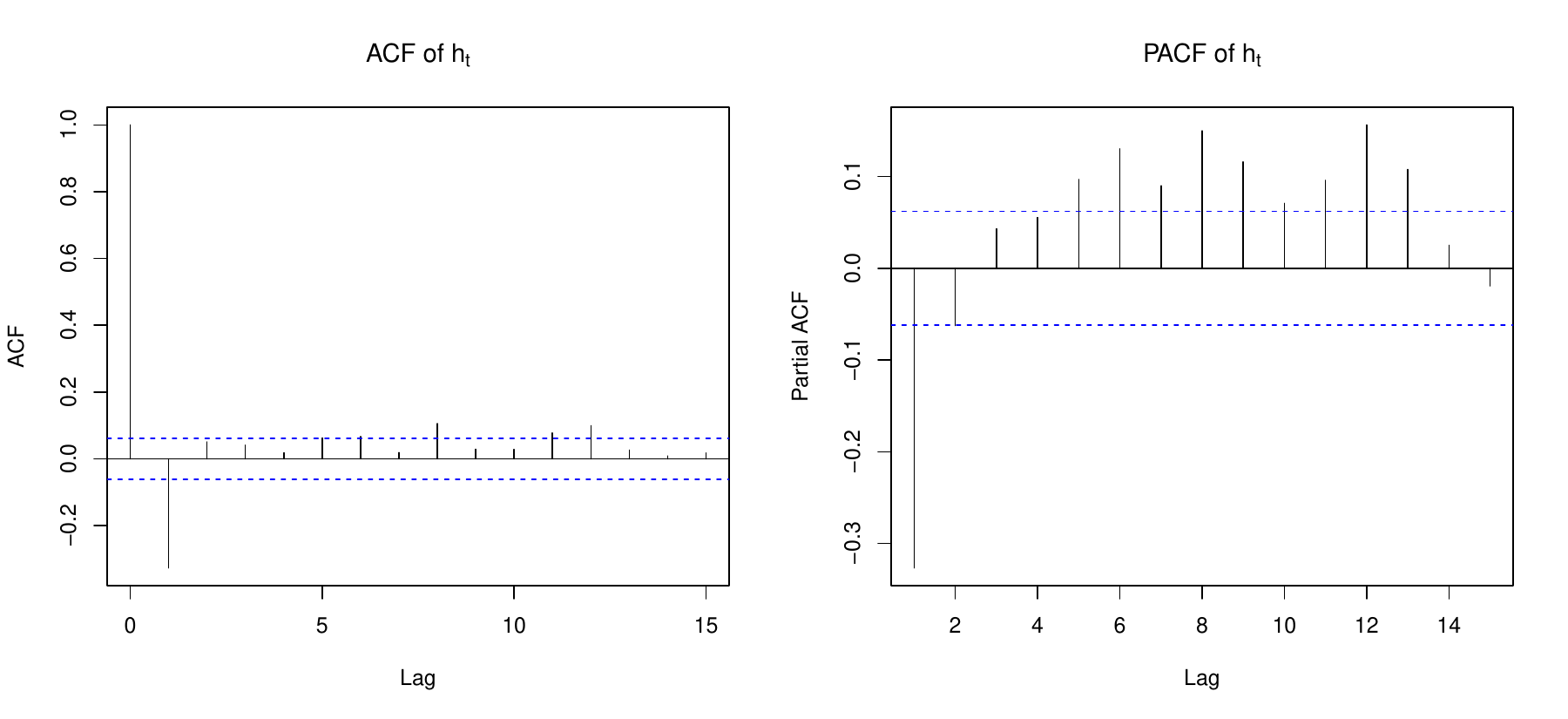}

    \label{fig:ma1_joint_hidden_state_acf_pacf}
\end{figure}
\medskip

\begin{figure}[H]
    \centering
        \caption{ACF and PACF of the Multitask Approach's $c_t$.}
    \includegraphics[width=\textwidth]{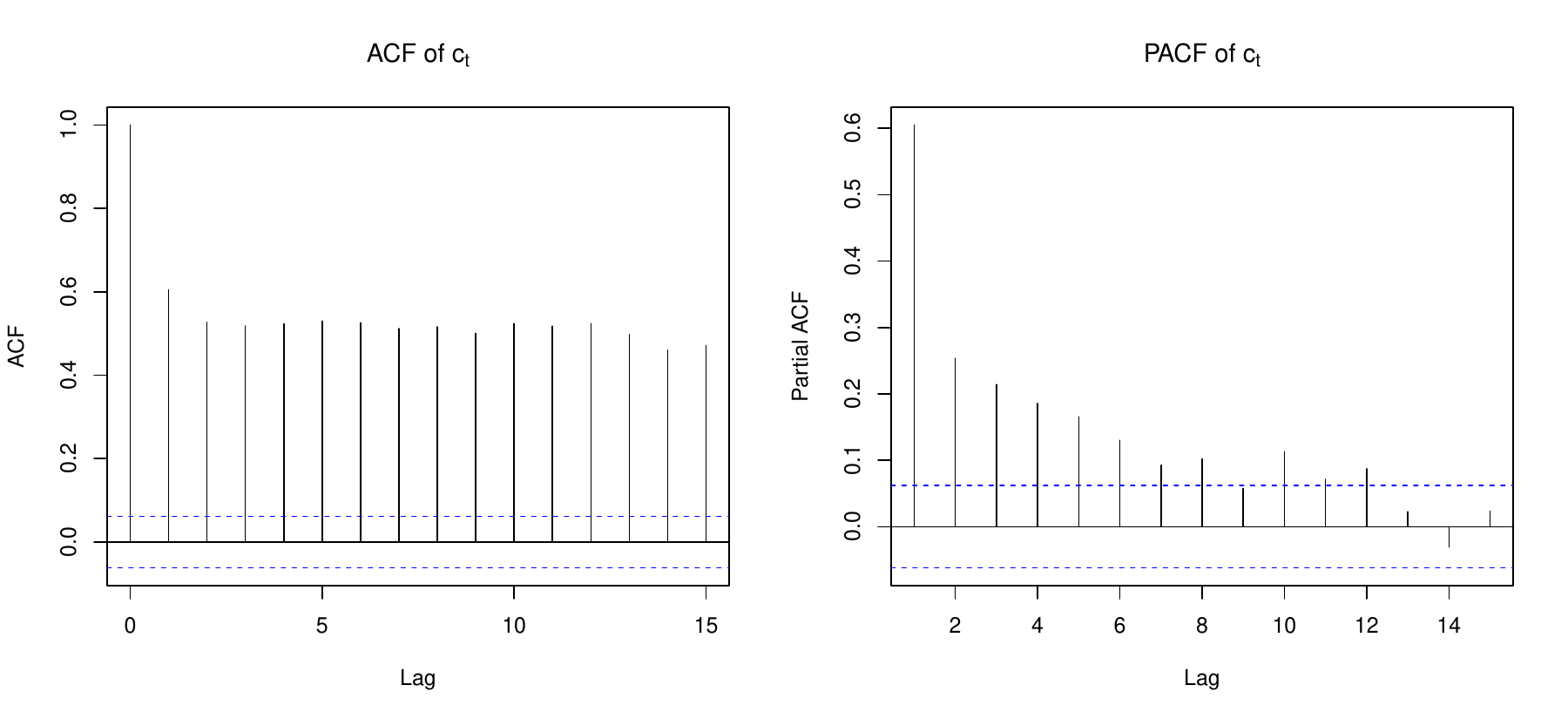}

    \label{fig:ma1_joint_cell_state_acf_pacf}
\end{figure}
\medskip

% BEGIN fixed-parameter recursive MA(1) results (Experiment 48)
\begin{figure}[H]
    \centering
        \caption{Realizations and Joint Online Prediction of the Recursive Approach}
    \includegraphics[width=\textwidth,height=0.78\textheight,keepaspectratio]{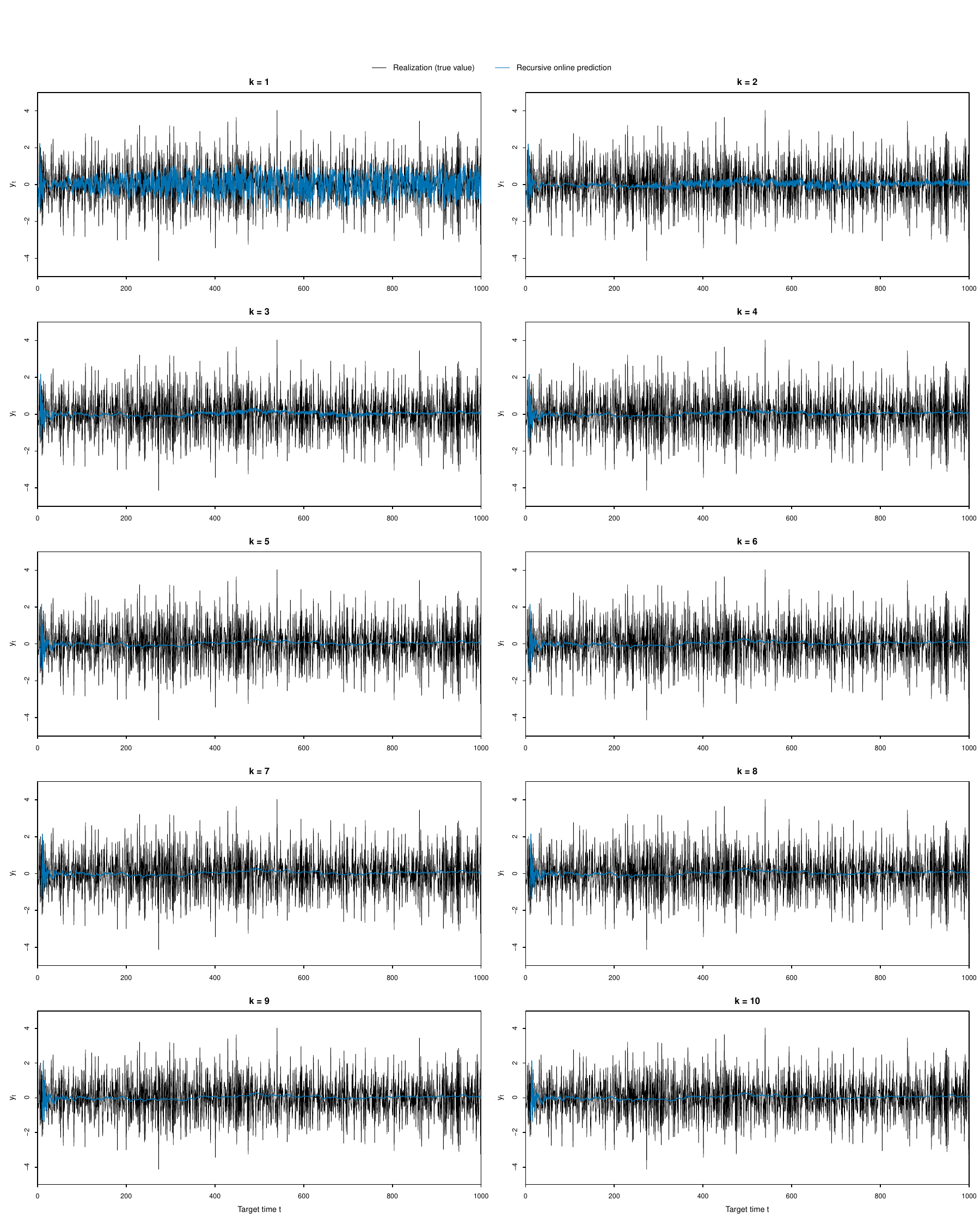}
    %\caption*{Realizations (black) and recursive online predictions (blue), $k=1,\ldots,10$. One shared unit ($c_t,h_t$) and $(\alpha,\beta)=(2.1,0.9)$; parameters remain fixed within each forecast rollout.}
    \label{fig:ma1_recursive_horizon_predictions}
\end{figure}

\end{document}